\documentclass[11pt,a4paper]{article}
\usepackage{jcappub}
\usepackage{bm}
\usepackage{graphicx}
\usepackage{multirow}
\usepackage{amsmath}
\usepackage{amssymb}
\usepackage{tabulary}
\usepackage{hyperref}
\usepackage{color}
\usepackage{xspace}
\usepackage{verbatim}
\usepackage{mathtools}
\usepackage{booktabs}
\usepackage[dvipsnames]{xcolor}
\usepackage{bbold}
\usepackage[caption=false]{subfig}
\allowdisplaybreaks

\newcommand{\vmin}{v_\mathrm{min}}

\newcommand{\vesc}{v_\mathrm{esc}}
\newcommand{\vchi}{\langle \mathbf{v}_\chi \rangle}
\newcommand{\vchihat}{\langle \hat{\mathbf{v}}_\chi \rangle}

\newcommand{\acos}{\cos^{-1}}

\newcommand{\kms}{\textrm{ km s}^{-1}}
\newcommand{\dbd}[2]{\ifmmode \frac{\textrm{d}#1}{\textrm{d}#2}\else $\textrm{d}#1/\textrm{d}#2$\fi}
\newcommand{\erf}{\mathrm{erf}}

\newcommand{\oper}[1]{\hat{\mathcal{O}}_{#1}}

\newcommand{\RE}{R_\oplus}
\newcommand{\pscat}{p_\mathrm{scat}}

\title{Signatures of Earth-scattering in the direct detection of Dark Matter}

\author[a]{Bradley J. Kavanagh,}
\emailAdd{bkavanagh@lpthe.jussieu.fr}
\affiliation[a]{LPTHE, CNRS, UMR 7589, 4 Place Jussieu, F-75252, Paris, France}

\author[b]{Riccardo Catena}
\emailAdd{catena@chalmers.se}
\affiliation[b]{Chalmers University of Technology, Department of Physics, SE-412 96 G\"oteborg, Sweden}

\author[c]{and Chris Kouvaris}
\emailAdd{kouvaris@cp3.sdu.dk}
\affiliation[c]{CP$^3$-Origins, University of Southern Denmark, Campusvej 55, DK-5230 Odense, Denmark}

\abstract{
Direct detection experiments search for the interactions of Dark Matter (DM) particles with nuclei in terrestrial detectors. But if these interactions are sufficiently strong, DM particles may scatter in the Earth, affecting their distribution in the lab. We present a new analytic calculation of this `Earth-scattering' effect in the regime where DM particles scatter at most once before reaching the detector. We perform the calculation self-consistently, taking into account not only those particles which are scattered away from the detector, but also those particles which are deflected towards the detector. Taking into account a realistic model of the Earth and allowing for a range of DM-nucleon interactions, we present the \textsc{EarthShadow} code, which we make publicly available, for calculating the DM velocity distribution after Earth-scattering. Focusing on low-mass DM, we find that Earth-scattering reduces the direct detection rate at certain detector locations while increasing the rate in others. The Earth's rotation induces a daily modulation in the rate, which we find to be highly sensitive to the detector latitude and to the form of the DM-nucleon interaction. These distinctive signatures would allow us to unambiguously detect DM and perhaps even identify its interactions in regions of the parameter space within the reach of current and future experiments. \\[.1cm]
{\footnotesize  \it Preprint: CP3-Origins-2016-050 DNRF90}}

\begin{document}
\maketitle

\section{Introduction}
\label{sec:introduction}

Dark matter (DM) has been a standing problem in modern physics for more than eighty years.  Although it is not yet clear if DM is in the form of new weakly interacting particles or more compact objects like primordial black holes, there is little doubt about its existence. The particle scenario for DM is also well motivated theoretically since almost all theories Beyond the Standard Model have particles that can play the role of DM. For this reason, a great deal of experimental effort has been dedicated to the discovery of DM through the production of these particles in colliders \cite{Mitsou:2013rwa}; the detection of the products of DM decay or annihilation \cite{Gaskins:2016cha}; and the direct detection of recoil events due to DM  scattering off Standard Model particles in underground detectors \cite{Goodman:1984dc, Drukier:1986tm}.

Direct detection constraints are usually derived based on the potential observation of nuclear recoils when DM particles scatter inside an underground detector. Such experiments give stringent constraints on DM-nucleon cross sections over a wide range of DM masses (see e.g.~Refs.~\cite{Agnese:2015nto,Angloher:2015ewa,Akerib:2016vxi, Tan:2016zwf}) and constraints can also be derived from searches for DM-electron interactions~\cite{Essig:2011nj}. However, there is a limit to the mass of DM particles which can be probed by underground detectors. Given that the speed of halo DM particles has an upper bound (set by the escape speed of the Galaxy), sufficiently light DM particles will not have enough energy to create a nuclear recoil above the energy threshold of the experiment, making the scattering event undetectable. With typical energy thresholds of $\mathcal{O}(1 \,\,\mathrm{keV})$, direct detection constraints are substantially weakened for DM masses below a few GeV.

This opens the possibility that if DM is sufficiently light, it could evade current constraints while still interacting strongly enough with nucleons or electrons to have an appreciable probability of scattering in the Earth. The underground scattering of DM particles with ordinary matter has the possibility to distort the DM density and velocity distribution near the Earth's surface, altering the recoil spectrum expected in direct detection experiments. 
 

Light DM is not the only scenario where the stopping effect of underground nuclei or electrons might be important. In two-component DM models, one can envision a dominant weakly interacting component and a small but nevertheless non-negligible  strongly interacting DM component. Since this subdominant component accounts for a tiny fraction of the DM relic abundance, it can easily evade direct detection constraints despite having potentially large DM-nucleon cross section. Such two-component DM models are well motivated since a small strongly interacting DM component could provide the seeds for building up the supermassive black hole of the Galaxy via gravothermal collapse~\cite{Pollack:2014rja}. 

The effect of Earth-scattering on DM particles for contact, long range or dipole interactions has been studied before in~Refs.~\cite{Collar:1992qc,Collar:1993ss,Hasenbalg:1997hs,Foot:2003iv,Zaharijas:2004jv,Sigurdson:2004zp,Mack:2007xj,Cline:2012is,Daci:2015hca,Lee:2015qva}. Additionally, the stopping power of nuclei, bound electrons and free electrons in metallic layers of the Earth was estimated in Ref.~\cite{Kouvaris:2014lpa}. This stopping effect of underground atoms can induce a diurnal modulation in the observed DM recoil signal~\cite{Collar:1992qc,Hasenbalg:1997hs,Kouvaris:2014lpa} (as well as a distinctive top-down asymmetry in directional detectors \cite{Kouvaris:2015laa}). This modulation has an easy explanation: as the Earth is moving with respect to the rest frame of the DM halo, DM particles cross the Earth in larger numbers from the direction opposite to the velocity of the Earth in the Galactic rest frame. Due to the rotation of the Earth around its own axis, the DM particles that come from this preferred direction travel different distances underground during the day in order to reach the detector. Since the probability of particles interacting is larger when they travel larger distances underground, this creates a variation in the DM signal with a period of one day. Similar diurnal modulation effects are produced also in the context of mirror DM~\cite{Foot:2011fh} and dissipative hidden sector DM~\cite{Foot:2014osa,Clarke:2015gqw}. 

The DAMA collaboration has searched for signs of such a diurnally modulated signal, with null results~\cite{Bernabei:2015nia}. If such a modulation due to Earth-scattering were observed in the future, however, it would allow us to unequivocally identify DM as the source of the signal. It may also provide hints about the strength and structure of DM-nucleon interactions. Thus, correctly accounting for these effects may be crucial for identifying and characterising DM.

Previous studies have focused primarily on the stopping effects of DM interactions in the Earth. DM particles which scatter are assumed to be deflected away from the detector or to lose energy without being deflected \cite{Bernabei:2015nia}, leading to an attenuation of the total DM flux. However, such an approach is not generally self-consistent. Any particle which interacts must necessarily be deflected from its trajectory. These particles do not disappear, but instead emerge from the Earth's surface on a different trajectory. Earth-scattering does not simply reduce the flux of DM particles but redistributes this flux, decreasing the density of DM at some points on the Earth while \textit{increasing it} at others. Any self-consistent calculation must conserve the total flux of DM particles and while early Monte-Carlo studies \cite{Collar:1992qc,Collar:1993ss,Hasenbalg:1997hs} did indeed account for this effect, it appears to have been largely neglected in the literature since.

In this paper, we study the deformation of the velocity distribution and recoil spectrum in the case where the DM-nucleon cross section is sufficiently large that DM particles undergo at most one underground scatter before reaching the detector. This `single-scatter' approximation allows us to evaluate the effects of Earth-scattering analytically, including contributions from both attenuation and deflection of particles. We study the size of the diurnal modulation of the DM signal for detectors at different latitudes, using a realistic model for the Earth and exploring a range of DM-nucleon contact interactions in the context of non-relativistic effective field theory \cite{Fitzpatrick:2012ix}. The \textsc{EarthShadow} code used to perform this study (as well as the numerical results produced) are made publicly available online \cite{EarthShadow}. The key result of the paper is illustrated in Fig.~\ref{fig:Rvt}, which shows the amplitude and phase of the daily modulation for low-mass DM in detectors at a number of underground labs. The modulation depends sensitively on the form of the DM-nucleon interaction, which may allow different  interactions to be distinguished if such a modulation is observed in future. 

The paper is organized as follows: In Sec.~\ref{sec:formalism}, we outline the non-relativistic effective field theory framework and enumerate the DM-nucleon interactions which we consider in this work. Sec.~\ref{sec:calculation} constitutes the main calculation of the paper: analytic expressions for the DM velocity distribution at the surface of the Earth when Earth-scattering is accounted for. In Sec.~\ref{sec:effects}, we present numerical results, showing the impact of Earth-scattering on the DM distribution for low- and intermediate-mass DM and for a selection of interaction types. In Sec.~\ref{sec:modulation}, we translate these results into predictions for the diurnal modulation. Finally, we discuss the implications of our results for future DM searches in Sec.~\ref{sec:discussion}, followed by a summary of our conclusions in Sec.~\ref{sec:conclusion}.

\section{NREFT formalism}
\label{sec:formalism}

In this section we briefly review the non-relativistic effective theory of DM-nucleon interactions.~This is the theoretical framework used here to calculate cross-sections, mean free paths and scattering probabilities for DM particles travelling through the Earth.~The theory was formulated in~\cite{Chang:2009yt,Fan:2010gt,Fitzpatrick:2012ix,Fitzpatrick:2012ib,Anand:2013yka}, and subsequently developed in~\cite{Menendez:2012tm,Cirigliano:2012pq,DelNobile:2013sia,Klos:2013rwa,Peter:2013aha,Hill:2013hoa,Catena:2014uqa,Catena:2014hla,Catena:2014epa,Gluscevic:2014vga,Panci:2014gga,Vietze:2014vsa,Barello:2014uda,Catena:2015uua,Schneck:2015eqa,Dent:2015zpa,Catena:2015vpa,Kavanagh:2015jma,D'Eramo:2016atc,Catena:2016hoj,Kahlhoefer:2016eds}.~In the remainder of this paper, we will refer to the above formalism as Non-Relativistic Effective Field Theory, or NREFT for short.

The relevant ``degrees of freedom'' in the effective theory of DM-nucleon interactions are four Hermitian operators~\cite{Fitzpatrick:2012ix}:~($i$ times) the momentum transfer operator,  $i\mathbf{\hat{q}}$, the transverse relative velocity operator, $\mathbf{\hat{v}}^{\perp}$, and the DM particle and nucleon spin operators, $\mathbf{\hat{S}}_\chi$ and $\mathbf{\hat{S}}_N$, respectively.~By construction, $\mathbf{\hat{q}}$ and $\mathbf{\hat{v}}^{\perp}$ obey the orthogonality condition $\mathbf{\hat{v}}^{\perp} \cdot \mathbf{\hat{q}}=0$.~In this context, the Hamiltonian density for DM-nucleon interactions, $\hat{\mathcal{H}}_{\chi N}$, is given by a linear combination of interaction operators, each of which is a Galilean invariant scalar combination of $i\mathbf{\hat{q}}$, $\mathbf{\hat{v}}^{\perp}$, $\mathbf{\hat{S}}_\chi$ and $\mathbf{\hat{S}}_N$.

\begin{table*}[t!]
    \centering
    \begin{tabular}{ll}
	\hline
    \toprule
    \toprule
     \toprule
        $\hat{\mathcal{O}}_1 = \mathbb{1}_{\chi N}$ & $\hat{\mathcal{O}}_9 = i{\bf{\hat{S}}}_\chi\cdot\left({\bf{\hat{S}}}_N\times\frac{{\bf{\hat{q}}}}{m_N}\right)$  \\
        $\hat{\mathcal{O}}_3 = i{\bf{\hat{S}}}_N\cdot\left(\frac{{\bf{\hat{q}}}}{m_N}\times{\bf{\hat{v}}}^{\perp}\right)$ \hspace{2 cm} &   $\hat{\mathcal{O}}_{10} = i{\bf{\hat{S}}}_N\cdot\frac{{\bf{\hat{q}}}}{m_N}$   \\
        $\hat{\mathcal{O}}_4 = {\bf{\hat{S}}}_{\chi}\cdot {\bf{\hat{S}}}_{N}$ &   $\hat{\mathcal{O}}_{11} = i{\bf{\hat{S}}}_\chi\cdot\frac{{\bf{\hat{q}}}}{m_N}$   \\                                                                             
        $\hat{\mathcal{O}}_5 = i{\bf{\hat{S}}}_\chi\cdot\left(\frac{{\bf{\hat{q}}}}{m_N}\times{\bf{\hat{v}}}^{\perp}\right)$ &  $\hat{\mathcal{O}}_{12} = {\bf{\hat{S}}}_{\chi}\cdot \left({\bf{\hat{S}}}_{N} \times{\bf{\hat{v}}}^{\perp} \right)$ \\                                                                                                                 
        $\hat{\mathcal{O}}_6 = \left({\bf{\hat{S}}}_\chi\cdot\frac{{\bf{\hat{q}}}}{m_N}\right) \left({\bf{\hat{S}}}_N\cdot\frac{\hat{{\bf{q}}}}{m_N}\right)$ &  $\hat{\mathcal{O}}_{13} =i \left({\bf{\hat{S}}}_{\chi}\cdot {\bf{\hat{v}}}^{\perp}\right)\left({\bf{\hat{S}}}_{N}\cdot \frac{{\bf{\hat{q}}}}{m_N}\right)$ \\   
        $\hat{\mathcal{O}}_7 = {\bf{\hat{S}}}_{N}\cdot {\bf{\hat{v}}}^{\perp}$ &  $\hat{\mathcal{O}}_{14} = i\left({\bf{\hat{S}}}_{\chi}\cdot \frac{{\bf{\hat{q}}}}{m_N}\right)\left({\bf{\hat{S}}}_{N}\cdot {\bf{\hat{v}}}^{\perp}\right)$  \\
        $\hat{\mathcal{O}}_8 = {\bf{\hat{S}}}_{\chi}\cdot {\bf{\hat{v}}}^{\perp}$  & $\hat{\mathcal{O}}_{15} = -\left({\bf{\hat{S}}}_{\chi}\cdot \frac{{\bf{\hat{q}}}}{m_N}\right)\left[ \left({\bf{\hat{S}}}_{N}\times {\bf{\hat{v}}}^{\perp} \right) \cdot \frac{{\bf{\hat{q}}}}{m_N}\right] $ \\                                                                               
    \bottomrule
     \bottomrule
     \bottomrule
     \hline
    \end{tabular}
    \caption{\textbf{Interaction operators relevant for the present analysis.}~The operator $\mathbb{1}_{\chi N}$ is the identity in the two-particle spin space, and $m_N$ is the nucleon mass.~All interaction operators have the same mass dimension.~In the above expressions, we omit the nucleon index $i$ for simplicity. Operator $\oper{1}$ corresponds to the standard spin-independent (SI) interaction, and $\oper{4}$ corresponds to the standard spin-dependent (SD).} 
    \label{tab:operators}
\end{table*}

Neglecting two-body DM-nucleon interactions, the Hamiltonian density for non-relativistic DM-nucleus scattering is:\footnote{Two-body corrections to Eq.~(\ref{eq:H_chiT}) have been estimated in chiral perturbation theory~\cite{Toivanen:2008zz,Cirigliano:2012pq,Menendez:2012tm,Klos:2013rwa,Hoferichter:2015ipa,Hoferichter:2016nvd}.}
\begin{eqnarray}
\hat{\mathcal{H}}_{\chi T} &\equiv & \sum_{i=1}^{A}  \hat{\mathcal{H}}_{\chi N}^{(i)} =  \sum_{i=1}^{A}  \sum_{\tau=0,1} \sum_{j} c_j^{\tau}\hat{\mathcal{O}}_{j}^{(i)} \, t^{\tau}_{(i)} \,,
\label{eq:H_chiT}
\end{eqnarray}
where $t^0_{(i)}=\mathbb{1}_{2\times 2}$, $t^1_{(i)}=\tau_3$, and $\tau_3$ is the third Pauli matrix.~The matrices $t^\tau_{(i)}$, $\tau=0,1$, act on the $i$-th nucleon isospin space, and $A$ is the mass number of the  target nucleus $T$.~Isoscalar and isovector coupling constants are denoted by $c_j^0$ and $c_j^1$, respectively, and are related to the coupling constants for protons and neutrons as follows:~$c^{p}_j=(c^{0}_j+c^{1}_j)/2$, $c^{n}_j=(c^{0}_j-c^{1}_j)/2$.~All coupling constants have dimension [mass]$^{-2}$.~The interaction operators $\hat{\mathcal{O}}_{j}^{(i)}$ in Eq.~(\ref{eq:H_chiT}) are Galilean invariant scalar combinations of $i\mathbf{\hat{q}}$, $\mathbf{\hat{v}}^{\perp}$, $\mathbf{\hat{S}}_\chi$ and $\mathbf{\hat{S}}_N$.~They are labelled by a nucleon index, $i$, and an interaction index, $j$.~Only fourteen independent Galilean invariant DM-nucleon interaction operators appear in Eq.~(\ref{eq:H_chiT}) if DM has spin less than or equal to 1/2~\cite{Anand:2013yka}.~For spin 1 DM, two additional interaction operators might be constructed~\cite{Dent:2015zpa}.~However, these are only relevant when the interference between the operators $\hat{\mathcal{O}}^{(i)}_4$ and $\hat{\mathcal{O}}^{(i)}_5$, and between $\hat{\mathcal{O}}^{(i)}_8$ and $\hat{\mathcal{O}}^{(i)}_{9}$ is not negligible, and will therefore not be considered further here.~We list the interaction operators relevant for the present analysis in Tab.~\ref{tab:operators}, omitting (from now onwards) the nucleon index $i$ for simplicity.~We point out that the operators $\oper{1}$ and $\oper{4}$ correspond to the familiar spin-independent (SI) and spin-dependent (SD) interactions respectively.

The differential cross-section for DM-nucleus scattering, ${\rm d}\sigma/{\rm d}q^2$, can be written as the sum of eight distinct terms~\cite{Anand:2013yka}:
\begin{align}
\frac{{\rm d}\sigma}{{\rm d}q^2} 
&=\frac{1}{(2J+1) v^2}\sum_{\tau,\tau'} \bigg[ \sum_{k=M,\Sigma',\Sigma''} R^{\tau\tau'}_k  \left(v_{T}^{\perp 2}, {q^2 \over m_N^2} \right) W_k^{\tau\tau'}(q^2) \nonumber\\
&+{q^{2} \over m_N^2} \sum_{k=\Phi'', \Phi'' M, \tilde{\Phi}', \Delta, \Delta \Sigma'} \hspace{-0.4 cm}R^{\tau\tau'}_k\left(v_{T}^{\perp 2}, {q^2 \over m_N^2}\right) W_k^{\tau\tau'}(q^2) \bigg] \,, 
\label{eq:dsigma} 
\end{align}
where $J$ is the spin of the target nucleus, $v$ the DM-nucleus relative velocity, $q$ the momentum transfer, $v_{T}^{\perp 2}=v^2-q^2/(4\mu_{T}^2)$, and $\mu_{T}$ the DM-nucleus reduced mass.~The eight ``DM-response functions'' $R^{\tau\tau'}_k$ depend on $q^2/m_N^2$, $v_{T}^{\perp 2}$, and the isoscalar and isovector coupling constants $c_j^\tau$.~We list them in the Appendix.~The eight nuclear response functions $W_k^{\tau\tau'}$ are quadratic in matrix elements of nuclear charges and currents generated in the scattering of DM by nuclei, and must be computed numerically.~For the elements considered in this investigation (see Tab.~\ref{tab:elements}) we adopt the nuclear response functions found in Ref.~\cite{Catena:2015vpa} using the {\sffamily NuShellX@MSU} shell-model code~\cite{Brown:2014} and phenomenological nucleon-nucleon interactions~\cite{Brown:2001zz}. For concreteness, we will assume isoscalar interactions ($c^p = c^n = c^0/2$) throughout this work.

\section{Earth-scattering calculation}
\label{sec:calculation}

With a formalism for DM-nucleus interactions in hand, we now present the main calculation of the Earth-scattering effect. If DM particles scatter with nuclei in the Earth as they travel underground, they will emerge near the surface of the Earth with a different energy and direction. The energy and direction of DM particles is encoded in the local velocity distribution $f(\mathbf{v})$, meaning that the effect of Earth-scattering will be to induce perturbations in $f(\mathbf{v})$, which will vary depending on the position of the detector on the Earth's surface. We will therefore write the perturbed velocity distribution as $\tilde{f}(\mathbf{v}, \gamma)$, where the angle $\gamma$ describes the position of the detector with respect to the average DM velocity. A more detailed definition of $\gamma$ is given in Sec.~\ref{sec:speeddist} and illustrated in Fig.~\ref{fig:gamma}.

In order to calculate this perturbed DM velocity distribution, we assume that the DM scatters at most once, which we will refer to as the `single scatter' approximation. This is roughly equivalent to assuming $R_\oplus \lesssim \lambda$, where $R_\oplus$ is the Earth's radius and $\lambda$ is the typical mean free path of the DM particles. In this case, the perturbed velocity distribution  contains two contributions:
\begin{equation}\label{eq:fpert}
\tilde{f}(\mathbf{v},\gamma) = f_{\rm A}(\mathbf{v},\gamma) + f_{\rm D}(\mathbf{v},\gamma)\,.
\end{equation}
Here, $f_{\rm A}$ is the \textit{attenuated} population of particles: those particles whose trajectories pass through the detector and which have not scattered before reaching the detector. Instead, $f_{\rm D}$ is the \textit{deflected} population of particles: those whose trajectories did not initially pass through the detector but which have scattered towards the detector during Earth-crossing. 

In our analysis, we neglect gravitational focusing of DM particles by the Earth, which may also lead to percent-level distortions in the local velocity distribution at low $v$ \cite{Lee:2013wza,Kouvaris:2015xga}. We also assume that the time scale over which a DM particle crosses the Earth is negligible compared with the Earth's rotational period; a DM particle with a typical speed of $\sim220 \kms$ will take only $\mathcal{O}(30 \,\, \mathrm{seconds})$ to cross the Earth. This means that we can assume that the detector has a fixed position in calculating the perturbed velocity distribution. 

With these caveats, we now proceed to describe the \textit{free} velocity distribution $f_0(\mathbf{v})$, which one would expect in the absence of Earth-scattering. We then calculate the two contributions to the perturbed distribution shown in Eq.~\ref{eq:fpert}: \textit{attenuation} and \textit{deflection}.

\subsection{Free velocity distribution}
\label{sec:speeddist}

We assume that the free DM velocity distribution is described by the Standard Halo Model (SHM) \cite{Green:2011bv}, which has the following analytic form in the laboratory frame:
\begin{equation} \label{eq:SHM}
f_0(\mathbf{v}) = \frac{1}{N} \exp \left[ -\frac{(\mathbf{v} - \vchi)^2}{2\sigma_v^2}   \right] \times \, \Theta (\vesc - \left|\mathbf{v} - \vchi\right|)\,.
\end{equation}
Here, the normalisation constant $N$ is given by:
\begin{equation}
N = (2\pi \sigma_v^2)^{3/2} \left( \erf \left( \frac{\vesc}{\sqrt{2}\sigma_v}\right) - \sqrt{\frac{2}{\pi}} \frac{\vesc}{\sigma_v} \exp \left( -\frac{\vesc^2}{2\sigma_v^2}   \right) \right)\,,
\end{equation}
and we assume $\sigma_v = 156 \kms$ for the velocity dispersion of the halo, and $\vesc = 533 \kms$ for the local escape speed in the Galactic frame \cite{Piffl:2013mla}. In the SHM, the average DM velocity in the Earth's frame arises from the motion of the Earth through the halo: $\vchi = -\mathbf{v}_e$. We assume a constant value of $v_e = 220 \kms$ for the Earth's speed.

It will be useful in the following sections to have an explicit coordinate expression for $f_0(\mathbf{v})$. In a coordinate system in which the detector lies along the positive $z$-axis (such as that illustrated in Figs.~\ref{fig:attenuation} and \ref{fig:deflection}) we can choose (without loss of generality) $\vchi$ to lie in the $x$-$z$ plane. We can then write the angle between a given DM velocity $\mathbf{v} = (v, \theta, \phi)$ and the average velocity $\vchi$ in terms of the polar coordinates as:
\begin{equation}\label{eq:gamma}
\mathbf{v} \cdot \vchi = v v_e \left( \sin\gamma \sin\theta \cos\phi + \cos\gamma \cos\theta \right)\,.
\end{equation}

\begin{figure}[t!]
\centering
	\includegraphics[width=0.30\textwidth]{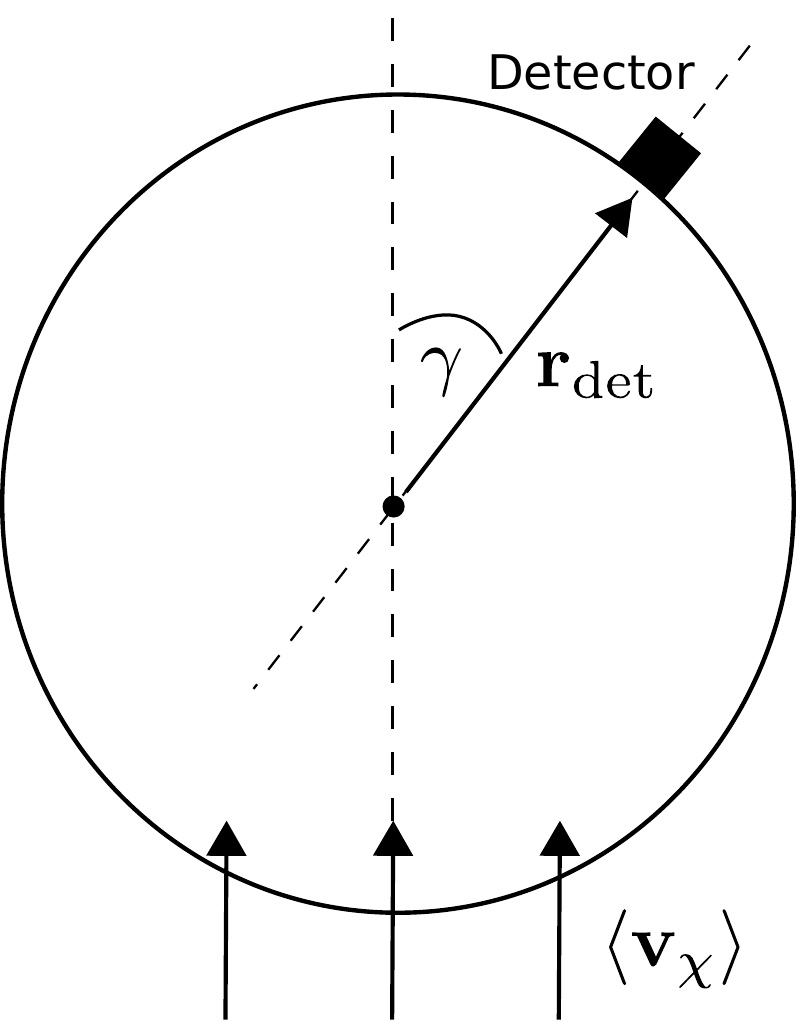}
    \caption{\textbf{Geometry of the detector position $\mathbf{r}_\mathrm{det}$ measured with respect to the mean DM velocity $\vchi$.} The angle between these two vectors is denoted $\gamma$. For $\gamma = 0$, the average DM particle experiences the maximal Earth-crossing distance before reaching the detector. Instead, for $\gamma = \pi$ the Earth crossing distance is minimal. We note that the \textit{true} DM velocities are distributed about the mean value $\vchi$ (according to Eq.~\ref{eq:SHM}). We also remind the reader that the flux of DM particles (before scattering) is spatially uniform, so DM particles enter the Earth's surface at all points (not only along the diameter). } \label{fig:gamma}
\end{figure}

The angle $\gamma$ should be interpreted as the angle between $\vchi$ and the position of the detector $\mathbf{r}_\mathrm{det}$ on the Earth's surface:
$\gamma = \acos \left(\vchihat \cdot \hat{\mathbf{r}}_\mathrm{det}\right)$. This is illustrated in Fig.~\ref{fig:gamma}. An angle of $\gamma = 0$ corresponds to a flux of DM particles which must (on average) cross the entire Earth before reaching the detector. Instead, $\gamma = \pi$ corresponds to the case where the majority of DM particles pass the detector \textit{before} crossing the Earth. We describe how to calculate the value of $\gamma$ for a given position on the Earth in Sec.~\ref{sec:modulation}. 


\subsection{Attenuation}

\begin{figure}[tb!]
\centering
	\subfloat[Attenuation]{\label{fig:attenuation}
	\includegraphics[width=0.33\textwidth]{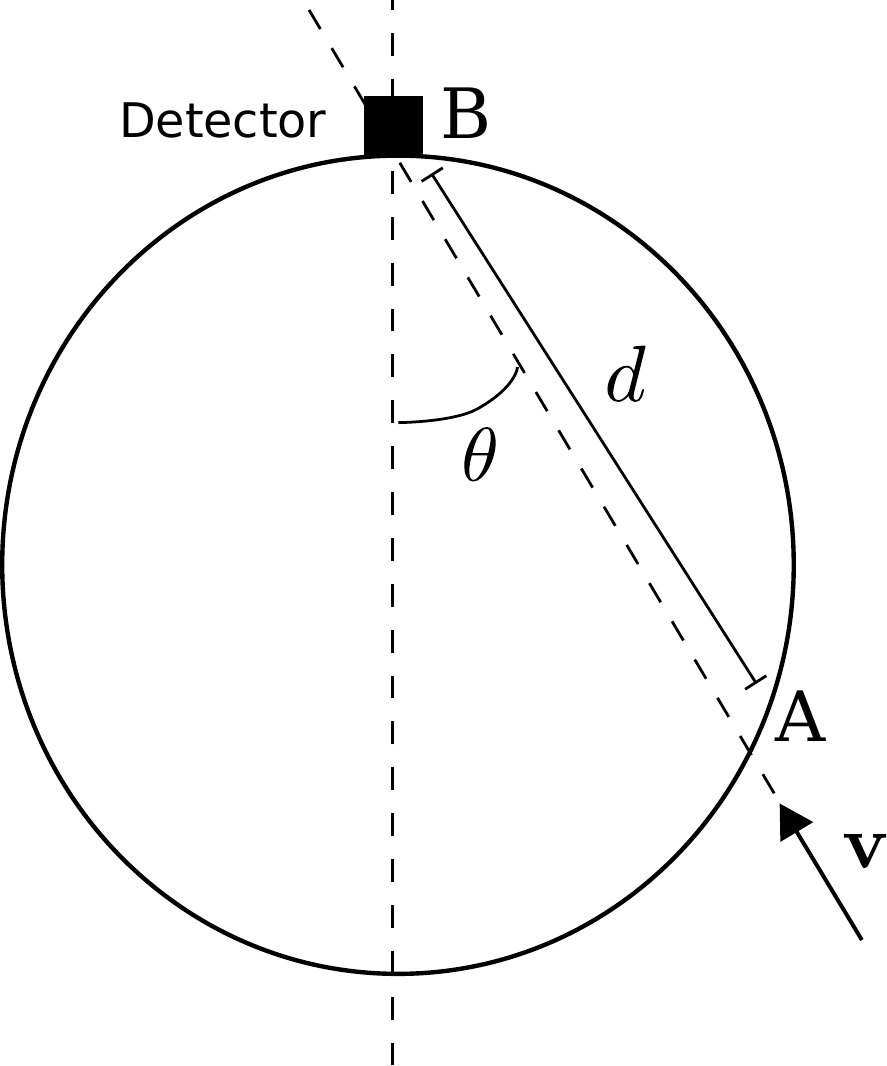}}\hspace{1cm}
	\subfloat[Deflection]{\label{fig:deflection}
	\includegraphics[width=0.30\textwidth]{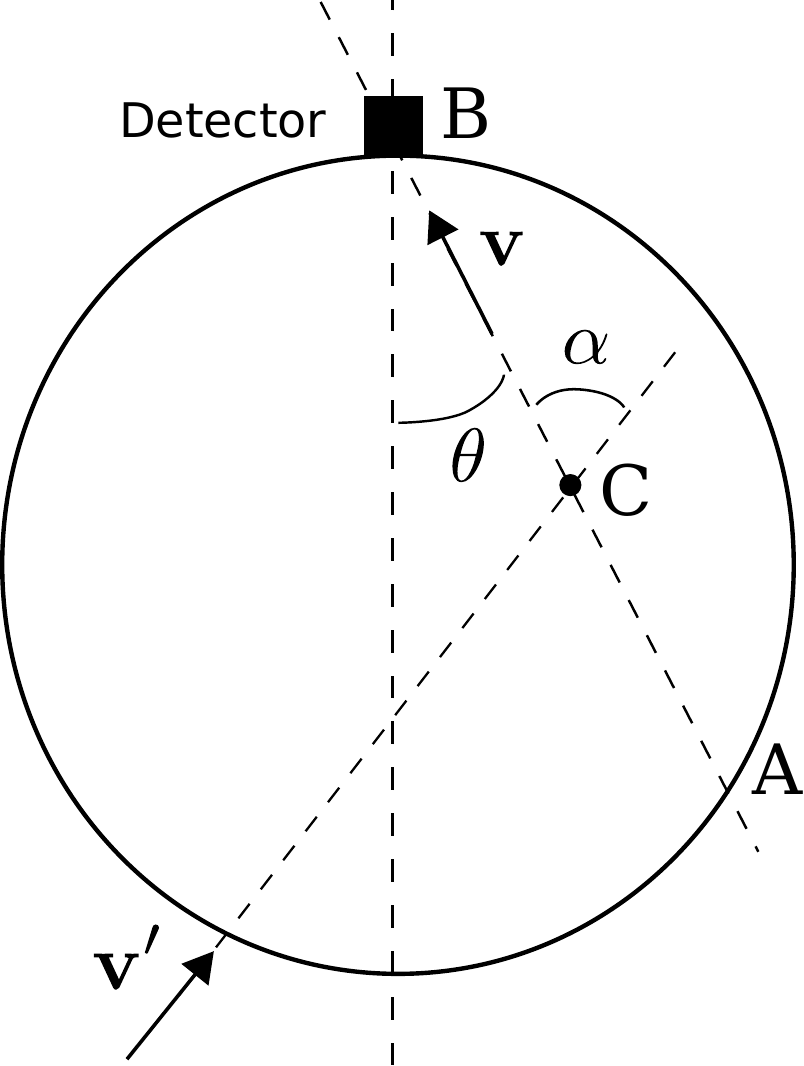}}
    \caption{\textbf{Geometry for the scattering of DM particles in the Earth.} \textbf{(a) Attenuation:} Particles with velocity $\mathbf{v} = (v, \theta,\phi)$ must cross the Earth along the trajectory $\mathrm{AB}$ without scattering in order to arrive at the detector. \textbf{(b) Deflection:} Particles with initial velocity $\mathbf{v}' = (v', \theta', \phi')$ will reach the detector with velocity $\mathbf{v} = (v, \theta, \phi)$ if they interact along the line $\mathrm{AB}$ and scatter through an angle $\alpha$.} \label{fig:geometry}
\end{figure}

We now calculate the effects of attenuation on the DM velocity distribution. In Fig.~\ref{fig:attenuation}, we show the scattering geometry for DM particles impinging on the detector. Particles with an initial velocity $\mathbf{v} = (v, \theta,\phi)$ will reach the detector with that same velocity $\mathbf{v}$ if they do not scatter during their passage through the Earth. If any such particle scatters, however, it will be deflected from the trajectory shown in Fig.~\ref{fig:attenuation} and will no longer reach the detector (assuming that the finite size of the detector can be neglected). Thus, the population of DM particles reaching the detector with velocity $\mathbf{v}$ will be depleted.

The survival probability for a particle with velocity $\mathbf{v}$ is given by:
\begin{align} \label{eq:survival}
\begin{split}
p_\mathrm{surv}(v) &= \exp \left[ - \int_{\mathrm{AB}} \frac{\mathrm{d}l}{\lambda(\mathbf{r}, v)} \right] = \exp \left[ -  \sum_{i}^{\mathrm{species}}   \sigma_i(v) \int_{\mathrm{AB}} n_i(\mathbf{r}) \mathrm{d}l \right]\,,
\end{split}
\end{align}
where the integral is over the path $\mathrm{A}\rightarrow \mathrm{B}$ from the surface of the Earth to the detector, as illustrated in Fig.~\ref{fig:attenuation}. We have also written the mean free path $\lambda$ in terms of the total interaction cross section with Earth species $i$ and the number density of that species: $\lambda(\mathbf{r}, v)^{-1} = \sum_{i}^{\mathrm{species}} \sigma_i (v) n_i(\mathbf{r})$. 

If the number density of particles in the Earth were uniform, then the integral over the DM path in Eq.~\ref{eq:survival} would simply be equal to the Earth-crossing distance, $d$, as shown in Fig.~\ref{fig:attenuation}. This is given by
\begin{align} \label{eq:dcross}
\begin{split}
d(\cos\theta) &= (\RE-l_D) \cos\theta + \sqrt{2\RE l_D - l_D^2 + (\RE-l_D)^2 \cos^2\theta} \\ 
&\approx 
\begin{cases} 
2 \RE \cos\theta & \quad \theta \in [0, \pi/2]\\
0 & \quad \theta \in [\pi/2, \pi]\,,
\end{cases}
\end{split}
\end{align}
where $\RE \approx 6371 \,\, \mathrm{km}$ is the Earth's radius and $l_D$ is the depth of the detector underground. The last line in Eq.~\ref{eq:dcross} is obtained in the limit $l_D \ll \RE < \lambda$. From now on, we assume that this inequality holds (i.e.~that a typical DM particle is unlikely to scatter in the shallow region of the Earth above the detector) and set $l_D$ to zero.

However, the Earth's density is not uniform, so we must account for the radial density profiles of each of the Earth elements $n_i(\mathbf{r}) = n_i(r)$. The distance $l$ from A to some point along the line AB can be written in terms of the distance $r$ of that point from the Earth's centre as,
\begin{equation}
l = \RE \cos\theta \pm \sqrt{r^2 - \RE^2\sin^2\theta}\,.
\end{equation}
With this, we can perform the integral along $\mathrm{AB}$ and calculate an \textit{effective} Earth-crossing distance, $d_\mathrm{eff}$:
\begin{align}\label{eq:deff}
\begin{split}
d_{\mathrm{eff},i}(\cos\theta) &= \frac{1}{\overline{n}_i}\int_{\mathrm{AB}} n_i(\mathbf{r}) \mathrm{d}l  = \ 2\int_{\RE \sin\theta}^{\RE} \frac{n_i(r)}{\overline{n}_i}  \frac{r \, \mathrm{d}r}{\sqrt{r^2 - \RE^2\sin^2\theta}}\,.
\end{split}
\end{align}
Here, we have defined the  number density averaged over the Earth's radius, $\overline{n}_i$:
\begin{align}
\label{eq:nbar}
\begin{split}
\overline{n}_i &= \frac{1}{\RE}\int_{0}^{\RE} n_i(r) \,\mathrm{d}r\,.
\end{split}
\end{align}

The velocity distribution of these \textit{attenuated} particles (i.e.~those which survive the Earth crossing) is therefore related to the free distribution by:
\begin{align}
\label{eq:attenuated}
\begin{split}
f_A(\mathbf{v}, \gamma) &= f_0(\mathbf{v}) \exp \left[ - \sum_{i}^\mathrm{species} \sigma_i(v) \,\overline{n}_i \,d_{\mathrm{eff},i}(\cos\theta) \right] =  f_0(\mathbf{v}) \exp \left[ - \sum_{i}^\mathrm{species} \frac{d_{\mathrm{eff},i}(\cos\theta)}{\overline{\lambda}_i(v)} \right]\,.
\end{split}
\end{align}
 Here, we have defined the average mean free path due to scattering with a given Earth species $i$ as  $\bar{\lambda}_i(v)= [\sigma_i (v) \bar{n}_i]^{-1}$. We consider the contribution of 8 elements, which are summarised in Table~\ref{tab:elements} with tabulated density profiles taken from Ref.~\cite{Lundberg:2004dn} (using data from Refs.~\cite{Geochemistry, Britannica}). We perform the integral in Eq.~\ref{eq:deff} numerically and tabulate the values of $d_\mathrm{eff}$ as a function of $\cos\theta$ for each element. We note that the majority of Earth elements have zero-spin, so we would not expect a large Earth-scattering effect for operators which couple predominantly to the nuclear spin.

\begin{table}[t!]\centering
\begin{tabular}{@{}lllllllll@{}} 
\toprule
\toprule
				 Element 	& A  & $m_A$ [GeV] & $\bar{n}$ [$\mathrm{cm}^{-3}$] & & Core & & Mantle\\
				\hline
	Oxygen & 16 &14.9 & $3.45 \times 10^{22}$ & & 0.0 & & 0.440 \\  
	Silicon & 28 & 26.1 & $1.77 \times 10^{22}$ & & 0.06 & & 0.210\\
	Magnesium & 24 & 22.3 & $1.17 \times 10^{22}$ & & 0.0 & & 0.228  \\
	Iron & 56 & 52.1 & $6.11 \times 10^{22}$ & & 0.855 & & 0.0626 \\
	Calcium & 40 & 37.2 & $7.94 \times 10^{20}$ & & 0.0 & & 0.0253  \\
	Sodium & 23 & 21.4 & $1.47 \times 10^{20}$ & & 0.0 & & 0.0027 \\
	Sulphur & 32 & 29.8 & $2.33 \times 10^{21}$ & & 0.019 & & 0.00025 \\
	Aluminium & 27 & 25.1 & $1.09 \times 10^{21}$ & &  0.0 & & 0.0235 \\
\bottomrule
\bottomrule
\end{tabular}
\caption{\textbf{Summary of Earth elements included in this analysis.}~Next to last and last columns report the mass fractions of each element in the Earth's core and mantle, respectively (values from Tab.~1 in Ref.~\cite{Lundberg:2004dn}). The core and mantle constitute roughly 32\% and 68\% of the Earth's total mass respectively.}
\label{tab:elements}
\end{table}

\subsection{Deflection}

In Fig.~\ref{fig:deflection}, we show the scattering geometry for particles which are deflected away from their initial trajectory and towards the detector. In order to arrive at the detector with velocity $\mathbf{v} = (v, \theta, \phi)$, DM particles with initial velocity $\mathbf{v}' = (v', \theta',\phi')$ must scatter somewhere along the line $\mathrm{AB}$ and be deflected through an angle $\alpha$. The contribution to the DM velocity distribution at velocity $\mathbf{v}$ is then obtained by integrating over the path $\mathrm{AB}$ and over the initial DM velocity distribution $f_0(\mathbf{v}')$.

\subsubsection{Contribution from a single interaction point}

We focus on an interaction region around the point $\mathrm{C}$ in Fig.~\ref{fig:deflection}. We assume this region has infinitesimal length $\mathrm{d}l$ (oriented along the line AB) and an arbitrary cross sectional area $\mathrm{d}S$ (perpendicular to the line AB). The path length of an incoming DM particle crossing this region is then $\mathrm{d}l/\cos\alpha$, meaning that the probability of scattering inside this region is\footnote{For brevity, we suppress the sum over the different elements in the Earth.}
\begin{equation}\label{eq:pscat}
\mathrm{d}\pscat = \frac{\mathrm{d}l}{\lambda(\mathrm{r}, v')\cos\alpha}\,.
\end{equation}
The rate of particles entering the interaction region and scattering into the direction $\mathbf{v}$ is then,
\begin{align}\label{eq:flux_in}
\begin{split}
\bigg[n_\chi  \,f_0(\mathbf{v}')  \,\mathbf{v}' \cdot \mathrm{d}\mathbf{S} \, \mathrm{d}^3\mathbf{v}' \bigg] \bigg[ \, \mathrm{d}\pscat \, P(\mathbf{v}' \rightarrow \mathbf{v})  \, \mathrm{d}^3\mathbf{v} \bigg]\,,
\end{split}
\end{align}
where the first bracket is the rate of particles entering the interaction region, with the surface vector $\mathrm{d}\mathbf{S}$ pointing along $\mathbf{v}$ (i.e. along the line from A to B). The number density of DM particles is denoted by $n_\chi$. By convention, we will keep $n_\chi$ constant before and after scattering, so that changes in the overall density are instead encoded in the velocity distribution. The second term in brackets is the probability of scattering from $\mathbf{v}'$ to $\mathbf{v}$ (i.e.~$\mathrm{d}\pscat$ is the probability of scattering, and $P(\mathbf{v}' \rightarrow \mathbf{v})$ is the probability of scattering into a particular velocity given that the particle has scattered).

The rate of deflected particles leaving the interaction region with velocity $\mathbf{v}$ is defined (in analogy to the incoming rate) to be,
\begin{equation}\label{eq:flux_out}
n_\chi f_D(\mathbf{v}, \gamma)  \,\mathbf{v}\cdot\mathrm{d}\mathbf{S} \, \mathrm{d}^3\mathbf{v}\,.
\end{equation}
The distribution $f_D(\mathbf{v})$ is not constrained to be normalised to unity. Instead, the overall normalisation is determined by setting equal the incoming and outgoing fluxes of particles in the scattering region. Equating Eqs.~\ref{eq:flux_in} and \ref{eq:flux_out}, we obtain an expression for the contribution to $f_D(\mathbf{v}, \gamma)$ from interactions at point $\mathrm{C}$,
\begin{equation}\label{eq:fD1}
f_D(\mathbf{v},\gamma) = \frac{\mathrm{d}l}{\lambda(\mathbf{r},v')}\frac{v'}{v} f_0(\mathbf{v}') P(\mathbf{v}' \rightarrow \mathbf{v}) \, \mathrm{d}^3 \mathbf{v}'\,,
\end{equation}
where we have used the fact that $\mathbf{v}' \cdot \mathrm{d}\mathbf{S} = v' \cos\alpha \, \mathrm{d}S$ and $\mathbf{v} \cdot \mathrm{d}\mathbf{S} = v\,\mathrm{d}S$.

\subsubsection{Contribution from all interaction points}

We now integrate over all points which give a contribution to the deflected velocity distribution. For a fixed final velocity $\mathbf{v}$, we must integrate over the line $\mathrm{AB}$ in Fig.~\ref{fig:deflection}. We assume that the DM particles scatter at most once, meaning that the initial velocity distribution $f(\mathbf{v}')$ is not distorted by the passage through the Earth. We therefore take $f(\mathbf{v}')$ to be spatially uniform. The integration over the path length $l$ along $\mathrm{AB}$ reduces to the same form given in Eq.~\ref{eq:deff}. The contribution of a given initial DM velocity $\mathbf{v}'$ to the deflected distribution is then:
\begin{equation}\label{eq:fD2}
f_D(\mathbf{v},\gamma) = \frac{\mathrm{d}_\mathrm{eff}(\cos\theta)}{\overline{\lambda}(v')}\frac{v'}{v} f_0(\mathbf{v}') P(\mathbf{v}' \rightarrow \mathbf{v}) \, \mathrm{d}^3 \mathbf{v}'\,.
\end{equation}

\subsubsection{Kinematics}

We now consider how to calculate $P(\mathbf{v}' \rightarrow \mathbf{v})$. The deflection angle $\alpha$ is fixed geometrically by the directions of DM particles incoming and outgoing from point $\mathrm{C}$:\footnote{We remind the reader that the primed angles $(\theta',\phi')$ describe the incoming DM particle direction (with the positive $z$-axis oriented from the centre of the Earth to the detector), while the unprimed angles $(\theta,\phi)$ describe the final DM direction (in the same coordinate frame).}   
\begin{equation}\label{eq:cosalpha}
\cos\alpha = \sin\theta \sin\theta' \cos(\phi - \phi') + \cos\theta \cos\theta'\,.
\end{equation}
For a given deflection angle $\alpha$, DM mass $m_\chi$ and target nuclear mass $m_A$, the ratio $v'/v$ is fixed by kinematics:
\begin{equation} \label{eq:kappa}
\frac{v'}{v} = \frac{m_\chi + m_A}{m_\chi \cos\alpha \pm \sqrt{m_A^2 - m_\chi^2 \sin^2\alpha}} \equiv \kappa^\pm(\alpha, m_\chi, m_A)\,.
\end{equation}
We note that for $m_\chi \leq m_A$, the only valid solution is $v' = \kappa^+ v$ for all values of $\alpha$. Instead, for $m_\chi > m_A$, we require $\cos\alpha > (1 - m_A^2/m_\chi^2)^{1/2}$, in which case both solutions are valid: $v' = \kappa^\pm v$.

We can now write the scattering probability as
\begin{align}\label{eq:pv}
\begin{split}
P(\mathbf{v}' \rightarrow \mathbf{v}) &= \frac{1}{v^2} \sum_{a=\pm} \delta(v - v'/\kappa^a)P^a(\hat{\mathbf{v}}) \,.
\end{split}
\end{align}
The factor of $1/v^2$ is required to ensure correct normalisation and we implicitly assume that the negative-sign solution in the sum is included only in the case that $m_\chi > m_A$. The term $P^\pm(\hat{\mathbf{v}})$ is now simply the probability distribution for the \textit{direction} of $\mathbf{v}$. The deflection of the DM particle is azimuthally symmetric, so we can write:
\begin{align}\label{eq:ppm}
\begin{split}
P^\pm(\hat{\mathbf{v}}) &=\frac{1}{2\pi}P^\pm(\cos\alpha)\,.
\end{split}
\end{align}

The probability distribution of $\cos\alpha$ is given by:
\begin{align}
\begin{split}
P^\pm(\cos\alpha) &=  \frac{1}{\sigma}\left.\frac{\mathrm{d}\sigma}{\mathrm{d}\cos\alpha}\right|_\pm\,,
\end{split}
\end{align}
where $\sigma$ is the total cross section. With this definition, the distribution of $\cos\alpha$ is normalised such that:
\begin{equation}
\int \left(P^+(\cos\alpha) + P^-(\cos\alpha) \right) \, \mathrm{d}\cos\alpha = 1\,,
\end{equation}
where the integral is over all kinematically allowed values of $\cos\alpha$, i.e.:
\begin{equation}\label{eq:cosalpharange}
\cos\alpha \in \begin{cases}
[-1, 1] & \quad \text{ for } m_\chi \leq m_A\,,\\
\left[\sqrt{1- m_A^2/m_\chi^2}, 1\right] & \quad \text{ for } m_\chi > m_A\,.
\end{cases}
\end{equation}
We again emphasise that the $P^-$ term is only included for $m_\chi > m_A$.

\begin{figure*}[t]
\includegraphics[width=0.32\textwidth]{{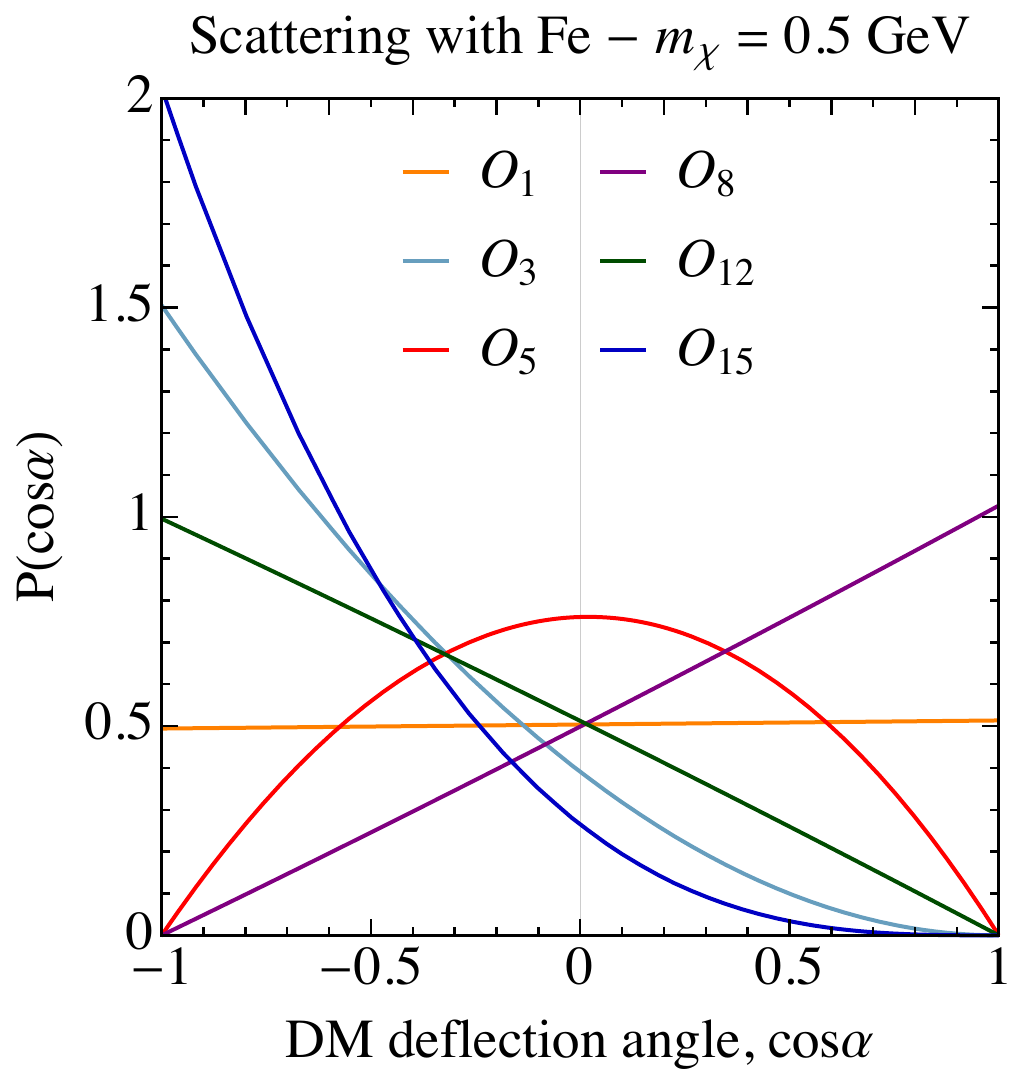}}
\includegraphics[width=0.32\textwidth]{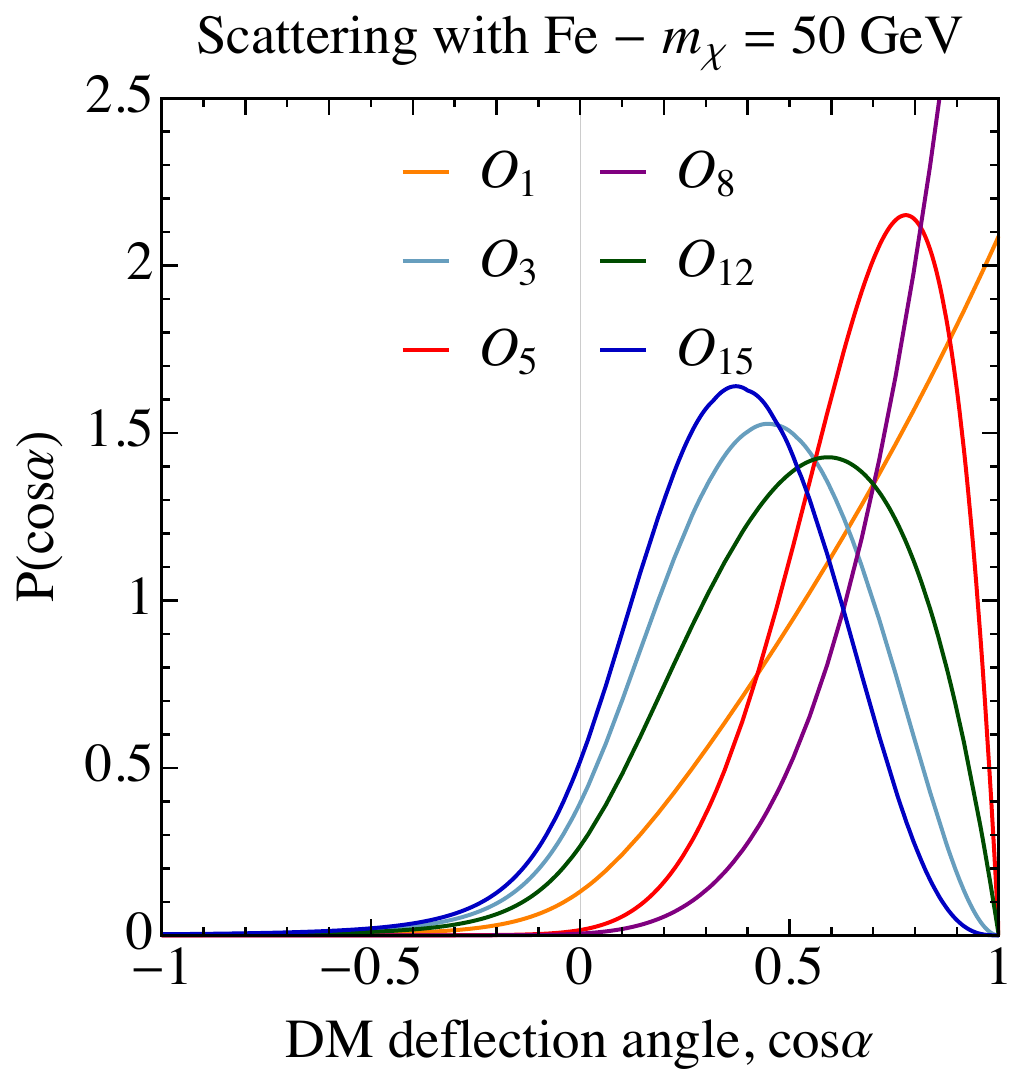}
\includegraphics[width=0.315\textwidth]{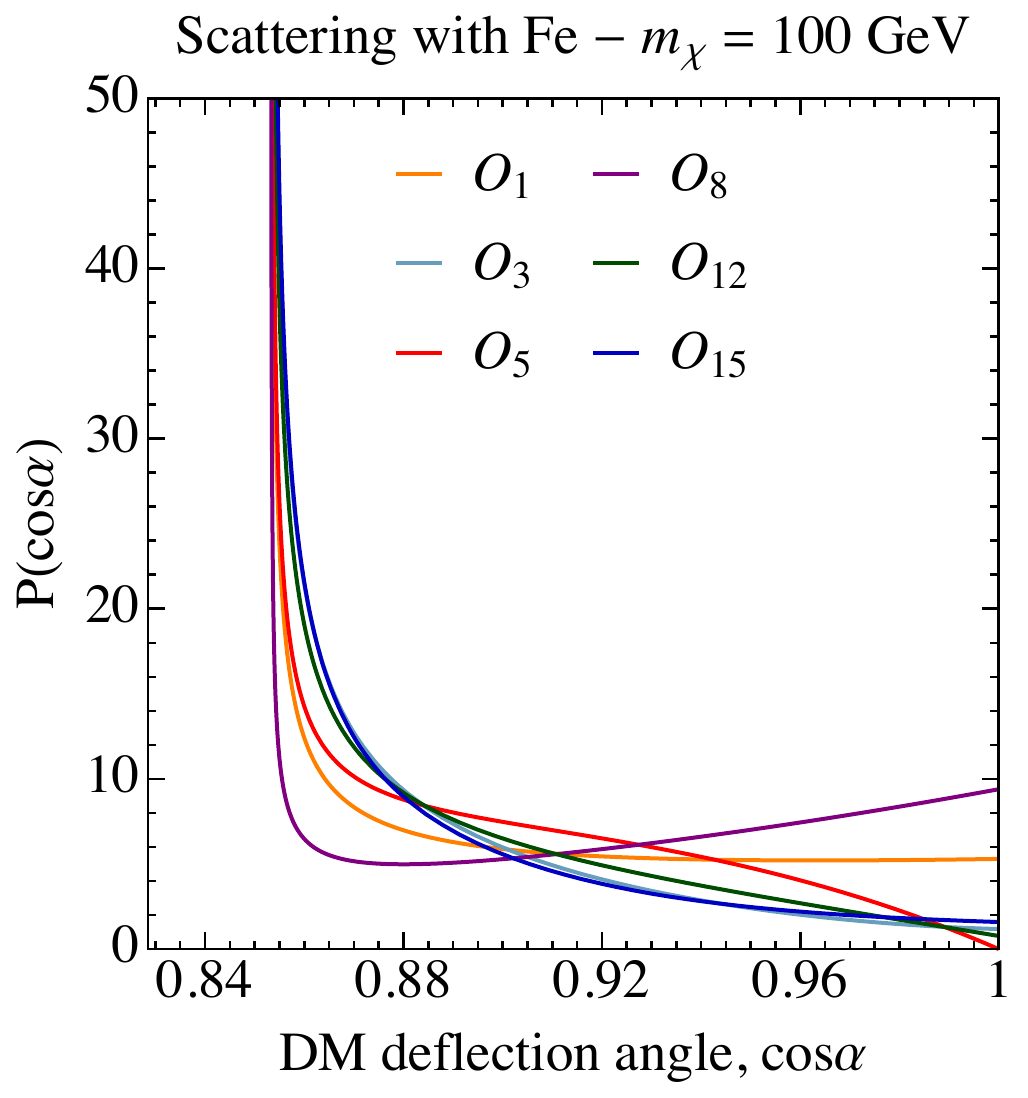}
\caption{\textbf{Probability distributions for the DM deflection angle $\alpha$.} Within each panel we show $P(\cos\alpha)$ for 6 different operators from Table~\ref{tab:operators}, while each panel shows a different DM particle mass: $m_\chi = $ 0.5 GeV (\textbf{left}),  $m_\chi = $ 50 GeV (\textbf{centre}), and $m_\chi = $ 100 GeV (\textbf{right}). We fix the DM speed at $v = 220 \kms$ and consider scattering with Iron nuclei ($m_\mathrm{Fe} \approx 52 \, \, \mathrm{GeV}$). DM particles which scatter through only a small angle relative to their incoming direction have $\cos\alpha \rightarrow 1$. Note the different scale on the x-axis in the right panel.}\label{fig:Pcosalpha}
\end{figure*}

The differential cross section with respect to $\cos\alpha$ can be obtained by a change of variables:
\begin{equation} \label{eq:dsigmadcosalpha}
\left. \dbd{\sigma}{\cos\alpha}\right|_\pm =  \left.\left(\dbd{E_R}{\cos\alpha} \dbd{\sigma}{E_R} \right)\right|_\pm \,.
\end{equation}
Again, from kinematics, we can calculate $E_R$ as a function of $\cos\alpha$:
\begin{align}\label{eq:ER}
\begin{split}
E^{\pm}_R(\cos\alpha) &= \frac{m_\chi^2 v'^2}{(m_\chi + m_A)^2}\left( m_\chi \sin^2\alpha  + m_A \mp \cos\alpha\sqrt{m_A^2 - m_\chi^2 \sin^2\alpha}\right)\,.
\end{split}
\end{align}
The derivative appearing in Eq.~\ref{eq:dsigmadcosalpha} then follows straightforwardly. It is now apparent why we have maintained the superscript $\pm$ throughout: the recoil energy as a function of $\cos\alpha$ depends on which of the two kinematic solutions is being considered. As a result, we obtain a different distribution for $\cos\alpha$ in each case. The distributions $P^\pm(\cos\alpha)$ can be calculated from a given differential cross section using the prescription we have just outlined. In Fig.~\ref{fig:Pcosalpha}, we plot the distributions of deflection angles for a number of the operators described in Sec.~\ref{sec:formalism}.

Substituting Eqs.~\ref{eq:pv} and \ref{eq:ppm} into Eq.~\ref{eq:fD2}, we obtain:
\begin{align}\label{eq:fD3}
\begin{split}
f_D(\mathbf{v},\gamma) &= \frac{1}{2\pi}\frac{\mathrm{d}_\mathrm{eff}(\cos\theta)}{\overline{\lambda}(v')}\frac{v'}{v^3} f_0(\mathbf{v}')\times \sum_{a=\pm} \delta(v - v'/\kappa^a)P^a(\cos\alpha) \, \mathrm{d}^3 \mathbf{v}'\,.
\end{split}
\end{align}
We can rewrite the $\delta$-function in terms of $v'$,
\begin{equation}
\delta(v - v'/\kappa^\pm) = \kappa^\pm \delta(v' - \kappa^\pm v)\,,
\end{equation}
and then perform the integral over all incoming velocities $\mathbf{v}'$:\footnote{We remind the reader that $\kappa^\pm$ depends on $\cos\alpha$, which in turn depends on the incoming and outgoing DM angles (see Eq.~\ref{eq:cosalpha}). This means that $\kappa^\pm$ must remain inside the integral over $\hat{\mathbf{v}}'$.}
\begin{equation}\label{eq:fD4}
f_D(\mathbf{v},\gamma) = \sum_{a=\pm} \int  \mathrm{d}^2\hat{\mathbf{v}}'  \frac{d_\mathrm{eff}(\cos\theta)}{\overline{\lambda}(\kappa^a v)} \frac{(\kappa^a)^4}{2\pi} f_0(\kappa^a v, \hat{\mathbf{v}}') P^a(\cos\alpha)\,.
\end{equation}

\subsubsection{Summing over elements}

We now reintroduce the sum over the different species in the Earth, to give the final deflected velocity distribution:
\begin{equation}\label{eq:fD4}
f_D(\mathbf{v},\gamma) = \sum_{i}^{\mathrm{species}}\sum_{a=\pm} \int  \mathrm{d}^2\hat{\mathbf{v}}' \,\, \frac{d_{\mathrm{eff},i}(\cos\theta)}{\overline{\lambda}_i(\kappa_i^a v)} \frac{(\kappa^a_i)^4}{2\pi} f_0(\kappa^a_i v, \hat{\mathbf{v}}') P^a_i(\cos\alpha)\,.
\end{equation}
We note that many of the terms in Eq.~\ref{eq:fD4} have now acquired an $i$ index: the kinematic term $\kappa^\pm$ depends on the target nuclear mass; the effective Earth-crossing distance $d_\mathrm{eff}$ depends on the density profile of the species; and both the mean free path $\overline{\lambda}$ and distribution of $\cos\alpha$ depend on the DM-nucleus cross section. We also emphasise that for some species, we will need to include both terms in the $a=\pm$ sum, while for others, only the $a=+$ term is required (depending on the DM mass). 

\subsection{DM speed distribution}

In order to explore the impact of Earth-scattering on event rates in (non-directional) direct detection experiments, we must calculate the DM \textit{speed} distribution at the detector, given by:\footnote{Note that we use the notation $f(\mathbf{v})$ for the full 3 dimensional velocity distribution and $f(v)$ for the distribution of the modulus $v = |\mathbf{v}|$. These two definitions are related by Eq. ~\ref{eq:speedpert}.}
\begin{align}\label{eq:speedpert}
\tilde{f}(v, \gamma) =  v^2 \int \mathrm{d}^2 \hat{\mathbf{v}}\,  \tilde{f}(\mathbf{v},\gamma)= v^2 \int \mathrm{d}^2 \hat{\mathbf{v}}\, \left(f_A(\mathbf{v},\gamma) + f_D(\mathbf{v},\gamma)\right) = f_A(v,\gamma) + f_D(v, \gamma)\,.
\end{align}
An analogous definition relates $f_0(\mathbf{v})$ and $f_0(v)$.

The \textit{attenuated} speed distribution is obtained straightforwardly by integrating over Eq.~\ref{eq:attenuated}:
\begin{align}\label{eq:speeddist_attenuation}
\begin{split}
f_A(v,\gamma) = v^2 \int_{0}^{2\pi} \mathrm{d}\phi \int_{-1}^1 \mathrm{d}\cos\theta \,\,f_0(v,\theta, \phi) \exp \left[ - \sum_{i}^\mathrm{species} \frac{d_{\mathrm{eff},i}(\cos\theta)}{\overline{\lambda}_i(v)} \right]\,.
\end{split}
\end{align}
The coordinate description for $f_0(v, \theta, \phi)$ for a given value of $\gamma$ is obtained from Eqs.~\ref{eq:SHM} and \ref{eq:gamma}. With this, it is straightforward to evaluate Eq.~\ref{eq:speeddist_attenuation} numerically for fixed $\gamma$.

Similarly, the \textit{deflected} speed distribution is given by
\begin{equation}\label{eq:fDfinal}
f_D(v, \gamma) = v^2 \sum_{i}^{\mathrm{species}}\sum_{a=\pm} \int  \mathrm{d}^2\hat{\mathbf{v}} \int  \mathrm{d}^2\hat{\mathbf{v}}'  \,\,\frac{d_{\mathrm{eff},i}(\cos\theta)}{\overline{\lambda}_i(\kappa_i^a v)} \frac{(\kappa^a_i)^4}{2\pi} f_0(\kappa^a_i v, \hat{\mathbf{v}}') P^a_i(\cos\alpha)\,.
\end{equation}
The angle $\phi$ enters only through the definition of $\cos\alpha$ (Eq.~\ref{eq:cosalpha}) in the factor $\cos(\phi - \phi')$. Because we are integrating over all values of $\phi$, we can make use of the shift symmetry of the integral and eliminate $\phi'$ from the expression for $\cos\alpha$:
\begin{equation}\label{eq:cosalpha2}
\cos\alpha = \sin\theta \sin\theta' \cos\phi + \cos\theta \cos\theta'\,.
\end{equation}
Now, the angle $\phi'$ enters only in the definition of the velocity distribution $f(\kappa^a_i v, \theta', \phi')$. We can therefore perform the integral over $\phi'$ independently of the other angular variables. The deflected speed distribution is then given by:
\begin{equation}\label{eq:speeddist_deflection}
f_D(v, \gamma) = v^2 \sum_{i}^{\mathrm{species}}\sum_{a=\pm} \int_{-1}^{1} \mathrm{d}\cos\theta \int_{0}^{2\pi} \mathrm{d}\phi \int_{-1}^{1}  \mathrm{d}\cos\theta'  \,\,\frac{d_{\mathrm{eff},i}(\cos\theta)}{\overline{\lambda}_i(\kappa_i^a v)} \frac{(\kappa^a_i)^4}{2\pi} f_0(\kappa^a_i v, \theta') P^a_i(\cos\alpha)\,,
\end{equation}
where we define $f(v', \theta') = \int_{0}^{2\pi} f(v', \theta', \phi') \, \mathrm{d}\phi'$. Again, we take the coordinate expression for $f(v', \theta', \phi')$ from Eqs.~\ref{eq:SHM} and \ref{eq:gamma} for a given value of $\gamma$. The integral over $\phi'$ can be performed numerically and tabulated as a function of $v'$ and $\theta'$. The 3-dimensional integral in Eq.~\ref{eq:speeddist_deflection} is then evaluated by Monte Carlo integration (as described in Sec.~\ref{sec:numerics}).

\subsection{Average scattering probability}
\label{sec:pscat}
With this framework, it is straightforward to calculate the average probability that DM particles will scatter at least once, assuming that they cross the Earth's surface. We average over both the velocity distribution of incoming DM particles and over the possible Earth-crossing trajectories of the incident particles. For particles entering the Earth's surface at a point $\mathbf{r}$ with a velocity $\mathbf{v}$, the probability of scattering at least once is given by
\begin{align} \label{eq:dpscat}
1- \exp\left[- \sum_{i}^\mathrm{species} \frac{d_{\mathrm{eff, i}}(-\hat{\mathbf{v}}\cdot\hat{\mathbf{r}})}{\bar{\lambda}_i(v)}\right]\,.
\end{align}
The rate of particles entering the Earth's surface at position $\mathbf{r}$ (with velocity $\mathbf{v}$) is
\begin{align}\label{eq:fluxelement}
f_0(\mathbf{v}) (-\mathbf{v}\cdot\hat{\mathbf{r}})  \, \mathrm{d}^3\mathbf{v} \, \mathrm{d}^2\mathbf{r} \qquad \text{for } \mathbf{v}\cdot \hat{\mathbf{r}} < 0\,,
\end{align}
where we restrict only to particles travelling \textit{inward} through the Earth's surface. Integrating Eq.~\ref{eq:fluxelement} over the Earth's surface and over DM velocities, we obtain the total rate of particles entering the Earth: $\pi \RE^2 \langle v\rangle$. If instead we weight the integral by the scattering probability in Eq.~\ref{eq:dpscat}, we obtain the average scattering probability for particles crossing the Earth:
\begin{align}\label{eq:pscat}
p_\mathrm{scat} &= 1- \frac{2}{\langle v \rangle}\int_{0}^1 \mathrm{d}\cos\theta \int_0^{\vesc  +v_e} \, \mathrm{d}v \, (v \cos\theta) f_0(v) \exp\left[- \sum_{i}^\mathrm{species} \frac{d_{\mathrm{eff, i}}(\cos\theta)}{\bar{\lambda}_i(v)}\right]\, \,,
\end{align}
where we have written $-\hat{\mathbf{v}} \cdot \hat{\mathbf{r}} = \cos\theta$.

\subsection{Numerical implementation}
\label{sec:numerics}

In order to compute the perturbed velocity distributions, speed distributions and average scattering probabilities, we have written the \textsc{EarthShadow} code, which we make publicly available \cite{EarthShadow}. Version 1.0 is available as a \textsc{Mathematica} module and accompanying notebook. This is supplemented with a number of data tables, including tabulated values of $\overline{n}$ and $d_\mathrm{eff}$ (as a function of $\cos\theta$) for each of the 8 elements in Table~\ref{tab:elements}. We also include tabulated values of the differential cross section $\mathrm{d}\sigma/\mathrm{d}E_R$ for each of these elements (as well as for Xenon) for the range of NREFT operators listed in Table~\ref{tab:operators}.

In order to calculate the speed distribution at the detector, we integrate using Quasi-Monte Carlo Integration \cite{Morokoff1995} using a minimum of 10000 samples of the integrand. For a given set of values for $m_\chi$, $\gamma$ and $v$, evaluation of $f_A$ takes roughly $0.5\,\mathrm{s}$ on a single core, while evaulation of $f_D$ takes roughly $10\,\mathrm{s}$. For $f_D$, we increase the number of integrand evaluations as a function of $m_\chi$, once $m_\chi$ is above the Oxygen mass. This is because the constraint $ \cos\alpha > (1- m_A^2/m_\chi^2)^{1/2}$ means that the integrand is zero over a wide range of angles $\{\theta, \,\theta',\, \phi,\, \phi' \}$. As a result, we perform 150000 samples of the integrand for a DM mass of 300 GeV. By varying the number of integrand samples, we have verified that for a given value of $v$, the error on $f_D(v, \gamma)$ is $\mathcal{O}(1\%)$. Note that $f_D$ is linear in $c^2$, the coupling of the DM particle to nucleons. This means that the deflected distribution does not need to be recalculated for different values of the couplings and can simply be rescaled. This is not true for the attenuated distribution (which is non-linear in $c^2$), though in that case the calculation is much faster.

\section{Effects on the DM speed distribution}
\label{sec:effects}

We now investigate the effects of Earth-scattering on the DM speed distribution for three of the operators described in Sec.~\ref{sec:formalism}. These operators are:
\begin{align}
\begin{split}
\hat{\mathcal{O}}_1 &= \mathbb{1}_{\chi N}\,,\\
\hat{\mathcal{O}}_8 &= {\bf{\hat{S}}}_{\chi}\cdot {\bf{\hat{v}}}^{\perp}\,,\\
\hat{\mathcal{O}}_{12} &= {\bf{\hat{S}}}_{\chi}\cdot \left({\bf{\hat{S}}}_{N} \times{\bf{\hat{v}}}^{\perp} \right)\,.
\end{split}
\end{align}
The operator $\oper{1}$ is the operator which mediates the familiar spin-independent (SI) interaction, while $\oper{8}$ and $\oper{12}$ are operators which are higher order in the DM-nucleon relative velocity.  A systematic analysis of all of the operators in Tab.~\ref{tab:operators} is beyond the scope of this work and we focus instead on these three operators because each one leads to distinctive behaviour in the deflection of scattered DM particles.

In the left panel of Fig.~\ref{fig:Pcosalpha}, we show the distribution of the deflection angle $\alpha$ for light DM particles scattering off Iron nuclei. For $\oper{1}$, we see that the deflection of scattered particles is isotropic. In the limit $m_\chi \ll m_A$, the differential scattering cross section for $\oper{1}$ is independent of the recoil energy. For light DM particles, the recoil energy is related to the deflection angle as $E_R = m_\chi^2 v'^2 (1-\cos\alpha)/m_A$ (see Eq.~\ref{eq:ER}). The uniform distribution of recoil energies therefore translates to a uniform distribution of deflection angles. Operator $\oper{8}$ instead has a differential cross section which peaks at small recoil energies, corresponding to deflections which are distributed preferentially in the forward direction ($\cos\alpha > 0$). In contrast, $\oper{12}$ leads to a cross section which increases with $E_R$, meaning that deflection of the DM particles in the backward direction ($\cos\alpha < 0$) is preferred.

\begin{figure*}[t]
\centering
\includegraphics[width=0.32\textwidth]{{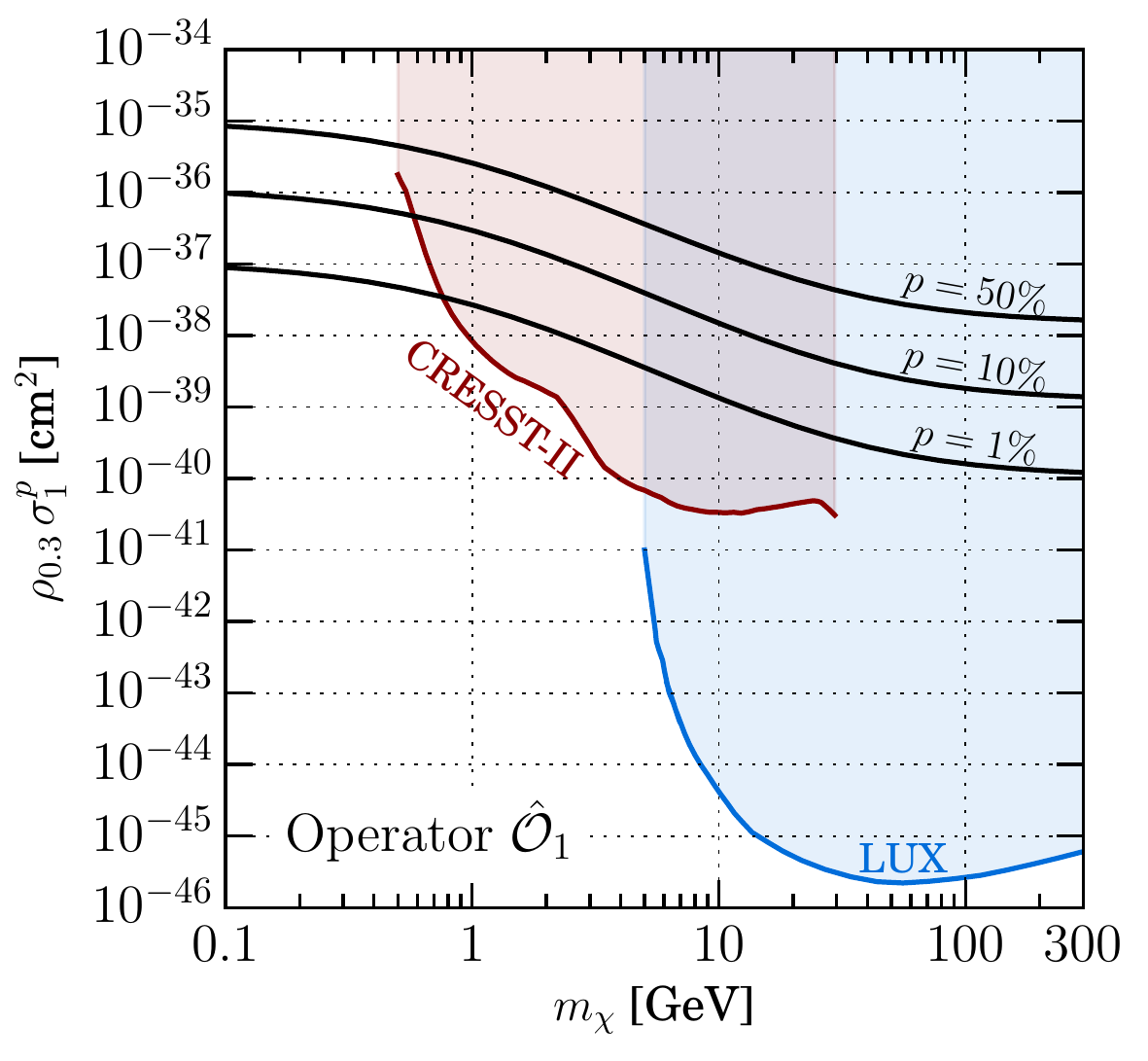}}
\includegraphics[width=0.32\textwidth]{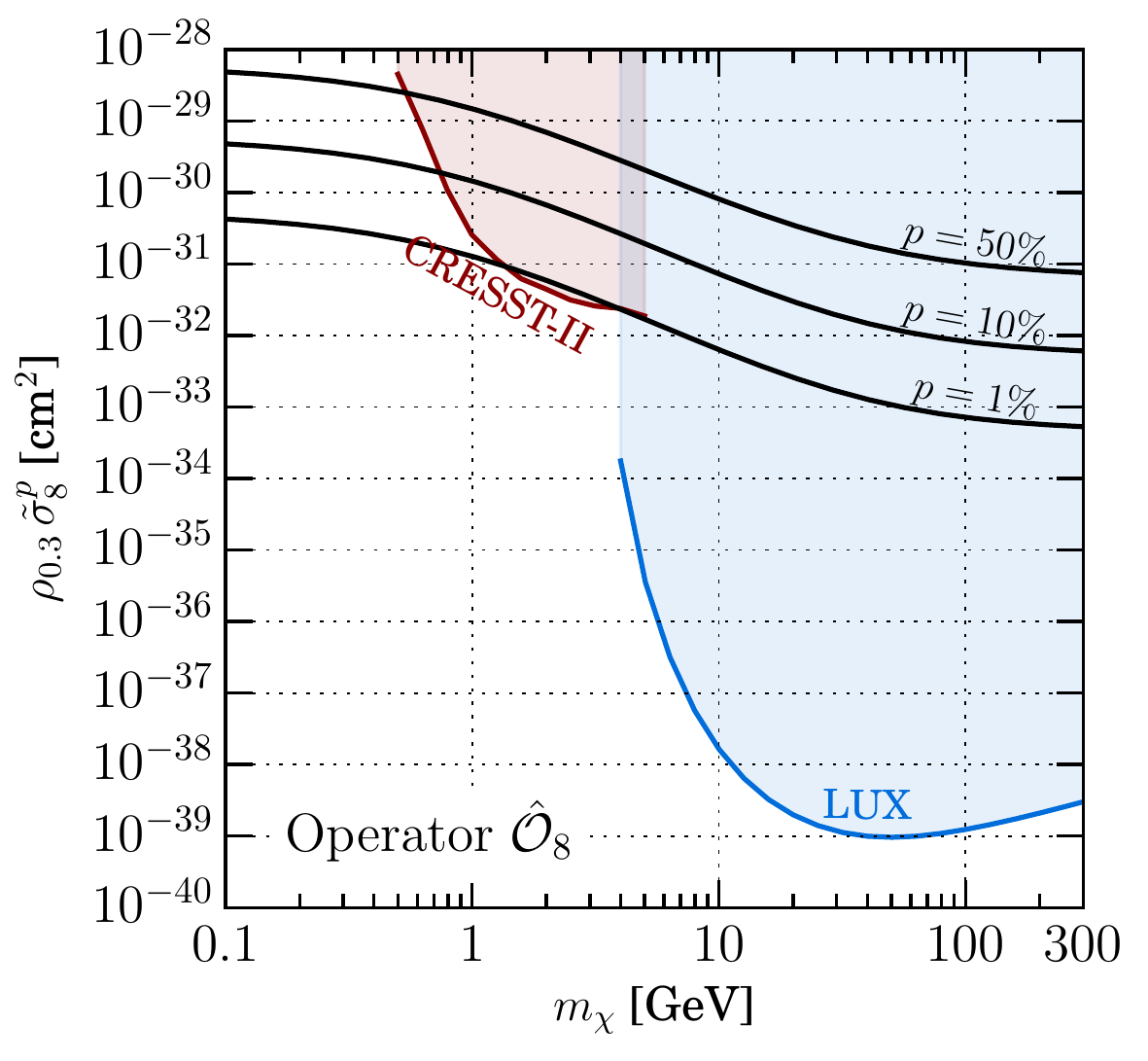}
\includegraphics[width=0.32\textwidth]{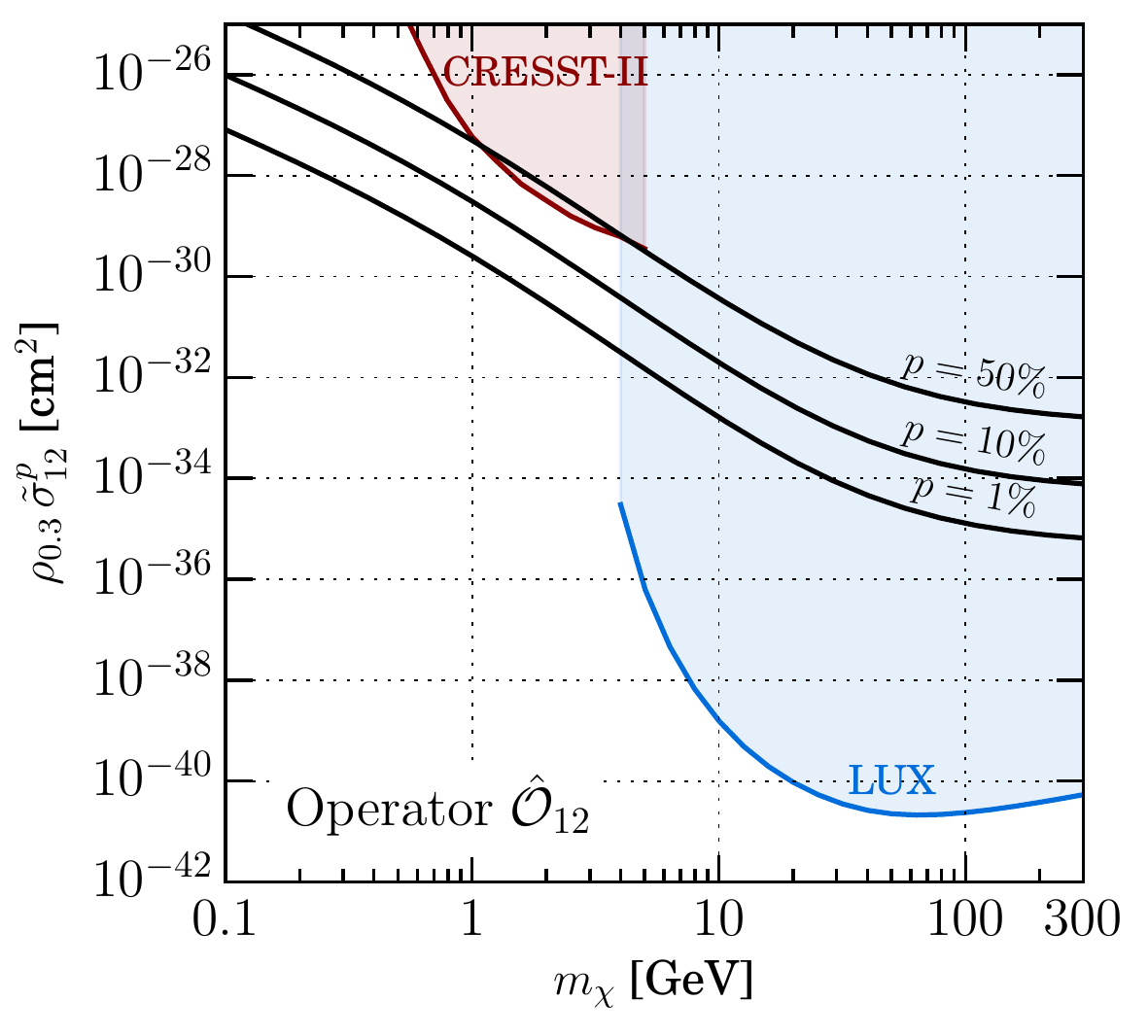}
\caption{\textbf{Current limits on the operators considered in this work.} We show limits for three of the operators described in Sec.~\ref{sec:formalism}: $\oper{1}$ (\textbf{left}), $\oper{8}$ (\textbf{middle}), and $\oper{12}$ (\textbf{right}). In each case, we show limits from CRESST-II~\cite{Angloher:2015ewa} and LUX~\cite{Akerib:2016vxi} on the product of the effective DM-proton cross section $\tilde{\sigma}_j^p$ (defined in Eq.~\ref{eq:sigeff}) and the local DM density $\rho_\chi$ (normalised such that $\rho_\chi = \rho_{0.3}/ 0.3 \,\,\mathrm{GeV}\,\,\mathrm{cm}^{-3}$). Limits for $\oper{8}$ and $\oper{12}$ are calculated as in Appendix~\ref{app:eventrates}. Also plotted are contours of the scattering probability $p$, defined in Eq.~\ref{eq:pscat}.} \label{fig:limits}
\end{figure*}

In Fig.~\ref{fig:limits}, we show current limits on DM-nucleon interactions for each of these three operators. At low mass, the most stringent limits come from the CRESST-II experiment~\cite{Angloher:2015ewa}, while at high mass we consider limits from the LUX WS2014-16 run~\cite{Akerib:2016vxi} (noting that similar limits are obtained by PANDA-X II \cite{Tan:2016zwf}). For operator $\oper{1}$, we present the limits on the standard DM-proton SI cross section, as reported by the collaborations. This is related to the operator coefficient $c_1^0$ as:
\begin{equation}
\sigma_1^p = \left(\frac{c_1^0}{2}\right)^2 \frac{\mu_{\chi p}^2}{\pi}\,,
\end{equation}
where $c_1^0/2$ is the coupling to protons (assuming isoscalar interactions). For operators $\oper{8}$ and $\oper{12}$, we define the effective cross sections (in analogy with the SI cross section) as:
\begin{equation}\label{eq:sigeff}
\tilde{\sigma}_j^p= \left(\frac{c_j^0}{2}\right)^2 \frac{\mu_{\chi p}^2}{\pi}\,.
\end{equation}
Approximate limits on $\tilde{\sigma}_p^j$ are calculated for CRESST-II and LUX as described in Appendix~\ref{app:eventrates}. Note that the limits on the cross sections are degenerate with the local DM density and in all cases we keep this fixed to $\rho_\chi = 0.3 \,\, \mathrm{GeV} \,\, \mathrm{cm}^{-3}$.

Also plotted in Fig.~\ref{fig:limits} are contours of equal scattering probability, $p_\mathrm{scat} = $ 1\%, 10\% and 50\%, calculated according to Eq.~\ref{eq:pscat}. With its low threshold, CRESST-II has sensitivity to DM particles with masses around 0.5 GeV. However, current constraints are weak enough that for all three operators, DM particles with such a low mass may still have a large enough interaction cross section to give a 10\% scattering probability in the Earth. For operator $\oper{12}$, the CRESST-II constraints are even weaker, as this operator gives rise to both spin-independent and spin-dependent interactions, while the majority of target nuclei in the CRESST-II experiment have zero spin.

In light of these constraints, we will focus on light DM with a mass of $m_\chi = 0.5 \,\, \mathrm{GeV}$. However, we will also briefly examine the case of heavier DM with a mass of $m_\chi = 50 \,\, \mathrm{GeV}$. We note that the SI cross section required to give a $10\%$ scattering probability for DM of mass 50 GeV is currently excluded by LUX by more than 6 orders of magnitude. Even so, it is instructive to explore how the high mass case differs from the light DM examples, bearing in mind that such high mass candidates could evade current constraints if they are (very) sub-dominant. In all cases, we fix the DM-nucleon couplings such that the average probability of scattering for DM particles is $p_\mathrm{scat} = 10\%$, as defined in Eq.~\ref{eq:pscat}. We then calculate the perturbed speed distribution $\tilde{f}(v, \gamma)$ for a range of values of $\gamma$, which defines the average direction of incoming DM particles relative to the detector position (see Fig.~\ref{fig:gamma}). We remind the reader that we do not require that the perturbed speed distribution be correctly normalised to unity. Thus, Earth-scattering may affect not only the shape but also the overall normalisation of $\tilde{f}(v, \gamma)$.

\subsection{Low Mass}

\begin{figure*}[t]
\includegraphics[width=0.32\textwidth]{{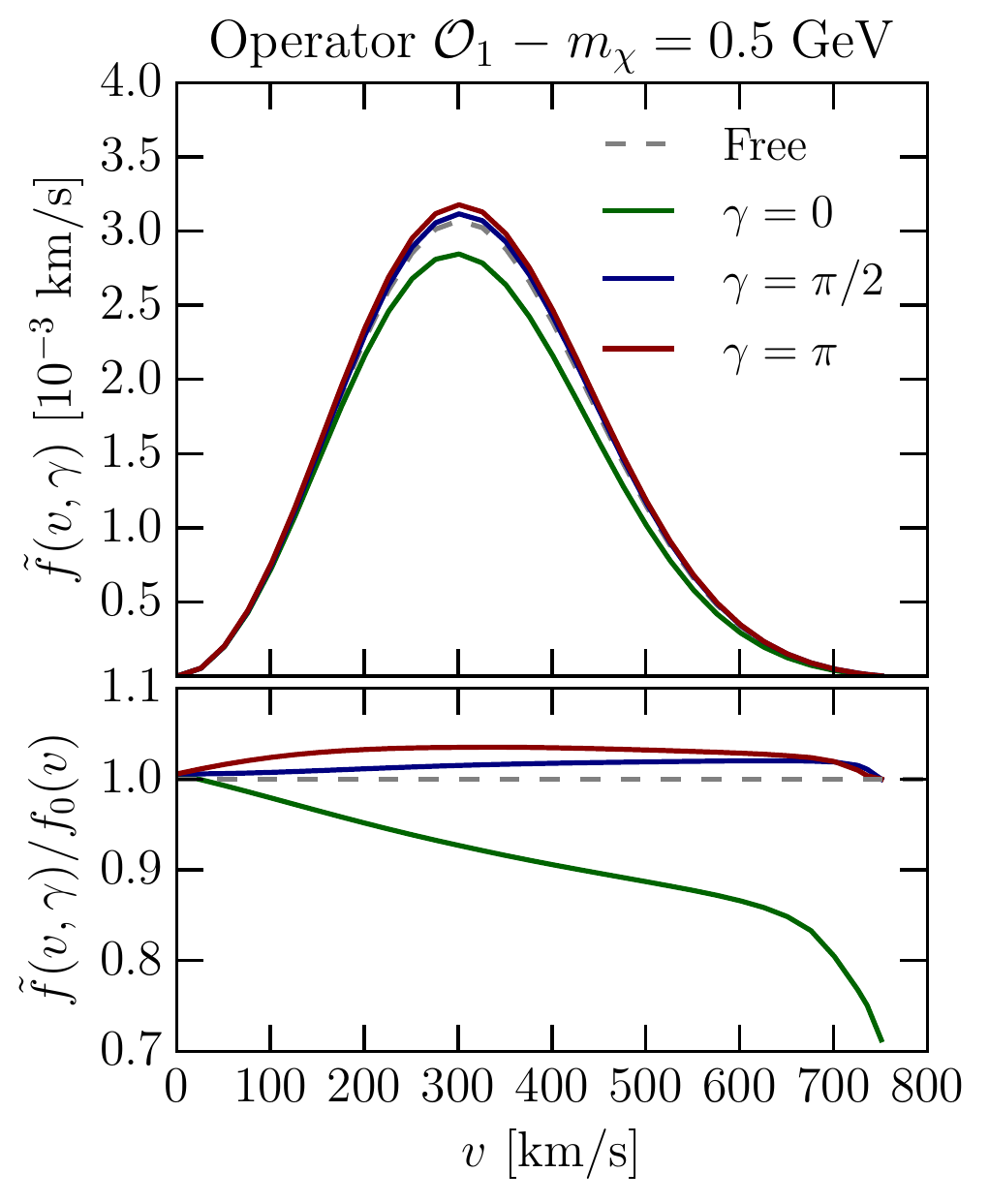}}
\includegraphics[width=0.32\textwidth]{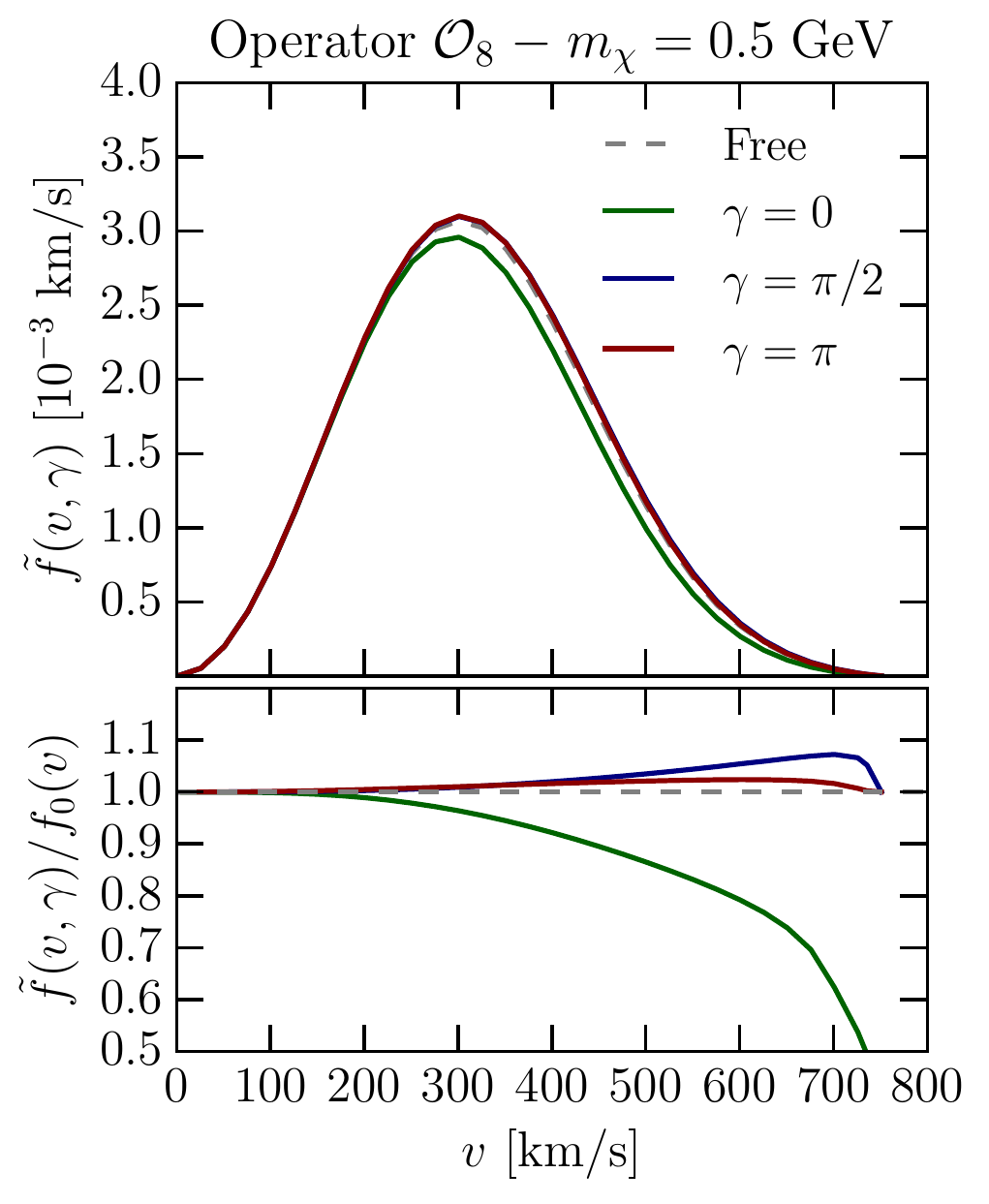}
\includegraphics[width=0.32\textwidth]{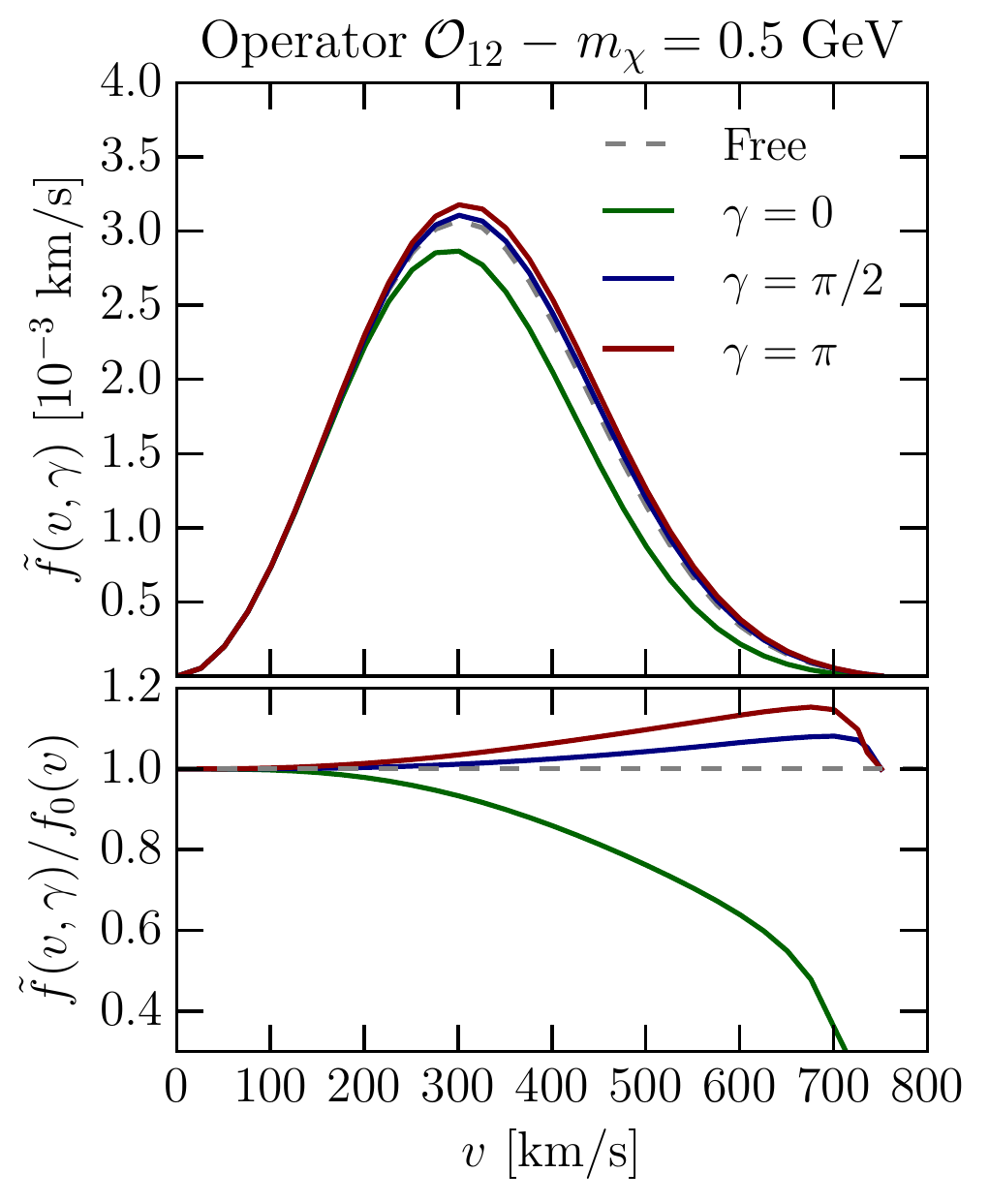}
\caption{\textbf{Perturbed DM speed distribution due to low-mass DM ($m_\chi = 0.5 \,\,\mathrm{GeV}$) scattering in the Earth.}The DM-nucleon couplings are normalised to give an average scattering probability of $p_\mathrm{scat} = 10\%$, as defined in Eq.~\ref{eq:pscat}. We show results for the three operators $\oper{1}$ (\textbf{left}), $\oper{8}$ (\textbf{middle}) and $\oper{12}$ (\textbf{right}). The dashed lines correspond to the free DM speed distribution without Earth-scattering $f_0(v)$ while the solid lines correspond to different average incoming DM direction (see Fig.~\ref{fig:gamma}): $\gamma = 0$ leads to maximal Earth-crossing, while $\gamma = \pi$ leads to minimal Earth-crossing before reaching the detector. Each solid line corresponds to a horizontal slice through Fig.~\ref{fig:SpeedDist-2D}.}\label{fig:SpeedDist-1D}
\end{figure*}

In Fig.~\ref{fig:SpeedDist-1D} we show the effects of Earth-scattering on the speed distribution for light DM ($m_\chi = 0.5 \,\, \mathrm{GeV}$) for three values of $\gamma$. As one might expect, for particles which must cross most of the Earth before reaching the detector ($\gamma = 0$, solid green)
, the predominant effect is that of attenuation, leading to a reduced DM population. For Operator $\oper{1}$ (left panel of Fig.~\ref{fig:SpeedDist-1D}), the size of this effect increases with increasing DM speed. In fact, the total scattering cross section for $\oper{1}$ is velocity-independent. However, the SHM velocity distribution (Eq.~\ref{eq:SHM}) becomes increasingly anisotropic as we increase $v$. For large $v$, more of the DM particles are travelling parallel to the mean DM velocity $\vchi$, meaning that the average Earth-crossing distance for particles to reach the detector increases. For Operators $\oper{8}$ and $\oper{12}$ (centre and right panels of Fig.~\ref{fig:SpeedDist-1D}), attenuation also increases as a function of $v$, though in this case the predominant cause is that the total cross section increases with the DM speed: $\sigma_{8, 12} \propto v^2$.


As we increase $\gamma$, the typical DM particle must travel through less of the Earth before reaching the detector. For $\gamma = \pi/2$ (solid blue line), particles travelling along $\hat{\mathbf{v}} = \vchihat$ must cross a negligibly small depth of the Earth before reaching the detector (typically on the order of a few kilometres). However, the distribution of DM velocities about the average means that some particles will still be travelling an appreciable distance through the Earth and therefore attenuation still has an effect. However, for all three operators in Fig.~\ref{fig:SpeedDist-1D} the effects of deflection towards the detector are more significant, leading to an \textit{increase} in the DM population for $\gamma = \pi/2$. For Operator $\oper{1}$, this increase is roughly a $2\%$ effect, increasing to $\sim 4\%$ for $\gamma = \pi$. In this latter case, the average DM particle arrives at the detector having only passed a small distance through the Earth (equal to $l_D$ the underground depth of the detector), meaning that the effects of attenuation are minimal. We note that at the highest DM speeds ($v = v_e + \vesc \approx 753 \kms$) the enhancement due to deflection reduces to zero. This is because the speed of DM particles is reduced on scattering. Particles at $v = v_e + \vesc$ must scatter down to smaller speeds, while there are no DM particles with larger speeds (i.e.~above the escape velocity) which can scatter down to $v = v_e + \vesc$ and give an enhancement.

For Operator $\oper{8}$, the enhancement in the DM speed distribution for $\gamma = \pi/2$ is greater than for $\gamma = \pi$. As we saw in Fig.~\ref{fig:Pcosalpha}, $\oper{8}$ leads to DM deflection preferentially in the forward direction. For $\gamma = \pi$, this means that particles are unlikely to scatter back once they have already passed  the detector. For $\gamma = \pi/2$, the effects of attenuation are relatively small, but there is a much greater probability of particles scattering towards the detector. For Operator $\oper{12}$, instead, the population of DM particles is once again increasing as a function of $\gamma$. Operator $\oper{12}$ favours backward scattering, meaning that the contribution to $\tilde{f}(v, \gamma)$ due to deflection is small for $\gamma = 0$ but maximal for $\gamma = \pi$. As attenuation becomes less important, deflection becomes more important, meaning that $\tilde{f}(v, \gamma)$ varies more rapidly as a function of $\gamma$.

\begin{figure}[h!]
\centering
\includegraphics[width=0.495\textwidth]{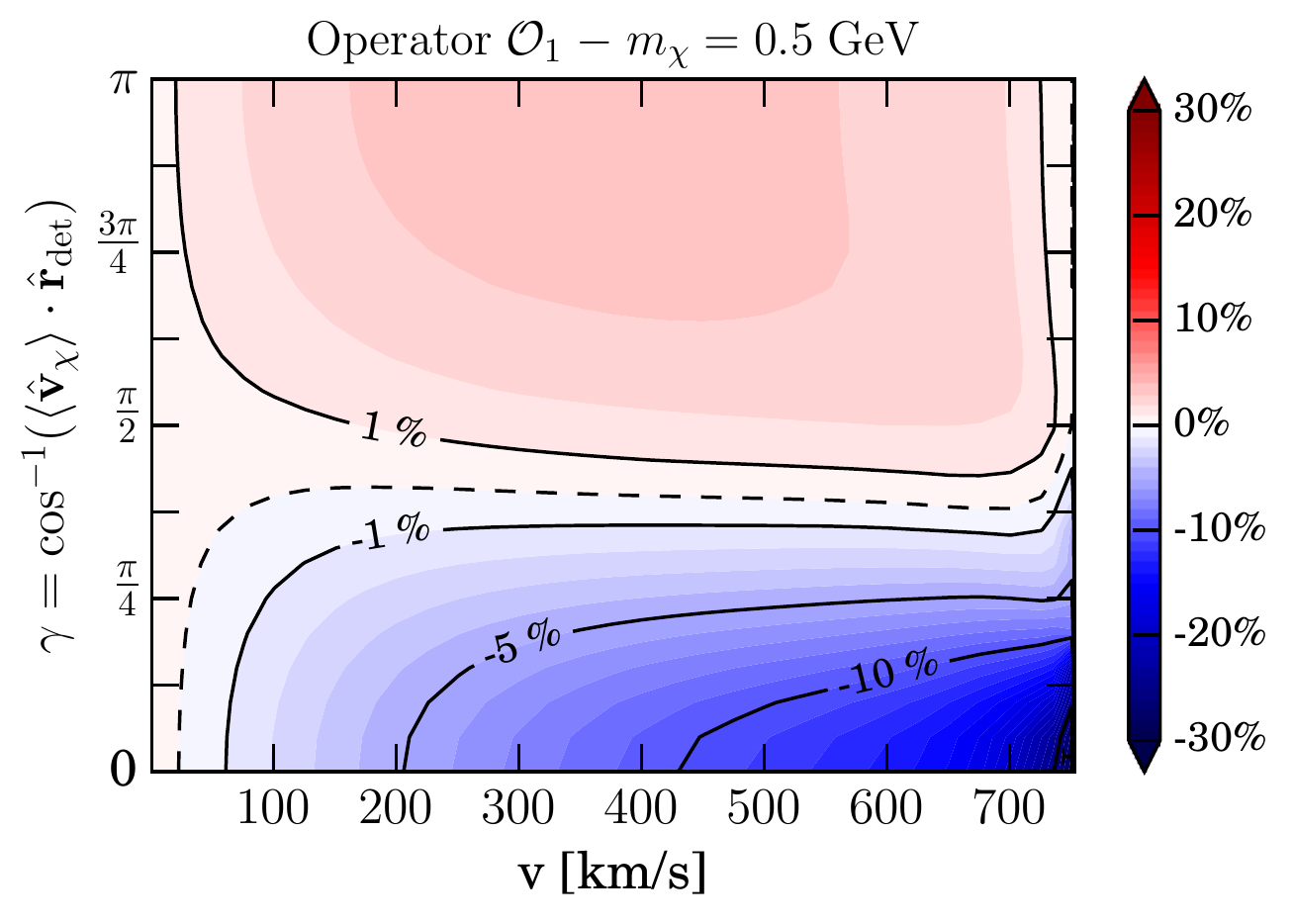}
\includegraphics[width=0.495\textwidth]{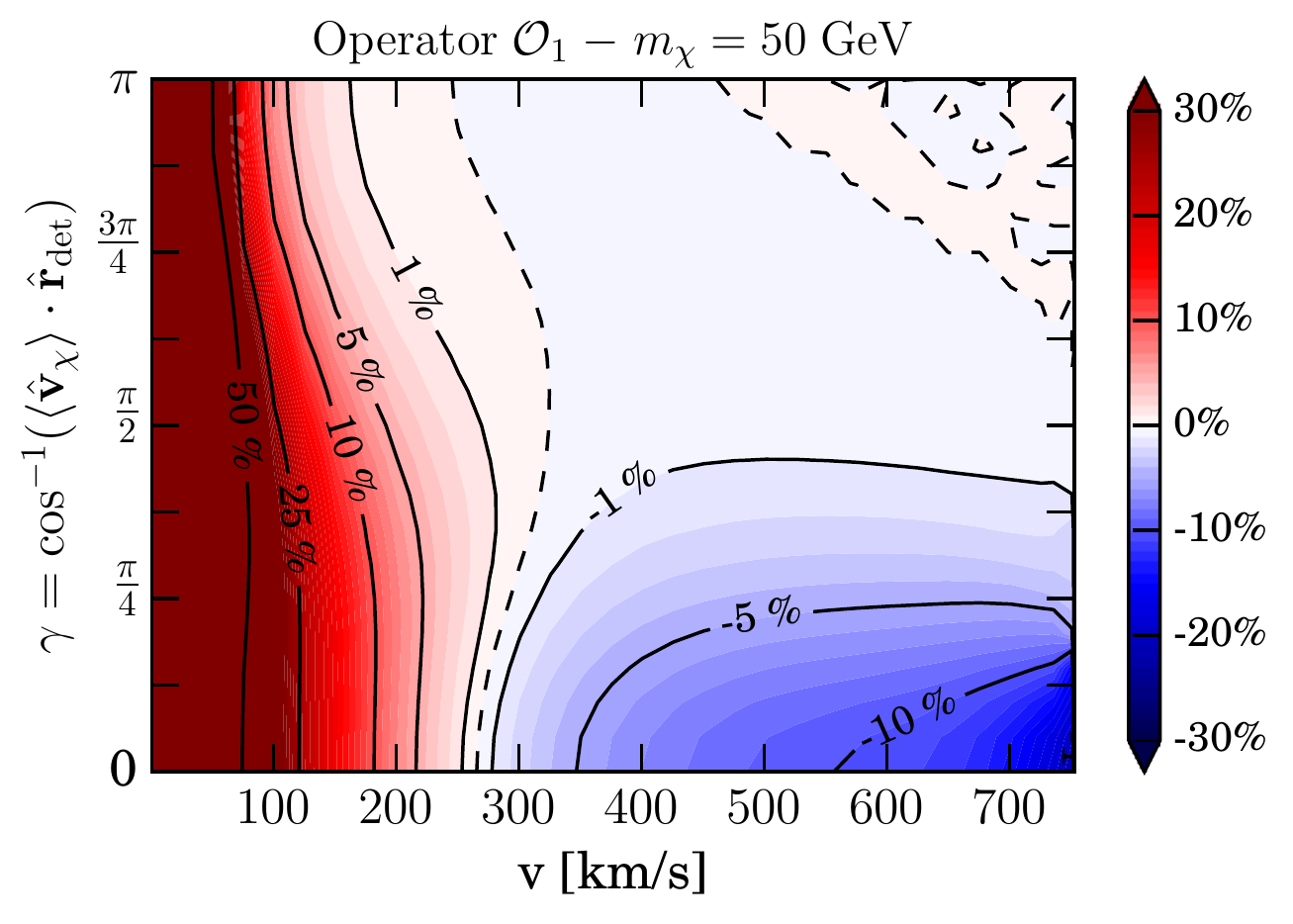}
\includegraphics[width=0.495\textwidth]{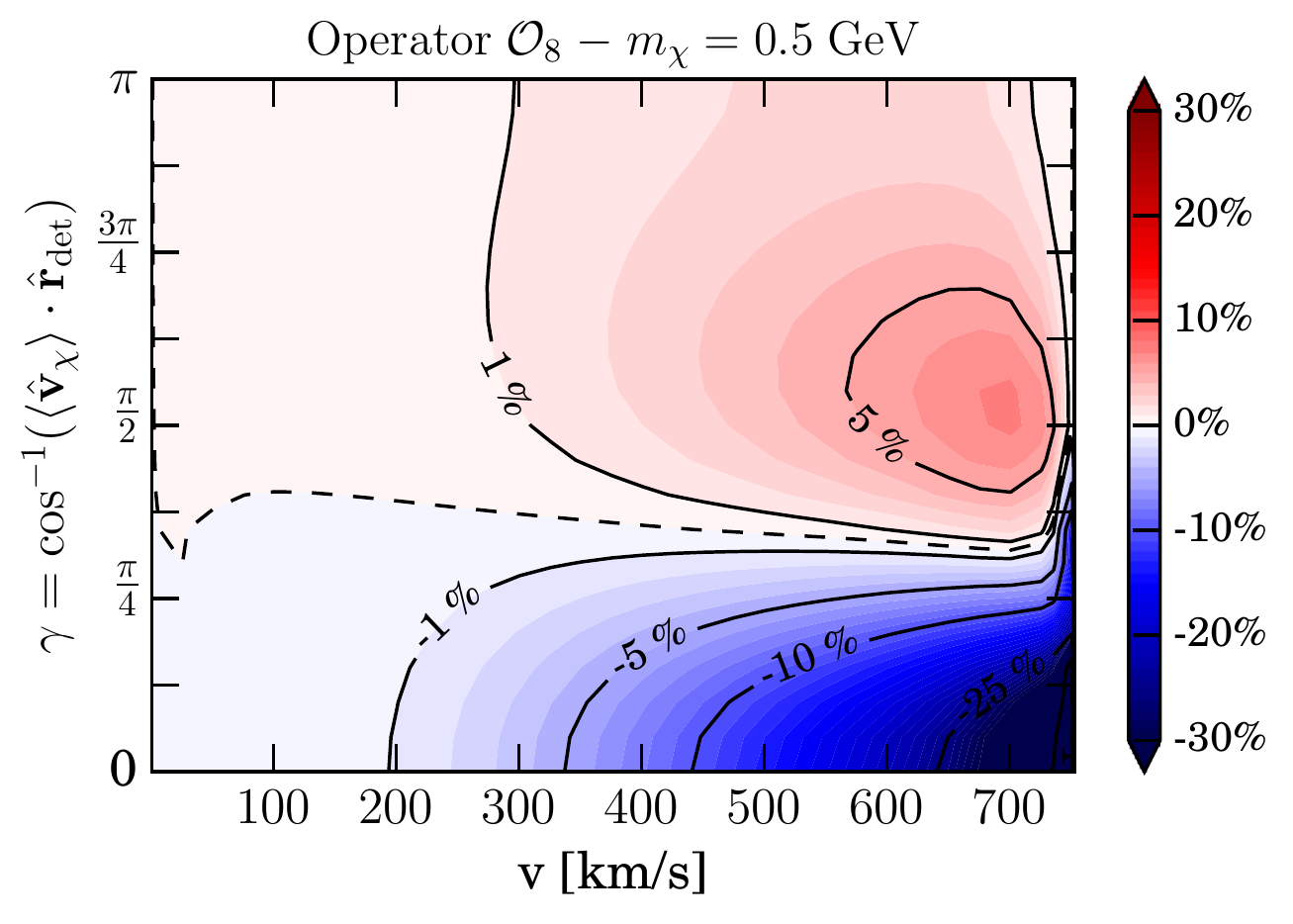}
\includegraphics[width=0.495\textwidth]{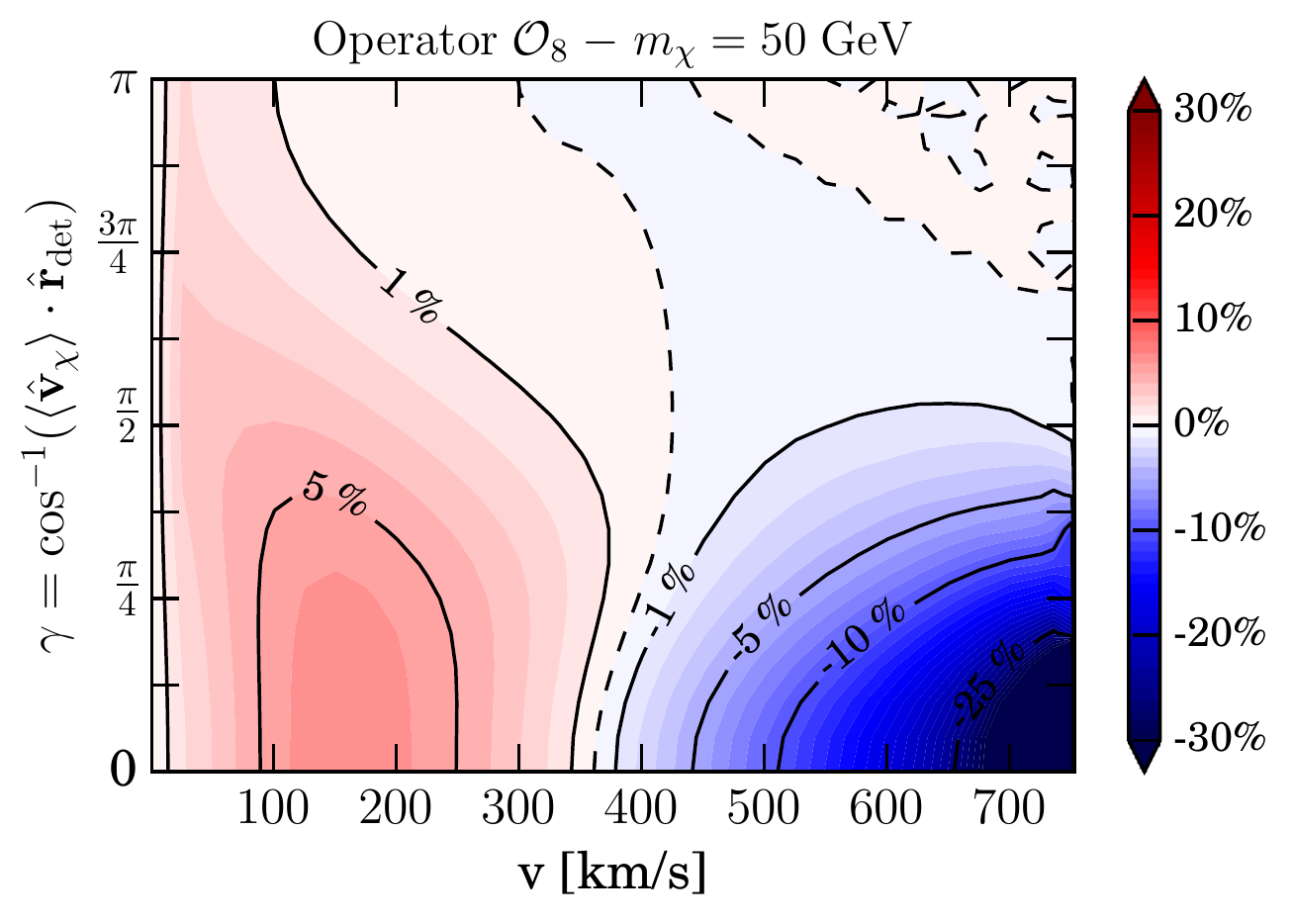}
\includegraphics[width=0.495\textwidth]{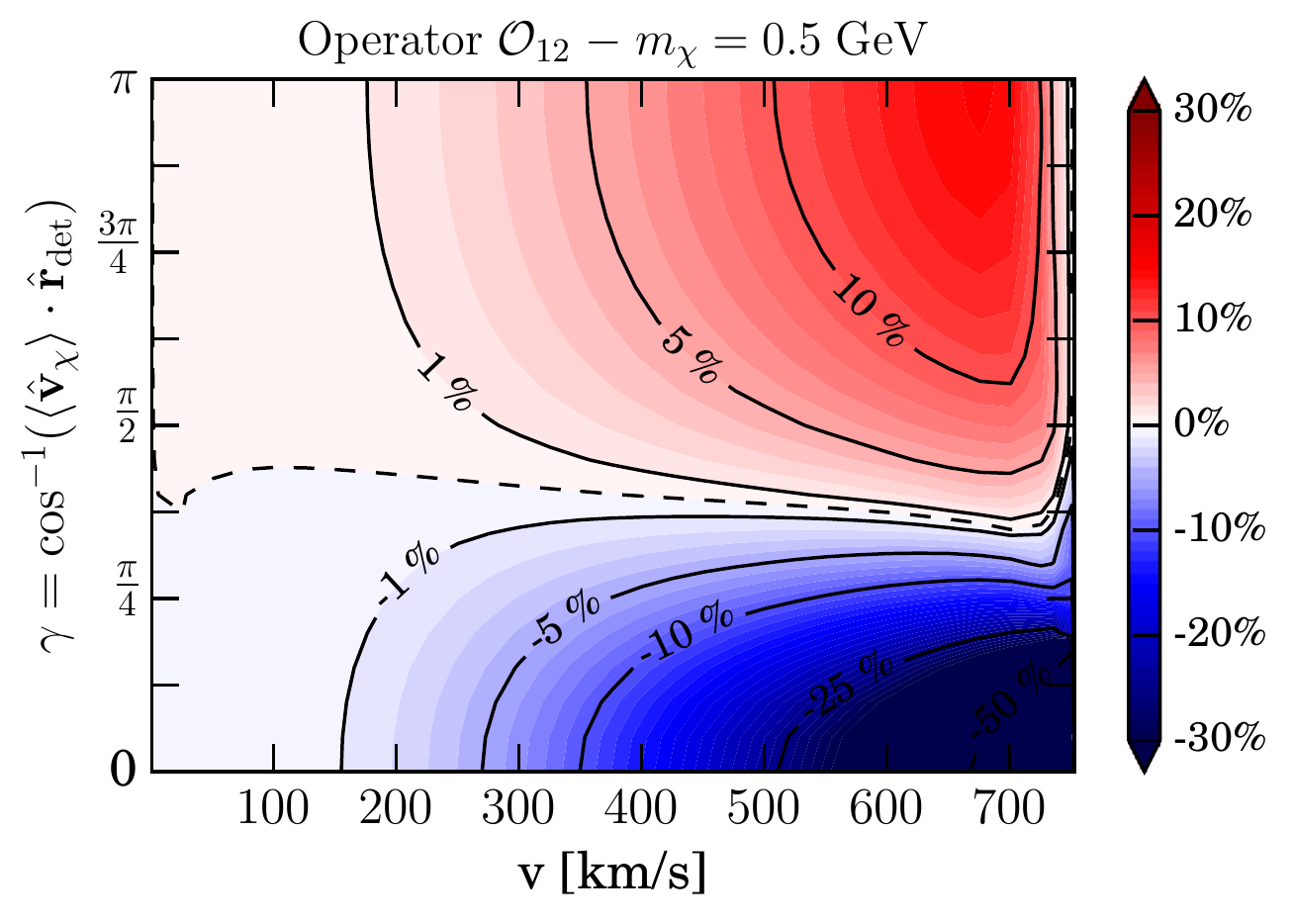}
\includegraphics[width=0.495\textwidth]{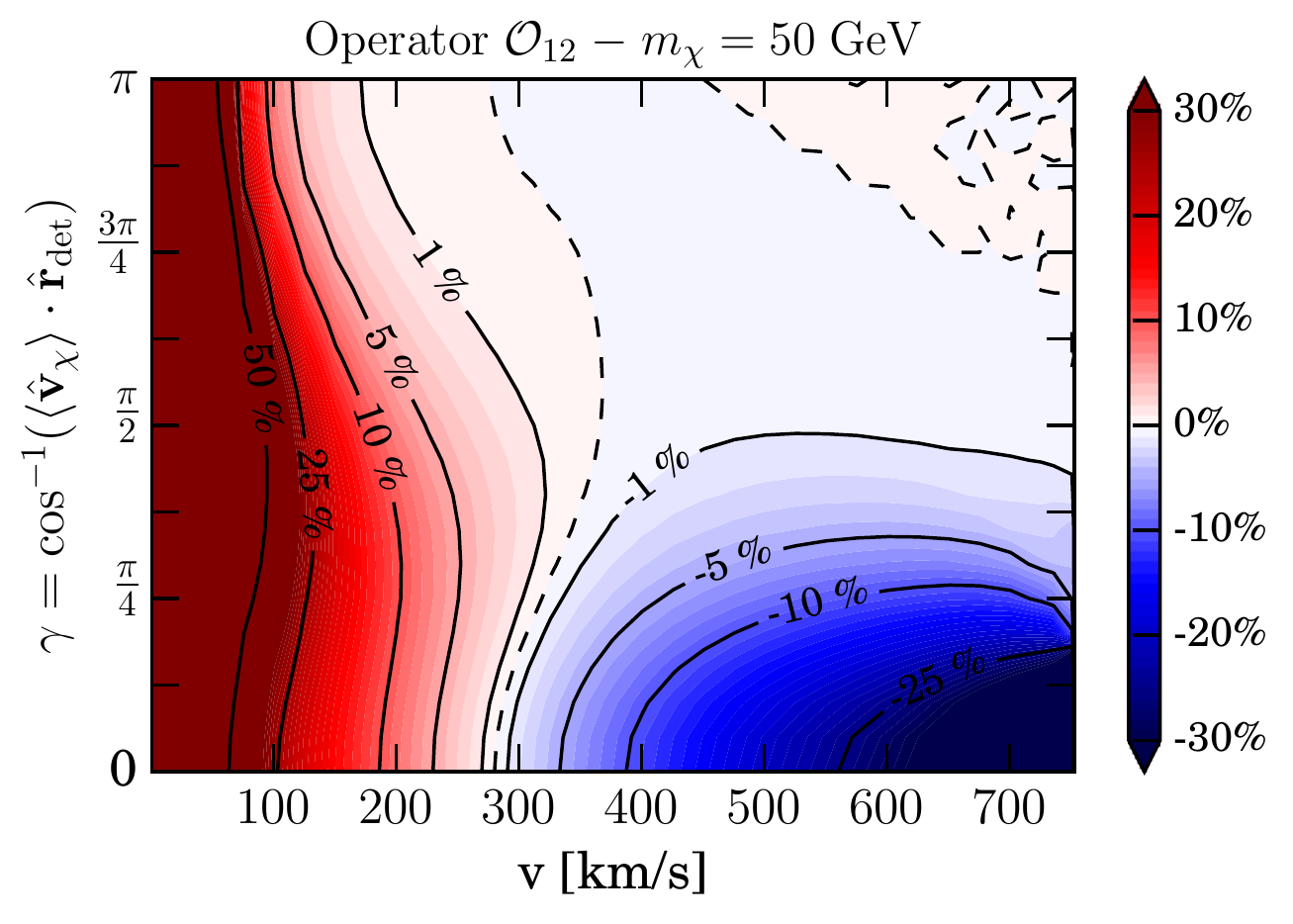}
\clearpage
\caption{\textbf{Percentage change in the speed distribution due to Dark Matter scattering in the Earth.} Results for the standard SI interaction (Operator $\oper{1}$) are shown in the top row, while results for two other DM-nucleon operators (see Sec.~\ref{sec:formalism}) are shown in the middle row (Operator $\oper{8}$) and bottom row (Operator $\oper{12}$). Results are shown for two DM masses: 0.5 GeV (\textbf{left column}) and 50 GeV (\textbf{right column}). In all cases, the DM-nucleon couplings are normalised to give an average scattering probability of $p_\mathrm{scat} = 10\%$, as defined in Eq.~\ref{eq:pscat}. In each panel, the $x$-axis shows the DM particle speed $v$ while the $y$-axis shows the angle $\gamma$ between the average incoming DM velocity $\vchihat$ and the detector position (see Fig.~\ref{fig:gamma}). The angle $\gamma = 0$ corresponds to maximal Earth-crossing before reaching the detector, while $\gamma = \pi$ corresponds to minimal Earth-crossing. The dashed contour corresponds to no change in the DM speed distribution. }\label{fig:SpeedDist-2D}
\end{figure}

We now extend the discussion to consider a wider range of values of $\gamma$. Figure~\ref{fig:SpeedDist-2D} shows the percentage difference between the free and perturbed speed distributions as a function of both $v$ and $\gamma$. In each plot, we show contours of $\pm 1\%$, $\pm 5\%$, $\pm 10\%$, $\pm 25\%$ and  $\pm 50\%$ change in the speed distribution. The dashed contours correspond to zero change in the speed distribution: $\tilde{f}(v ,\gamma) = f_0(v)$.

In the left column of Fig.~\ref{fig:SpeedDist-2D}, we note immediately that (for a given operator) the effects of depletion appear to be much larger than the effects of enhancement. For operator $\oper{1}$, compare the maximum enhancement of $\mathcal{O}(3\%)$ with the maximum depletion of $\mathcal{O}(-25\%)$. The attenuation of particles is relevant only when the DM particles must cross a large fraction of the Earth before reaching the detector (i.e.~for small $\gamma$). These scattered particles will of course lead to an enhanced flux at other points on the surface of the Earth. For operator $\oper{1}$ the deflection of low mass DM particles is isotropic, meaning that this enhancement due to deflection is distributed almost uniformly across the Earth's surface. Particles which are significantly depleted from a small region of the surface with $\gamma \rightarrow 0$ are redistributed over the entire Earth, giving a small enhancement at any given point.

In fact, to first order in $\RE/\lambda$, the effects of attenuation and deflection should be equal. This means that the total rate $\Gamma_\mathrm{out}$ of DM particles passing outward through the surface of the Earth (taking into account both of these contributions) should be the same as the total inward flux $\Gamma_\textrm{in}$. We could in principle check this by integrating the perturbed velocity distribution over all positions $\mathbf{r}$ on the Earth's surface:\footnote{Note that here we have written the perturbed distribution in terms of the position vector $\mathbf{r}$ on the surface of the Earth rather than explicitly in terms of the angle $\gamma$.}
\begin{align}
\begin{split}
\Gamma_\mathrm{in} &= \int_{\mathbf{v}\cdot \mathbf{r} < 0} \mathrm{d}^2\mathbf{r} \int \mathrm{d}^3\mathbf{v} \,f_0(\mathbf{v})\, (-\mathbf{v}\cdot \mathbf{r})\,,\\
\Gamma_\mathrm{out} &= \int_{\mathbf{v}\cdot \mathbf{r} > 0} \mathrm{d}^2\mathbf{r} \int \mathrm{d}^3\mathbf{v} \,\tilde{f}(\mathbf{v}, \mathbf{r}) \,(\mathbf{v}\cdot \mathbf{r})\,.
\end{split}
\end{align}
In this work, we have calculated the perturbed speed distribution $\tilde{f}(v, \mathbf{r})$ (defined in Eq.~\ref{eq:speedpert}) rather than the perturbed velocity distribution $\tilde{f}(\mathbf{v}, \mathbf{r})$. We can however re-express $\Gamma_\mathrm{out}$ in terms of the \textit{speed} distribution:
\begin{align}
\begin{split}
\Gamma_\mathrm{out} &= \int_{\mathbf{v}\cdot \mathbf{r} > 0} \mathrm{d}^2\mathbf{r} \int \mathrm{d}^3\mathbf{v} \,\tilde{f}(\mathbf{v}, \mathbf{r}) \,(\mathbf{v}\cdot \mathbf{r}) +\int_{\mathbf{v}\cdot \mathbf{r} < 0} \mathrm{d}^2\mathbf{r} \int \mathrm{d}^3\mathbf{v} \,\tilde{f}(\mathbf{v}, \mathbf{r})\, (-\mathbf{v}\cdot \mathbf{r})\\ 
&\qquad- \int_{\mathbf{v}\cdot \mathbf{r} < 0} \mathrm{d}^2\mathbf{r} \int \mathrm{d}^3\mathbf{v} \,\tilde{f}(\mathbf{v}, \mathbf{r})\, (-\mathbf{v}\cdot \mathbf{r})\\
&= \int \mathrm{d}^2\mathbf{r} \int \mathrm{d}^3\mathbf{v} \,\tilde{f}(\mathbf{v}, \hat{\mathbf{r}}) \,|\mathbf{v}\cdot \mathbf{r}| - \int_{\mathbf{v}\cdot \mathbf{r} < 0} \mathrm{d}^2\mathbf{r} \int \mathrm{d}^3\mathbf{v} \,\tilde{f}(\mathbf{v}, \mathbf{r})\, (-\mathbf{v}\cdot \mathbf{r})\\
&= \left[ 2\pi \RE^2 \int_{-1}^{1} \mathrm{d}\cos\gamma \int_{0}^{v_e + \vesc} \mathrm{d}v \, \tilde{f}(v, \gamma) \,v \,|\cos\gamma|\right] - \Gamma_\textrm{in}\,.
\end{split}
\end{align}
Here, we have made use of the fact that scattering in the Earth can only affect the distribution of particles with velocities pointing away from the Earth (so $\tilde{f}(\mathbf{v}, \mathbf{r}) = f_0(\mathbf{v})$ for $\mathbf{v}\cdot \mathbf{r} < 0$). As in Sec.~\ref{sec:pscat}, the rate of DM particles flowing inward through the Earth's surface can be calculated straightforwardly from the equation for $f_0(\mathbf{v})$ (or from geometric arguments), giving $\Gamma_\mathrm{in} = \pi \RE^2 \langle v \rangle$. Thus, we can calculate $\Gamma_\mathrm{out}$ from the distributions presented in this section and compare with $\Gamma_\mathrm{in}$. The results of this comparison are shown in Tab.~\ref{tab:fluxes}. In the case of operator 1, the depletion due to attenuation and the enhancement due to deflection cancel to within about $10\%$, meaning that the total DM particle flux through the Earth is conserved to better than $1\%$.\footnote{We work to first order in $\RE/\lambda$, so we would expect errors of the order $p_\mathrm{scat}^2$ (or $1\%$ for $p_\mathrm{scat} = 10\%$).}

\begin{table}[t!]\centering
\begin{tabular}{@{}lllll@{}}
\toprule
\toprule
				 DM mass [GeV] 	& Operator  & $\Delta \Gamma_\mathrm{out}^\mathrm{Atten.}/\Gamma_\mathrm{in}$ & $\Delta \Gamma_\mathrm{out}^\mathrm{Defl.}/\Gamma_\mathrm{in}$ & $\Gamma_\mathrm{out}/\Gamma_\mathrm{in}$	\\ \hline
				 0.5 & $\oper{1}$ &  $-7.8\%$ & $+7.0\%$ & $99.2\%$ \\
				 0.5 & $\oper{8}$ & $-8.0\%$ & $+7.3\%$ & $99.2\%$ \\
				 0.5 & $\oper{12}$ & $-7.8\%$ & $+7.2\%$ & $99.4\%$ \\
				 50 & $\oper{1}$ & $-7.5\%$ & $+7.3\%$ & $99.9\%$ \\
				  50 & $\oper{8}$ & $-8.0\%$ & $+8.4\%$ & $100.4\%$ \\
				  50 & $\oper{12}$ & $-7.3\%$ & $+6.6\%$ & $99.3\%$ \\
\bottomrule
\bottomrule
\end{tabular}
\caption{\textbf{Rate of DM particles passing inward ($\Gamma_\mathrm{in}$) and outward ($\Gamma_\mathrm{out} $) through the Earth's surface once scattering has been accounted for.} In all cases, we normalise the average scattering probability to 10\%. We show the percentage change in the outward flow rate $\Delta\Gamma_\mathrm{out}$ due to attenuation (`Atten.') and deflection (`Defl.') separately. In the rightmost column, we give the total outward flow rate including both effects. Conservation of particle number implies that $\Gamma_\mathrm{out} = \Gamma_\mathrm{in}$.}
\label{tab:fluxes}
\end{table}

We now turn to operator $\oper{8}$ (middle left panel of Fig.~\ref{fig:SpeedDist-2D}). As we have already discussed, operator $\oper{8}$ leads to deflection predominantly in the forward direction. This means that the enhancement due to deflected particles is maximal around $\gamma = 0$, though this is not observable due to the strong attenuation. At $\gamma = \pi$, the attenuation effect is negligible, but (unlike for $\oper{1}$) very few particles are scattered back, leading to only a small enhancement due to deflection. The largest enhancement therefore occurs between these two extrema, around $\gamma = \pi/2$. This enhancement is maximised at a speed of $v \sim 700 \kms$. For operator $\oper{8}$, the scattering cross section increases with $v$, meaning that particles just below the escape velocity are most likely to scatter. These particles lose a small amount of energy when they are deflected, leading to a peak around $v \sim 700 \kms$. In this case too, the DM particle flux is well conserved (see Tab.~\ref{tab:fluxes}). 

Finally, for operator $\oper{12}$, the enhancement in the DM speed distribution peaks sharply towards $\gamma = \pi$ with a strong attenuation effect for $\gamma = 0$. Particles are deflected preferentially in the backwards direction, meaning that the enhancement due to deflection provides little replenishment of the DM flux for $\gamma = 0$. The calculation of the attenuation effects is exact, but calculation of the deflection relies on the first-order, `single scatter' approximation we have employed. With attenuation effects of $\mathcal{O}(50\%)$, we may be concerned that this approximation will break down. However, such large scattering probabilities occur only over a small range of angles (say, $\gamma < \pi/4$). In fact, for a ring on the Earth's surface defined by a given value of $\gamma$, the surface area of that ring scales as $\sin\gamma$. This means that the DM wind sees only a small area of the Earth's surface with a small value of $\gamma$ or, physically, that only a small fraction of particles will cross the full diameter of the Earth. Most particles will traverse the Earth on trajectories with much smaller crossing distances, for which the linear approximation holds. The success of the linear approximation can be verified explicitly by calculating the DM number density at the surface after scattering and, as before, the attenuation and deflection effects cancel to a high degree ($\Gamma_\mathrm{out}/\Gamma_\mathrm{in} = 99.4\%$).

\subsection{High mass}

We now explore the effects of Earth-scattering for higher mass DM particles. In the right column of Fig.~\ref{fig:SpeedDist-2D}, we show the percentage change in the speed distribution for DM with mass $m_\chi = 50 \, \, \mathrm{GeV}$, interacting through operators $\oper{1}$, $\oper{8}$ and $\oper{12}$.

As for light DM, the greatest effects of attenuation are experienced by the fastest DM particles when they must cross the maximal distance through the Earth to reach the detector ($\gamma = 0$), with negligible attenuation for large values of $\gamma$. However, in contrast to the light DM case, there is no enhancement in the speed distribution for speeds above $\sim300 \kms$. At 50 GeV, the DM particles have a mass within a factor of a few of the Earth species which they scatter off. This means that the transfer of kinetic energy is significantly more efficient. Particles with high speeds which scatter in the Earth are therefore down-scattered to much smaller speeds, leading to a substantial enhancement for $v < 100 \kms$. 

As shown in Fig.~\ref{fig:Pcosalpha}, the deflection of particles is increasingly focused in the forward direction as the DM mass is increased. At a given value of $v$, this leads to a slight increase in the enhancement for $\gamma = 0$ compared with $\gamma = \pi$. However, from kinematics, the more energy a DM particle loses in a collision, the larger the deflection angle must be. Forward deflection (which leads to an enhancement at $\gamma = 0$) is more likely than backward deflection (enhancement at $\gamma = \pi$), but backward deflection leads to a larger energy loss of the DM particles. At very low speeds, these two effects balance and for $\oper{1}$ (top right panel) and $\oper{12}$ (bottom right panel) there is a large enhancement for all values of $\gamma$.

For operator $\oper{8}$ (middle right panel), there is instead little enhancement for large $\gamma$ and for small $v$. As can be seen from the middle panel of Fig.~\ref{fig:Pcosalpha}, the deflection of DM particles for this operator is even more strongly focused in the forward direction, with a negligible probability of backward deflection. This also means that the probability of losing a large amount of energy is small which, coupled with the increasing cross section as a function of $v$, leads to a negligible enhancement at low $v$.

We note that even though there is substantial enhancement in the DM speed distribution at low speeds for $\oper{1}$ and $\oper{12}$, this does not spoil the validity of the single-scatter approximation. These low speed particles give little contribution to the flux of DM leaving the Earth. As such, the total number of DM particles is well preserved (Tab.~\ref{tab:fluxes}): as for low mass DM, the rate of DM particles leaving the Earth matches the rate of particles entering to an accuracy of better than $1\%$.

We briefly comment on the effects of Earth-scattering for ultra-heavy DM. For DM particles much heavier than the nuclei in the Earth ($m_\chi \gg m_A$), the scattering kinematics requires that $\cos\alpha \rightarrow 1$ (see Eq.~\ref{eq:cosalpharange} and the right panel of Fig.~\ref{fig:Pcosalpha}) and therefore that $\kappa \rightarrow 1$ (i.e.~the DM particles lose no energy through scattering). Applying these limits in Eq.~\ref{eq:speeddist_deflection}, we see that the deflected DM population is always equal to the population of particles lost to attenuation (to first order in $\RE/\lambda$). Physically, ultra-heavy DM particles are not deflected by scattering in the Earth and so they continue to the detector unaffected. We therefore expect the effects of Earth-scattering to diminish as the DM mass tends to infinity, though this conclusion could change in the presence of multiple scatters or long-range interactions.

\section{Modulation signatures}
\label{sec:modulation}

We now explore how the effects of Earth-scattering on the DM speed distribution translate into effects on the direct detection event rate. The effect will vary not only as a function of the detector latitude, but also over the course of a day, as the detector moves due to the Earth's rotation. This will give rise to interesting modulation signatures. 

We first consider the calculation of the DM-induced direct detection event rate as a function of $\gamma$.  For low-mass DM, we will examine the event rate in a mock detector inspired by CRESST-II \cite{Angloher:2015ewa}. Calculation of the CRESST-II event rate is detailed in Appendix~\ref{app:CRESST-II} (including the contribution from DM scattering on both Oxygen and Calcium nuclei in the detector). For simplicity, we will compare the total number of expected signal events with and without the effects of Earth-scattering. Note that factors such as the local DM density and the detector exposure simply affect the overall normalisation of the rate and will cancel in the ratio.

\begin{figure*}[t!]
\centering
\includegraphics[width=0.45\textwidth]{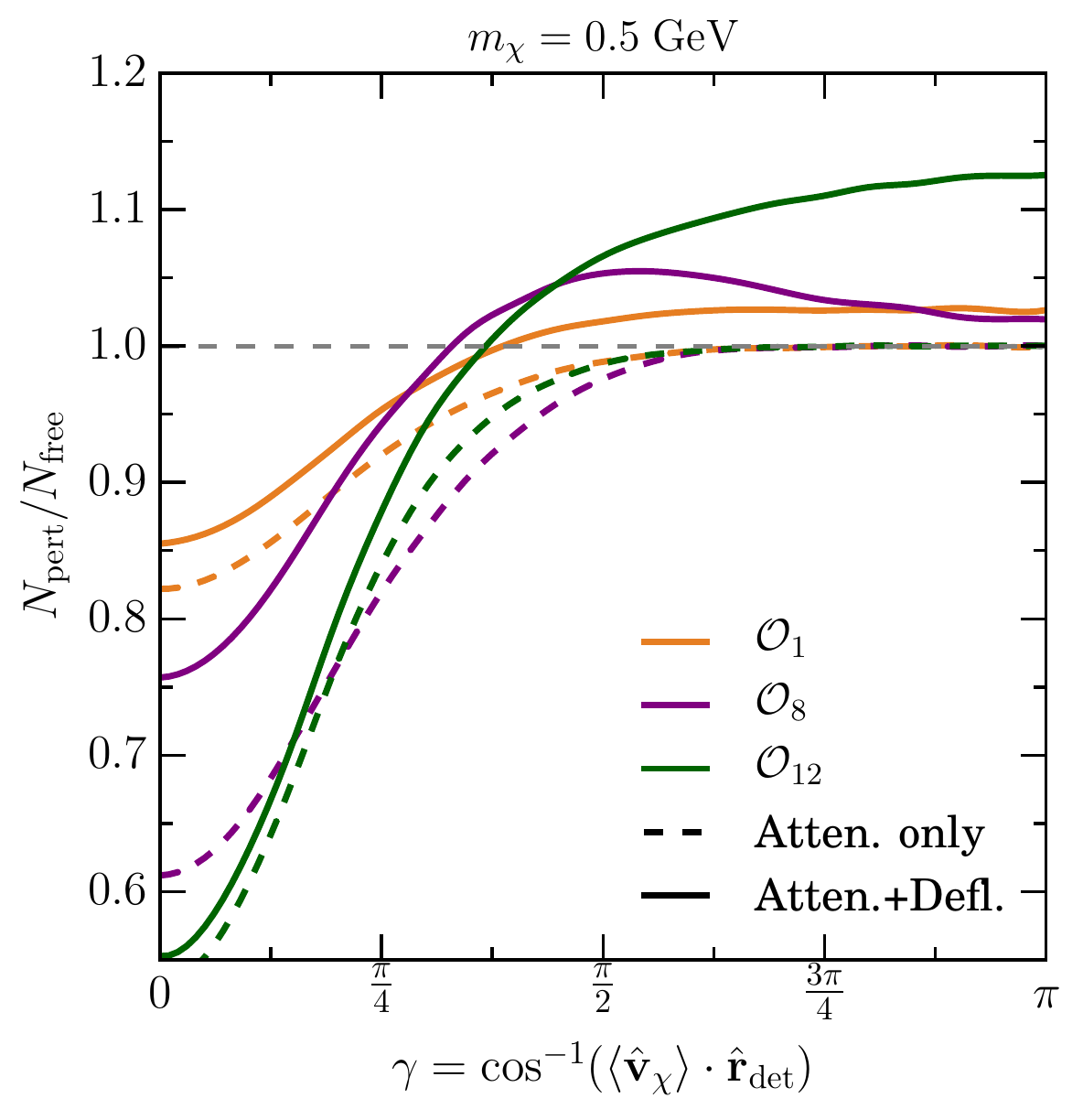}
\includegraphics[width=0.45\textwidth]{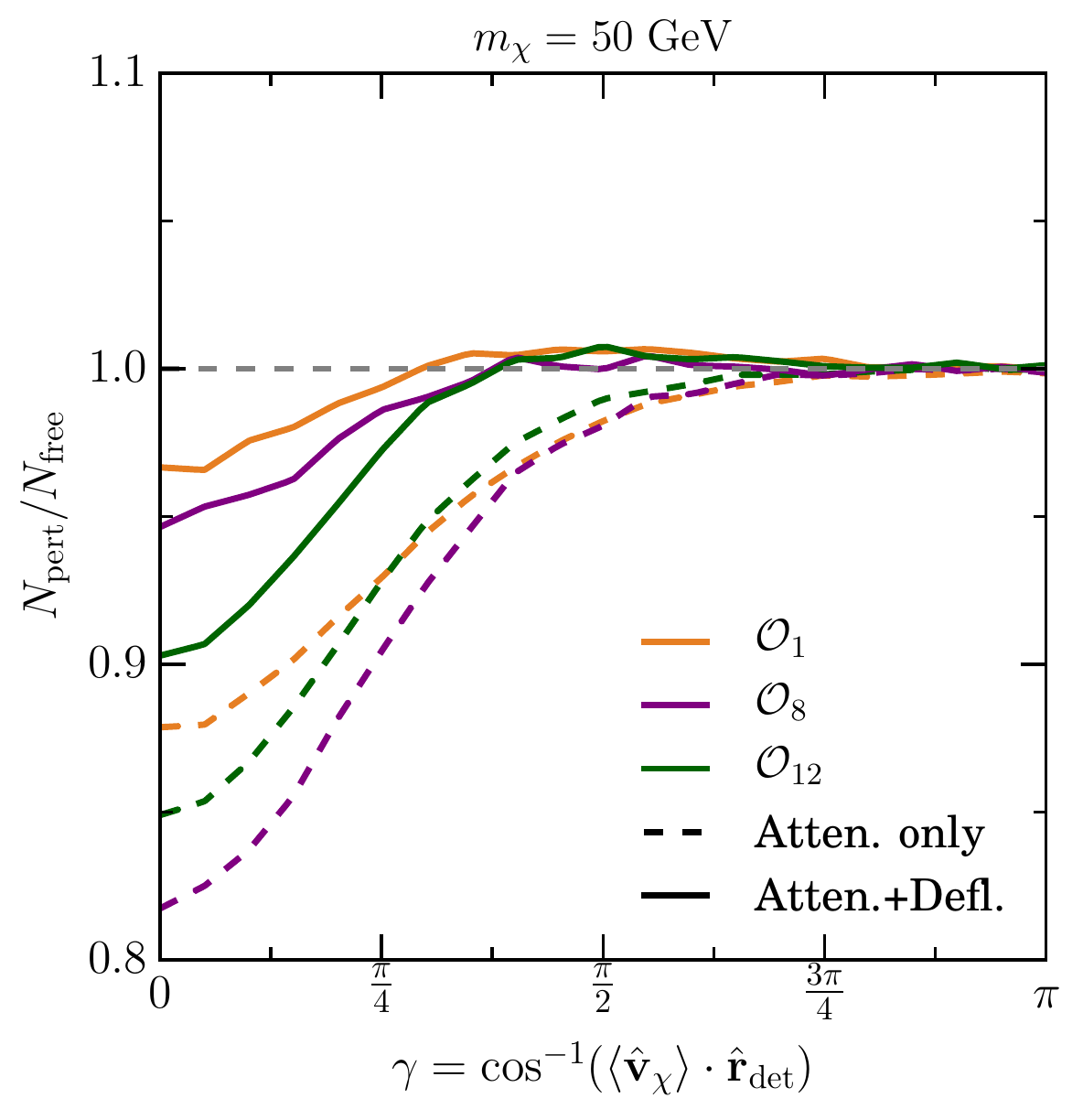}
\caption{\textbf{Ratio of the number of direct detection signal events with Earth-scattering $N_\mathrm{pert}$ and without Earth-scattering $N_\mathrm{free}$.} In the left panel, we show results for $m_\chi = 0.5 \, \, \mathrm{GeV}$ (in a CRESST-II-like detector \cite{Angloher:2015ewa}) and in the right panel for $m_\chi = 50 \, \, \mathrm{GeV}$ (in a LUX-like detector \cite{Akerib:2016vxi}). In all cases, we normalise the DM-nucleon couplings to give average scattering probability in the Earth of 10\%. Dashed lines show the results when only the effects of attenuation are included in the calculation. The angle $\gamma = 0$ ($\gamma = \pi$) corresponds to maximal (minimal) Earth-crossing before reaching the detector (see Fig.~\ref{fig:gamma}). }\label{fig:RvGamma}
\end{figure*}

In Fig.~\ref{fig:RvGamma}, we show the ratio of the number of events expected with Earth-scattering $N_\mathrm{pert}$ to the number expected without Earth-scattering $N_\mathrm{free}$ in this CRESST-II-like detector for the three operators discussed in Sec.~\ref{sec:speeddist}. Solid lines include the effects of both attenuation and deflection while dashed lines show the results when only attenuation is accounted for. For operator $\oper{1}$, we see that deflection leads to an enhancement of a few percent in the event rate, relative to the case of attenuation-only. The size of this effect is independent of $\gamma$, which is consistent with the isotropic deflection of low mass DM for $\oper{1}$. Including both effects, there is a 15\% variation in the total direct detection event rate over the full range of $\gamma$.

For operator $\oper{12}$, the size of the effect is much larger, with $N_\mathrm{pert}/N_\mathrm{free}$ ranging from 0.55 to 1.12 depending on the value of $\gamma$. For light DM, only the particles with the largest speeds have enough kinetic energy to contribute to the recoil rate. As can be seen from the bottom right panel of Fig.~\ref{fig:SpeedDist-2D}, the effect on the speed distribution is greatest at high speeds, where the cross section is enhanced ($\sigma_{12} \sim v^2$). We also note that the effect of deflection on the direct detection rate is very small for $\gamma = 0$, but gives a more than $10\%$ enhancement to the rate at $\gamma = \pi$. Once again, this is consistent with the observations in the previous section: $\oper{12}$ favours backward deflection, giving a large contribution when the DM particles transit the Earth after having passed the detector.

Finally, for $\oper{8}$ the enhancement due to deflection is maximised at $\gamma = 0$ due to preferentially forward scattering. As we increase $\gamma$, both attenuation and deflection effects reduce in size. However, the effects of deflection decrease more slowly, resulting in a peak in the event rate around $\gamma = \pi/2$. Depending on the range of $\gamma$ probed by a given detector, this may result in a phase shift in the daily modulation of the DM signal for $\oper{8}$ relative to operators $\oper{1}$ and $\oper{12}$.

We comment briefly on the modulation for the higher mass 50 GeV particle (right panel of Fig.~\ref{fig:RvGamma}). In this case, the only observable effect is a reduction in the signal rate for small $\gamma$. The reason for this can be seen in the right column of Fig.~\ref{fig:SpeedDist-2D}; the only substantial enhancement in the velocity distribution is at low speeds, which fall below the 1.1 keV threshold of the LUX experiment. We note that in this case, the main consequence of including deflected DM particles is that they partially replenish those particles lost to attenuation, reducing the size of the modulation effect.

\subsection{Daily modulation}

We now consider how this modulation as a function of $\gamma$ translates into a modulation as a function of time. To do this, we need to calculate $\gamma$ for a given time and detector latitude. First, we define a coordinate system in which the positive $z$-direction points along the Earth's North pole. The position of a detector at latitude $\theta_l$ is then\footnote{We assume by convention that latitudes in the Northern hemisphere are positive, while latitudes in the Southern hemisphere are negative.}
\begin{equation}
\hat{\mathbf{r}}_\mathrm{det} = \left(\cos\theta_l \cos\omega t,\, \cos\theta_l \sin \omega t, \,\sin\theta_l\right)\,,
\end{equation}
where $\omega = 2\pi/\mathrm{day}$ is the angular velocity of the Earth's rotation. Here, we define $t = 0$ as the time at which the detector position is maximally aligned with the Earth's velocity. In this coordinate system, the Earth's velocity with respect to the Galactic rest-frame can be written
\begin{equation}
\hat{\mathbf{v}}_e = \left(\sin\alpha , 0, \cos\alpha\right)\,,
\end{equation}
where the angle $\alpha$ varies between $36.3^\circ$ and $49.3^\circ$ over the course of a year \cite{Kouvaris:2014lpa}. For concreteness, in this work, we fix the angle $\alpha$ to $42.8^\circ$. The average DM velocity is simply $\vchi = -\mathbf{v}_e$, so we obtain: 
\begin{align}
\begin{split}
\gamma &= \acos (\vchihat \cdot \hat{\mathbf{r}}_\mathrm{det})\\
&=\acos\left(-\cos\theta_l \cos\omega t \sin \alpha - \sin\theta_l\cos\alpha\right)\,.
\end{split}
\end{align}
With this expression, we can directly map the results of Fig.~\ref{fig:RvGamma} onto the rate as a function of time (for a given detector latitude). 

\begin{figure*}[t!]
\centering
\includegraphics[trim={0 0 2.1cm 0} ,clip ,width=0.46\textwidth]{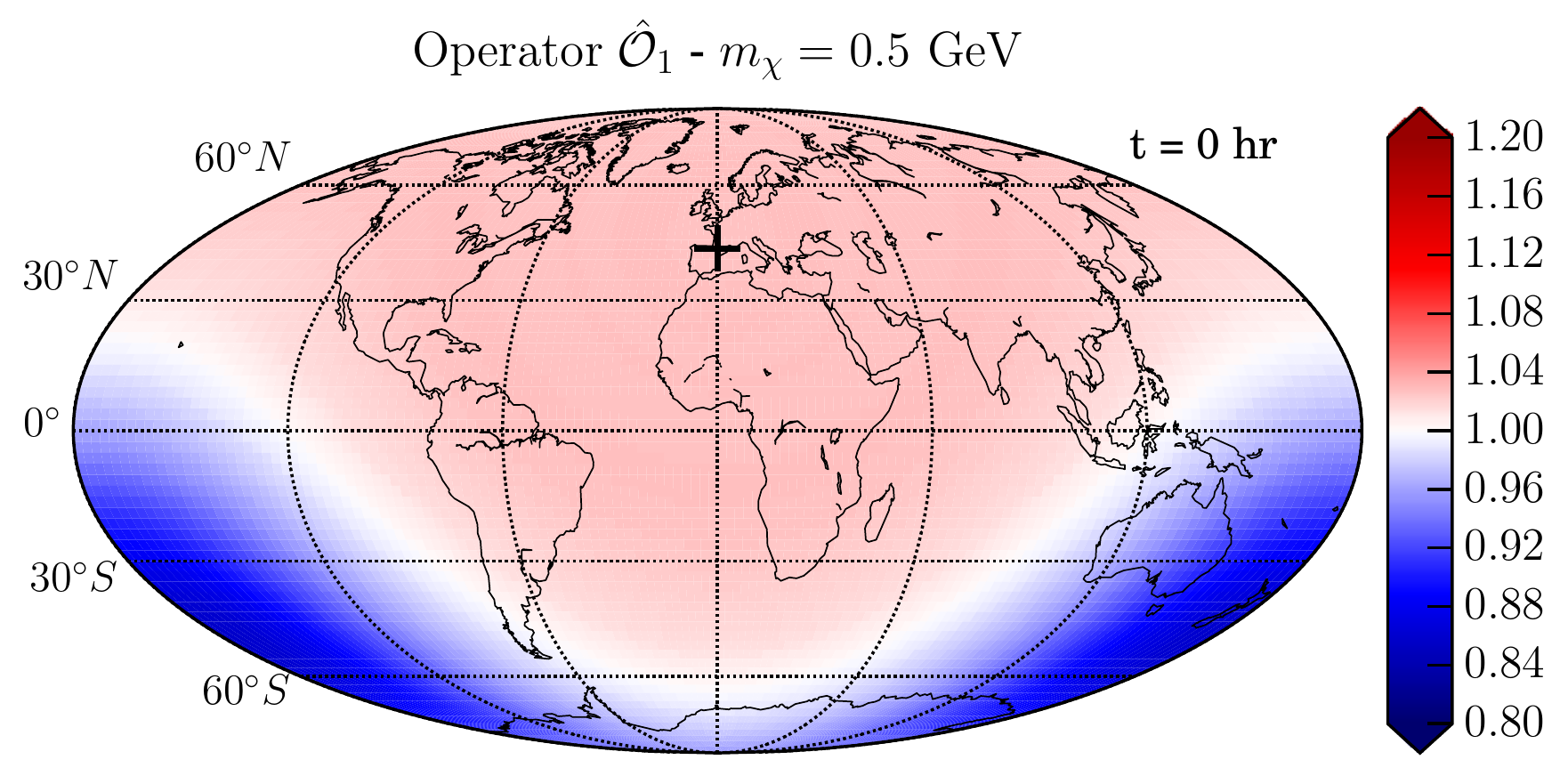}
\includegraphics[width=0.522\textwidth]{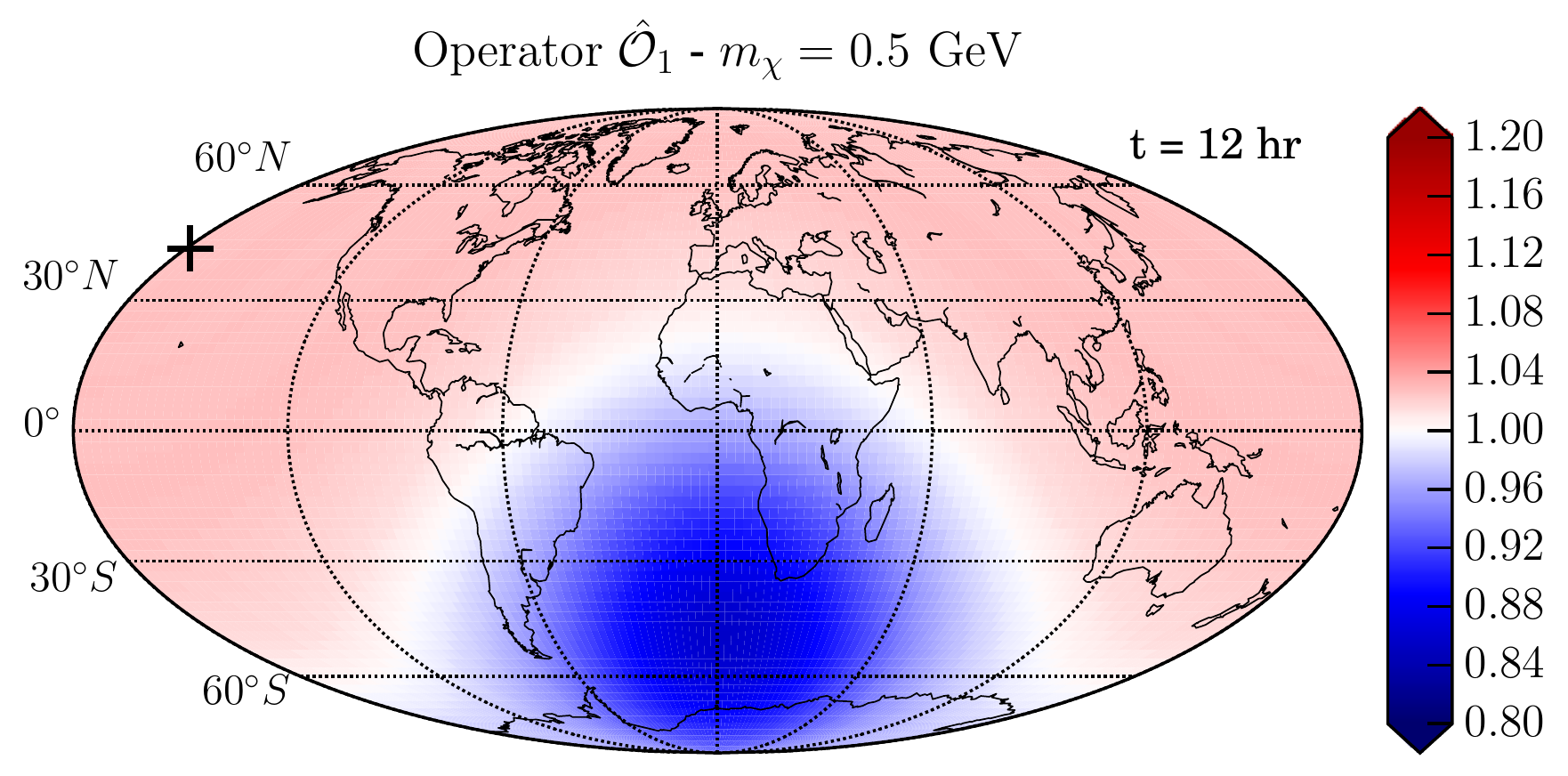}
\caption{\textbf{Earth-scattering effects over the surface of the Earth.} Relative enhancement in the event rate in a CRESST-II-like detector \cite{Angloher:2015ewa} due to the effects of Earth scattering. We assume a DM mass of $m_\chi = 0.5 \,\,\mathrm{GeV}$, interacting through the standard SI operator $\oper{1}$ (with normalisation fixed such that $p_\mathrm{scat} = 10\%$). The black cross shows the point on the Earth at which the average DM particle would appear to be coming from directly overhead. There is a 12 hour time difference between the left and right panels. Animations available online at \href{https://github.com/bradkav/EarthShadow/tree/master/videos}{github.com/bradkav/EarthShadow}.} \label{fig:EarthMap}
\end{figure*}

In Fig.~\ref{fig:EarthMap}, we show the ratio of the rate with Earth-scattering to the rate without Earth-scattering over the surface of the Earth, for 0.5 GeV DM particles interacting through the operator $\oper{1}$. The black cross (at a latitude of $~42.8^\circ \,\mathrm{N}$) shows the point on the Earth at which DM particles appear to be coming from directly overhead. As expected the maximum reduction in the expected event rate (dark blue) occurs on the opposite side of the Earth; particles emerging from this dark blue region have crossed almost the entire diameter. We notice also that the red region of the maps is much larger than the blue region.\footnote{We use an equal-area Mollweide projection to produce the maps in Fig.~\ref{fig:EarthMap}, so such a comparison between areas is reasonable.} As discussed in Sec.~\ref{sec:effects}, the effects of attenuation are large but focused only over a small area of the Earth. The two panels of Fig.~\ref{fig:EarthMap} compare the effects of Earth-scattering at 12 hour intervals. As the Earth rotates, the apparent source of the DM wind travels across the sky and the pattern of Earth-scatter effects rotates across the surface of the Earth. Animations showing the modulation signal over the entire surface of the Earth, as well as at a selection of underground laboratories, can be found online accompanying the \textsc{EarthShadow} code \cite{EarthShadow}.

\begin{figure*}[t!]
\centering
\includegraphics[width=0.45\textwidth]{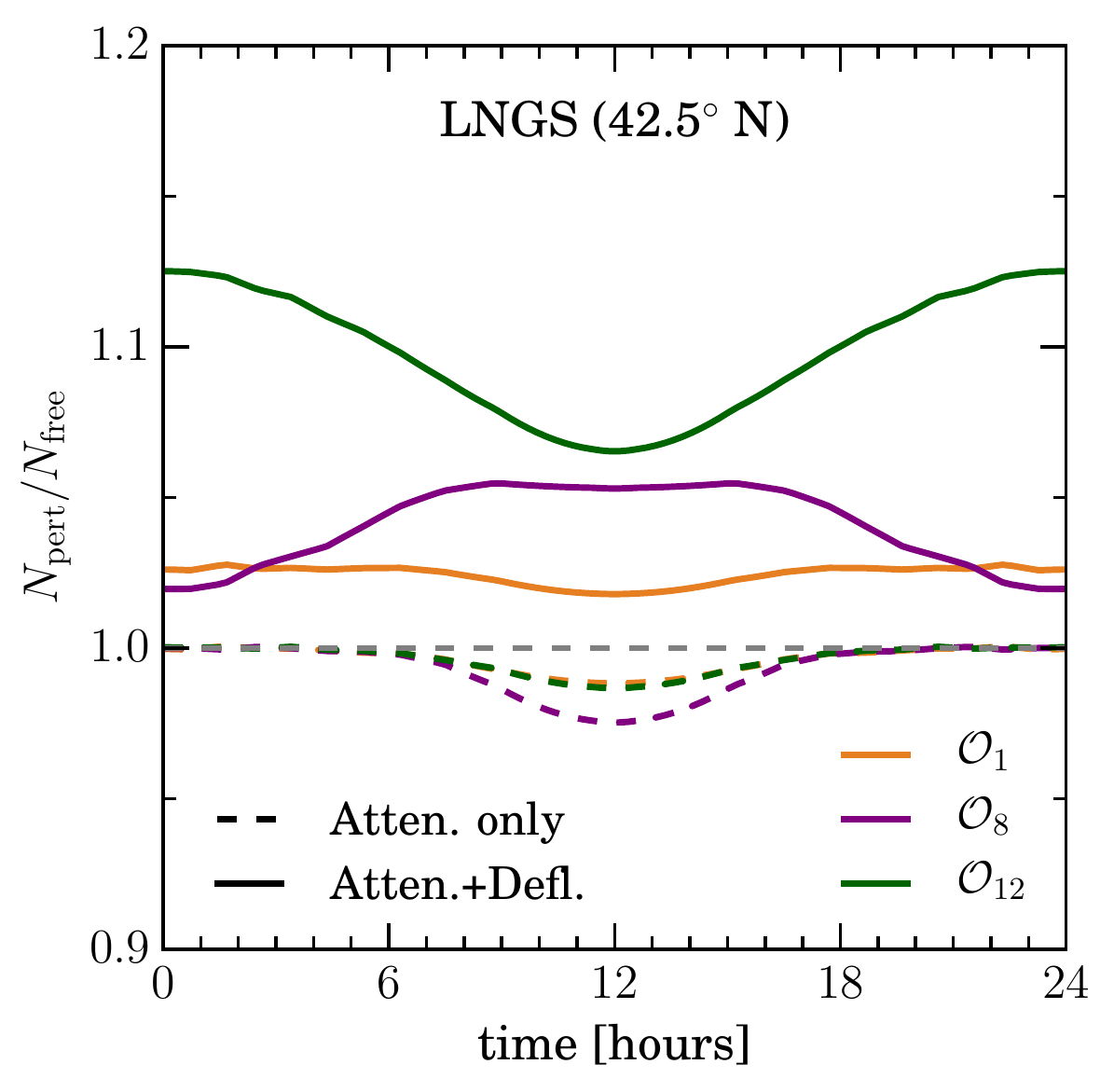}
\includegraphics[width=0.45\textwidth]{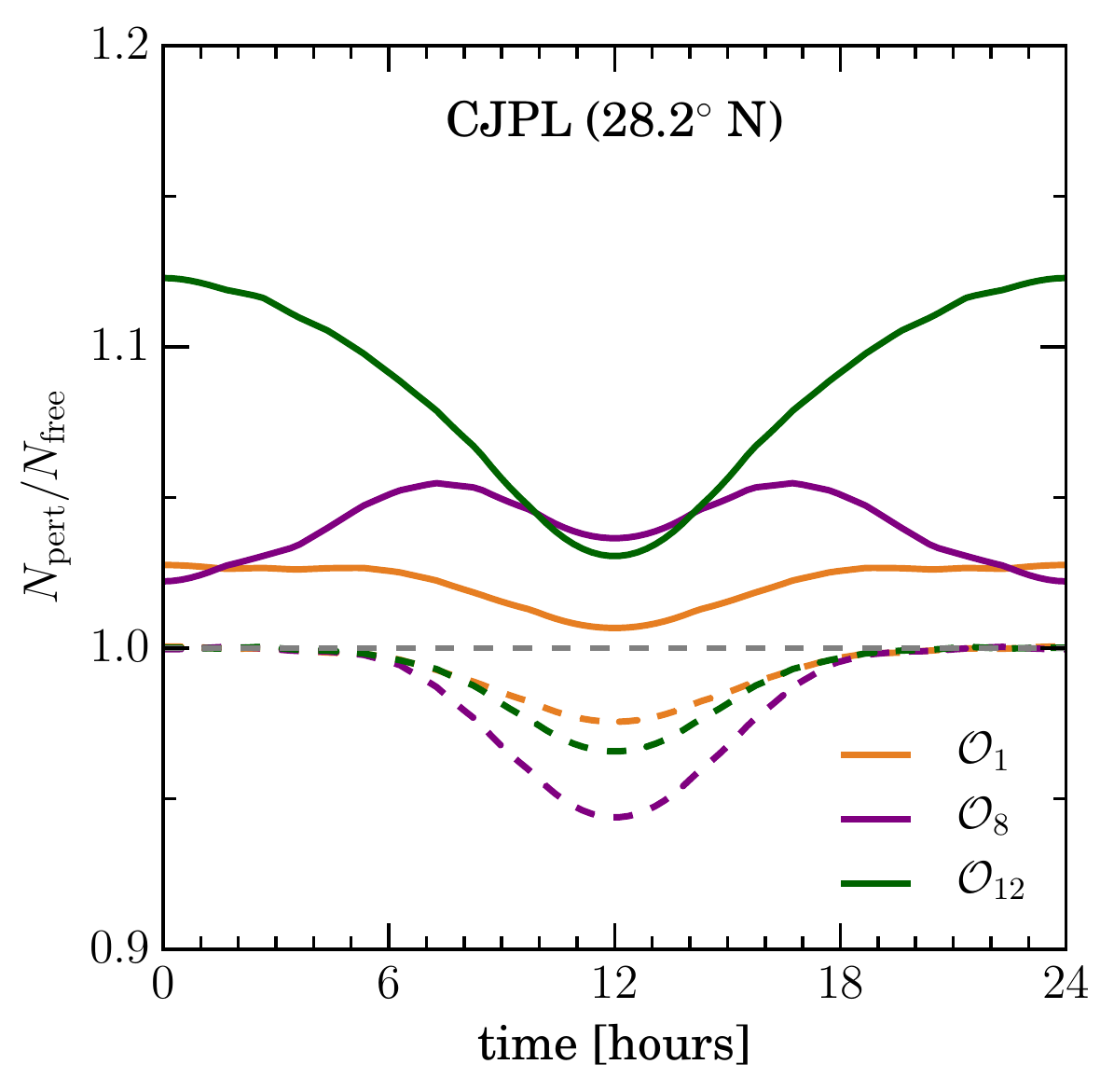}

\includegraphics[width=0.45\textwidth]{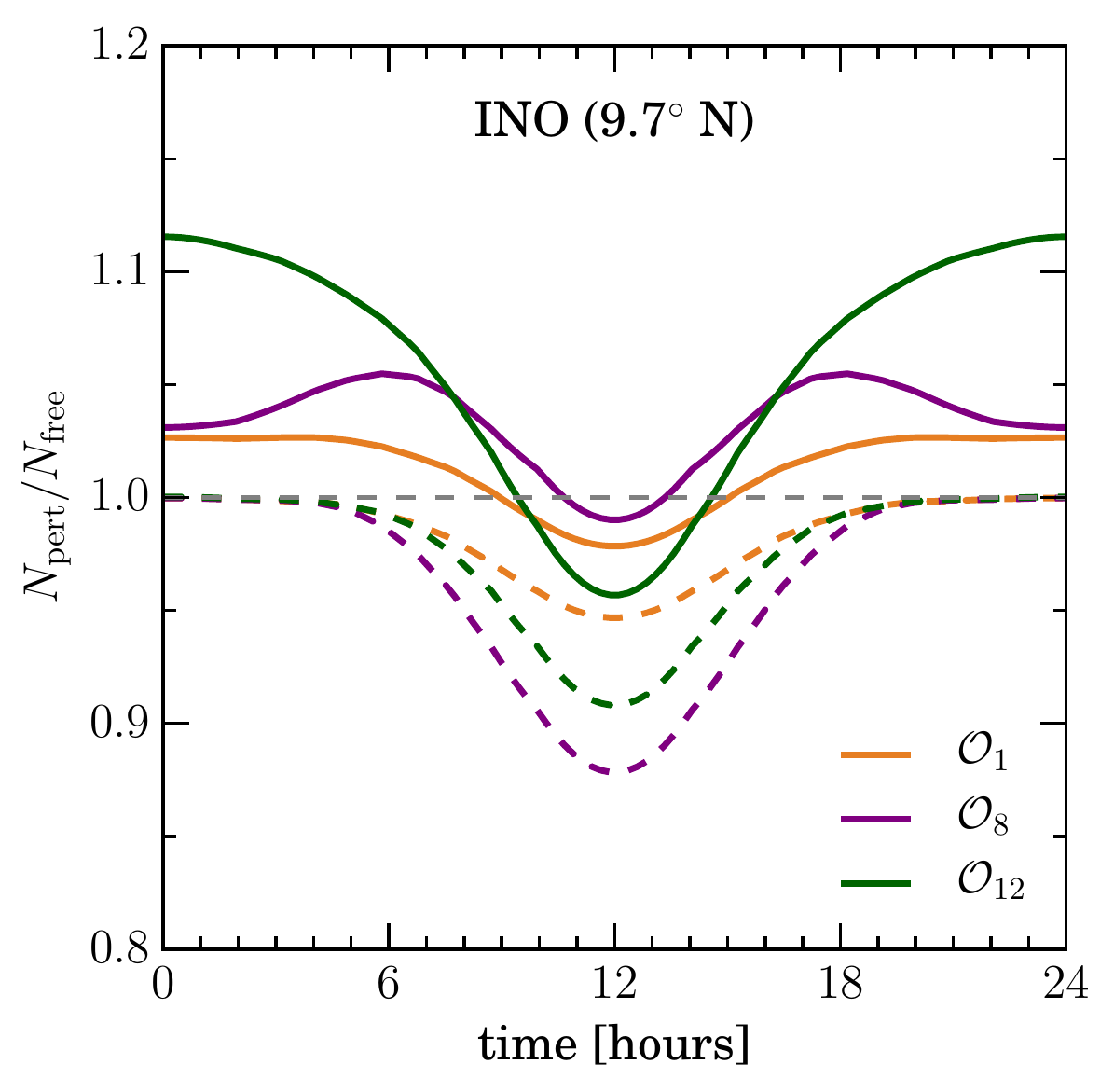}
\includegraphics[width=0.45\textwidth]{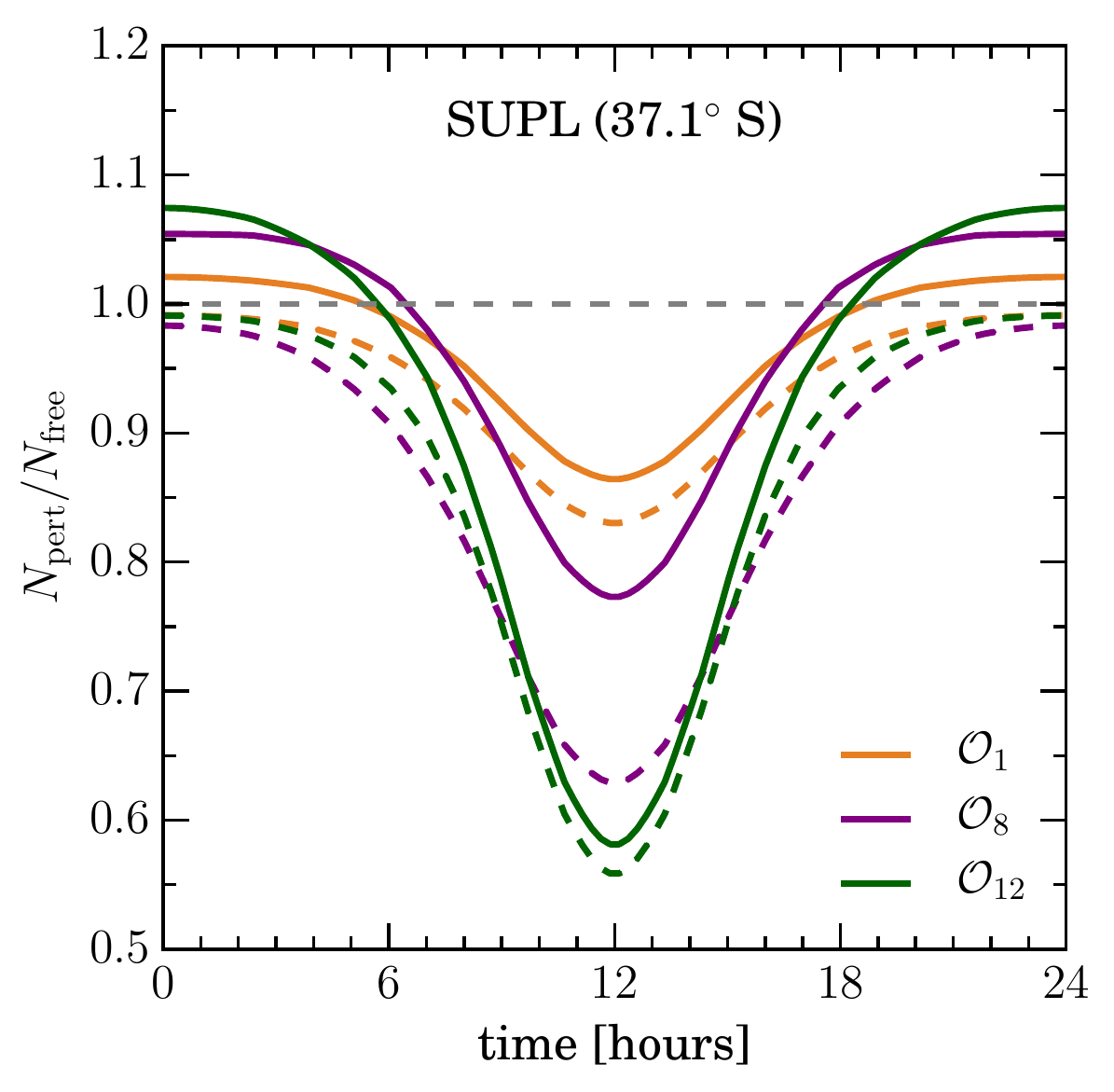}

\caption{\textbf{Daily modulation in the event rate.} Ratio of numbers of events in a CRESST-II-like detector \cite{Angloher:2015ewa} with and without the effects of Earth scattering for a DM particle of mass $m_\chi = 0.5 \,\,\mathrm{GeV}$. We fix the average scattering probability at $p_\mathrm{scat} = 10\%$. Dashed lines show the modulation when only attenuation is included, while solid lines show the effect when both attenuation and deflection are included. Each panel shows results for a lab in a different location (latitude given in parentheses).}\label{fig:Rvt}
\end{figure*}

In Fig.~\ref{fig:Rvt}, we show the ratio of the rate with and without Earth-scattering for detectors in 4 locations on the Earth. For the \href{https://www.lngs.infn.it/en}{LNGS lab} in Italy (top left, $\theta_l = 42.5^\circ \, \mathrm{N}$) Earth-scattering always leads to a net increase in the DM signal. For $\oper{1}$, however, there is no appreciable modulation, only a few percent increase in the total rate, regardless of the time. As is clear from Fig.~\ref{fig:EarthMap}, the enhancement due to Earth-scattering is roughly constant at high latitudes. At LNGS, we have $\theta_l \approx \alpha$, meaning that $\gamma$ is always relatively large. In other words, the DM wind appears to be coming from directly above for most of the day, leading to minimal attenuation. The main contribution then is a constant enhancement due to isotropic deflection of the DM particles over the Earth's surface.

Instead, a modulation of a few percent is observed for $\oper{8}$. At $t \sim 0 \,\,\mathrm{hr}$, the DM wind comes from directly overhead ($\gamma = \pi$) and the enhancement is at a minimum due to the limited deflection of DM particles back towards the detector. For $\oper{12}$, the phase of the modulation is reversed; the predominantly backwards deflection enhances the rate when DM particles come from directly overhead. 

As we move towards the Equator, the incoming direction of the DM wind varies more rapidly over the course of a day. In addition, the size of the attenuation effect becomes larger, as the underground distance travelled by DM particles at $t = 12 \, \, \mathrm{hr}$ becomes larger. For the CJPL lab \cite{Li:2014rca} in China (top right, $\theta_l = 28.2^\circ \,\mathrm{N}$) we see a more pronounced modulation than at the more northerly LNGS lab. At the India-based Neutrino Observatory \cite{Indumathi:2015hfa}, under construction in India (bottom left, $\theta_l = 9.7^\circ \, \mathrm{N}$) this modulation is more pronounced still. Most notable in this case is that for $\oper{8}$, the diurnal modulation is no longer purely sinusoidal. For this operator, DM particles predominantly scatter forwards, so we would expect a maximum enhancement due to deflection when the Earth-crossing distance is maximal ($t = 12 \, \, \mathrm{hr}$). However, at this time, the attenuation effect is also maximal and dominates the deflection effect. We therefore observe a minimum in the rate at $t = 12 \,\, \mathrm{hr}$ as well as at $t = 0 \, \, \mathrm{hr}$.

In the Southern hemisphere, the SUPL lab \cite{Urquijo:2016dxd} in Australia (bottom right, $\theta_l = 37.1^\circ \, \mathrm{S}$) is located such that at certain times of day DM particles must cross most of the Earth before reaching the detector ($\theta_l \approx - \alpha$). It therefore probes the low $\gamma$ regime, with the dominant effect being attenuation of the DM flux. The daily modulation observed here has the largest amplitude, in the range $10-30\%$ depending on the interaction. Attenuation dominates only over a small area of the Earth (blue regions in Fig.~\ref{fig:EarthMap}), but if the detector falls within this region, the effects can be substantial. While attenuation dominates, we note that the inclusion of the deflected DM population is not negligible in this case, as it counteracts this attenuation effect and in some cases even leads to a net enhancement.

\section{Discussion}
\label{sec:discussion}

We have demonstrated the effects of Earth-scattering for low mass DM on the expected event rate at direct detection experiments. Contrary to the common lore, `Earth-shadowing' in fact \textit{increases} the direct detection rate over much of the Earth's surface, as illustrated in Fig.~\ref{fig:EarthMap}. This has a large impact on the expected modulation signatures for detectors in different locations, as shown in Fig.~\ref{fig:Rvt}. In the Northern hemisphere (such as at LNGS), there is a net increase in the direct detection rate, but a relatively small modulation over the course of a day. Instead, labs in the Southern Hemisphere (such as SUPL) experience the largest modulation, due to strong attenuation effects. We have also demonstrated that the properties of this modulation depend sensitively on the interaction of DM with nuclei. Different interactions generally lead to the deflection of scattered DM particles in different directions, which has the possibility to alter both the amplitude and phase of the modulation. Laboratories closer to the Equator (such as the Indian-based Neutrino Observatory) typically have substantial contributions from both deflection and attenuation over the course of a day leading to more complicated time variation of the signal. 

The observation of such a modulation would be an indication of DM with relatively strong interactions with nucleons. The location-dependence of the modulation would then act as a useful verification of the DM origin of the signal. In addition, if the detailed time variation of the signal could be characterised, this may help us to distinguish between different DM-nucleon interactions. We note also that the size of the modulation depends only on the DM interaction cross section and not on the local DM density. Thus, the observation of a modulation due to Earth-scattering would allow us to determine both quantities separately.

We have focused here on a benchmark point just below the current CRESST-II limits, giving a scattering probability of 10\% in the Earth. This parameter space will soon be explored by CRESST-III  \cite{Strauss:2016sxp} as well as other low-threshold experiments such as SuperCDMS SNOLAB \cite{Agnese:2016cpb} and DAMIC \cite{Aguilar-Arevalo:2016ndq}. We encourage experimental collaborations to set limits not only on the standard spin-independent interactions but also on the more general operators of the NREFT. As we showed for Operator $\oper{12}$, the limits may be comparatively weaker than in the SI case, opening a wider parameter space for interesting Earth-scattering effects.

Of course, if a DM signal is observed by upcoming experiments, the number of events required to detect and characterise a modulation signature will depend on the size of the effect. From a theoretical perspective, it will be necessary to compare the Earth-scattering effect with diurnal modulation from other sources, such as gravitational focusing \cite{Kouvaris:2015xga} or time-variation of the detector velocity due to the Earth's rotation \cite{Civitarese:2016uuc}. Experimentally, it would be necessary to account for the various detector responses and their possible time variation (although in contrast to annual modulation, diurnal modulations set much less stringent requirements on detector stability). Exploring the prospects for confirming a diurnal signal in a given experiment  is beyond the scope of the current study and we leave this for future work.

We have focused on light (0.5 GeV) DM particles interacting through the three operators $\oper{1}$, $\oper{8}$ and $\oper{12}$. Of course, the space of  DM models is much wider than this, including not only the full range of DM-nucleon contact operators from NREFT, but also long-range and DM-electron interactions. The \textsc{EarthShadow} code we have developed can be used to calculate the Earth-scattering effects for any DM mass in the range $0.1-300$~GeV and for any NREFT interaction. Furthermore, the framework we have developed can be extended to accommodate more general interactions or to investigate the \textit{directional} signatures which arise from Earth-scattering. We encourage the reader to use these tools to further explore the DM-interaction parameter space for other interesting signatures.

Though we have considered arbitrary DM-nucleon interactions, the calculations we have performed are not valid for arbitrary interaction strengths. We have worked in the  `single scatter' regime, where we account for only a single scatter in the deflected DM population. As the DM-nucleon interaction strength increases, this approximation will no longer be valid. It may be possible to account for additional scatters analytically but this is likely to become rapidly intractable. An alternative approach may be to perform calculations in a `many-scatter' or `diffusion' regime in which the DM particles are not assumed to follow ballistic trajectories but instead interact so strongly that they are best described as diffusing through the Earth. A final possibility for extending these calculations would be the use of Monte-Carlo simulations,  which would allow a smooth interpolation between the `single-scatter' and `diffusion' regimes. This final possibility will be explored in an upcoming publication \cite{Emken}. The calculations presented in this work will then act as a crucial validation check for the development of such simulations.

\section{Summary}
\label{sec:conclusion}

In this work, we have presented a new analytic calculation of the Earth-scattering effect and its impact on the DM speed distribution at the Earth's surface. Working in the `single scatter' regime, we have demonstrated that our calculations are self-consistent, in that they conserve the flux of DM particles through the Earth. Though our calculations are valid for all DM masses and for a wide range of DM-nucleon interactions, we have focused on low mass (0.5 GeV) DM and a small subset of the NREFT operators introduced in Ref.~\cite{Fitzpatrick:2012ix}. The conclusions of this study are summarised as follows:
\begin{itemize}

\item Earth-scattering substantially reduces the speed distribution when DM particles must cross a large fraction of the Earth to reach the detector. This effect dominates for latitudes in the range $36^\circ\,\mathrm{S} - 49^\circ \,\mathrm{S}$. The rotation of the Earth then leads to a diurnal modulation with a large amplitude ($10-30\%$) for detectors in the Southern Hemisphere.

\item At other latitudes (and particularly in the Northern hemisphere), Earth-scattering typically leads to a net \textit{increase} in the DM speed distribution. In this case, the overall rate is slightly enhanced, but the modulation amplitude is typically smaller ($1-10\%$).

\item Different DM-nucleon interactions cause DM particles to be deflected in different directions after they have scattered in the Earth. The result is that the expected phase and amplitude of the modulation depends on the type of interaction which is assumed for the DM particle. 
\end{itemize}

The benchmark parameters assumed in this work are not excluded by current experiments, but could be explored with an increase in exposure or by upcoming detectors. If we are lucky enough to detect DM which interacts strongly  with nuclei, the characteristic diurnal modulation described here would act as unequivocal confirmation of the signal.  Such a signal would also allow us to determine the local DM density and cross section independently. The dependence of the modulation on the specific form of the DM-nucleon interaction may even provide insights into the particle identity of DM. Though we have not explored all possible DM parameters and interactions, we have demonstrated that signatures of Earth-scattering would have profound consequences if they were to be observed in the future.

\acknowledgments

BJK is supported by the European Research Council ({\sc Erc}) under the EU Seventh Framework Programme (FP7/2007-2013)/{\sc Erc} Starting Grant (agreement n.\ 278234 --- `{\sc NewDark}' project). BJK also acknowledges the hospitality of the Institut d'Astrophysique de Paris, where part of this work was completed. CK is partially funded by the Danish National Research Foundation, grant number DNRF90 and by the Danish Council for Independent Research, grant number DFF 4181-00055.

\appendix
\section{Dark matter response functions}
\label{sec:appDM}
In this appendix we define the functions $R_k^{\tau\tau'}$, $k=M,\Sigma',\Sigma'',\Phi'', \Phi'' M, \tilde{\Phi}', \Delta, \Delta \Sigma'$ appearing in Eq.~(\ref{eq:dsigma}).~We list them below using the same notation adopted in the body of the paper:
\begin{eqnarray}
 R_{M}^{\tau \tau^\prime}\left(v_{T}^{\perp 2}, {q^2 \over m_N^2}\right) &=& c_1^\tau c_1^{\tau^\prime } + {J_\chi (J_\chi+1) \over 3} \Bigg[ {q^2 \over m_N^2} v_{T}^{\perp 2} c_5^\tau c_5^{\tau^\prime } +  v_{T}^{\perp 2}c_8^\tau c_8^{\tau^\prime} + {q^2 \over m_N^2} c_{11}^\tau c_{11}^{\tau^\prime } \Bigg] \nonumber \\
 R_{\Phi^{\prime \prime}}^{\tau \tau^\prime}\left(v_{T}^{\perp 2}, {q^2 \over m_N^2}\right) &=& {q^2 \over 4 m_N^2} c_3^\tau c_3^{\tau^\prime } + {J_\chi (J_\chi+1) \over 12}   \left( c_{12}^\tau-{q^2 \over m_N^2} c_{15}^\tau\right)\hspace{-0.1 cm}\left( c_{12}^{\tau^\prime }-{q^2 \over m_N^2}c_{15}^{\tau^\prime} \right)  \nonumber \\
 R_{\Phi^{\prime \prime} M}^{\tau \tau^\prime}\left(v_{T}^{\perp 2}, {q^2 \over m_N^2}\right) &=&  c_3^\tau c_1^{\tau^\prime } + {J_\chi (J_\chi+1) \over 3}  \left( c_{12}^\tau -{q^2 \over m_N^2} c_{15}^\tau \right) c_{11}^{\tau^\prime } \nonumber \\
  R_{\tilde{\Phi}^\prime}^{\tau \tau^\prime}\left(v_{T}^{\perp 2}, {q^2 \over m_N^2}\right) &=&{J_\chi (J_\chi+1) \over 12}  \left[ c_{12}^\tau c_{12}^{\tau^\prime }+{q^2 \over m_N^2}  c_{13}^\tau c_{13}^{\tau^\prime}  \right] \nonumber \\
   R_{\Sigma^{\prime \prime}}^{\tau \tau^\prime}\left(v_{T}^{\perp 2}, {q^2 \over m_N^2}\right)  &=&{q^2 \over 4 m_N^2} c_{10}^\tau  c_{10}^{\tau^\prime } +
  {J_\chi (J_\chi+1) \over 12} \Bigg[ c_4^\tau c_4^{\tau^\prime}  + {q^2 \over m_N^2} ( c_4^\tau c_6^{\tau^\prime }+c_6^\tau c_4^{\tau^\prime }) \nonumber\\
 &+& {q^4 \over m_N^4} c_{6}^\tau c_{6}^{\tau^\prime } + v_{T}^{\perp 2} c_{12}^\tau c_{12}^{\tau^\prime }+{q^2 \over m_N^2} v_{T}^{\perp 2} c_{13}^\tau c_{13}^{\tau^\prime } \Bigg] \nonumber \\
    R_{\Sigma^\prime}^{\tau \tau^\prime}\left(v_{T}^{\perp 2}, {q^2 \over m_N^2}\right)  &=&{1 \over 8} \left[ {q^2 \over  m_N^2}  v_{T}^{\perp 2} c_{3}^\tau  c_{3}^{\tau^\prime } + v_{T}^{\perp 2}  c_{7}^\tau  c_{7}^{\tau^\prime }  \right]+ {J_\chi (J_\chi+1) \over 12} \Bigg[ c_4^\tau c_4^{\tau^\prime} + {q^2 \over m_N^2} c_9^\tau c_9^{\tau^\prime } \nonumber\\
       &+&\hspace{-0.1 cm}{v_{T}^{\perp 2} \over 2} \hspace{-0.1 cm} \left(\hspace{-0.05 cm}c_{12}^\tau-{q^2 \over m_N^2}c_{15}^\tau \hspace{-0.05 cm}\right) \hspace{-0.1 cm}\left( \hspace{-0.05 cm}c_{12}^{\tau^\prime }-{q^2 \over m_N^2}c_{15}^{\tau \prime} \hspace{-0.05 cm}\right)
       + {q^2 \over 2 m_N^2} v_{T}^{\perp 2} c_{14}^\tau c_{14}^{\tau^\prime } \Bigg] \nonumber \\
     R_{\Delta}^{\tau \tau^\prime}\left(v_{T}^{\perp 2}, {q^2 \over m_N^2}\right)&=&  {J_\chi (J_\chi+1) \over 3} \left[ {q^2 \over m_N^2} c_{5}^\tau c_{5}^{\tau^\prime }+ c_{8}^\tau c_{8}^{\tau^\prime } \right] \nonumber \\
 R_{\Delta \Sigma^\prime}^{\tau \tau^\prime}\left(v_{T}^{\perp 2}, {q^2 \over m_N^2}\right)&=& {J_\chi (J_\chi+1) \over 3} \left[c_{5}^\tau c_{4}^{\tau^\prime }-c_8^\tau c_9^{\tau^\prime} \right]. \nonumber\\
 \label{eq:R}
\end{eqnarray}
In all numerical applications, we set $J_\chi=1/2$, where $J_\chi$ is the DM particle spin.

\section{Direct detection event rates and limits}
\label{app:eventrates}

In this appendix, we outline the calculation of event rates in the CRESST-II \cite{Angloher:2015ewa} and LUX \cite{Akerib:2016vxi} experiments and describe the procedure for calculating limits on the NREFT operator couplings which are presented in Fig.~\ref{fig:limits}.

\subsection{Event rates}

In a direct detection experiment, the differential recoil rate due to DM particles impinging on a target particle $T$ is \cite{Cerdeno:2010jj}:

\begin{align}
\begin{split}
\frac{\mathrm{d}R_T}{\mathrm{d}E_R} = X_T \frac{\rho_\chi}{m_\chi m_T} \int_{\vmin}^\infty v f(v) \frac{\mathrm{d}\sigma_T}{\mathrm{d}E_R} \, \mathrm{d}v\,,
\end{split}
\end{align}
where $\rho_\chi$ is the local DM density and $v_\mathrm{min}$ is the minimum DM speed required to excite a nuclear recoil of energy $E_R$. This is given by
\begin{equation}
v_\mathrm{min} = \sqrt{\frac{m_T E_R}{2 \mu_{\chi T}^2}}\,,
\end{equation}
where $\mu_{\chi T} = m_\chi m_T/(m_\chi + m_T)$ is the reduced mass of the DM-nucleus system. The mass fraction of target particles $T$ in the detector is denoted $X_T$.

The distribution of \textit{measured} recoil energies $E'$ is then given by
\begin{align}
\begin{split}
\frac{\mathrm{d}\tilde{R}}{\mathrm{d}E'} = \int_{0}^\infty  \epsilon(E_R, E') \frac{\mathrm{d}R}{\mathrm{d}E_R} \,\mathrm{d}E_R\,,
\end{split}
\end{align}
where $\epsilon(E_R, E')$ is the probability of measuring an energy $E'$ given a recoil of energy $E_R$ and characterises the detector response. The total number of events observed in a detector is then
\begin{equation}\label{eq:Nsig}
N_\mathrm{sig} =   \int_{E_\mathrm{th}}^{E_\mathrm{max}}    \frac{\mathrm{d}\tilde{R}}{\mathrm{d}E'} \,\mathrm{d}E'\,,
\end{equation}
where we have included all events between the threshold energy $E_\mathrm{th}$ and the maximum energy of the analysis $E_\mathrm{max}$. 

\subsection{LUX}
\label{app:LUX}

The LUX detector \cite{Akerib:2016vxi} uses liquid Xenon as the target material, for which we assume natural isotopic abundances. We neglect finite energy resolution of the detector and consider recoils with energies in the range $E_R \in [1.1, \,60] \,\, \mathrm{keV}$. We take the nuclear recoil detection efficiency as the mean value from Fig.~2 of Ref.~\cite{Akerib:2016vxi}.

For Fig.~\ref{fig:limits}, we calculate approximate limits from the LUX WS2014-16 run. The data from this run is shown in Fig.~1 of Ref.~\cite{Akerib:2016vxi}. We include all events (both filled and unfilled circles) below the 90\% contour of the nuclear recoil band (upper dashed red curve). We estimate the total number of events observed in this region as $N_\mathrm{obs} \approx 120$.

The expected background in this region is estimated using the fitted background counts in Table~1 of Ref.~\cite{Akerib:2016vxi}. The expected nuclear recoil background is taken to be 90\% of the $^8\mathrm{B}$ and PTFE surface counts. The 10\% electron recoil contour (lower dashed blue curve) roughly coincides with the 90\% nuclear recoil contour, so we estimate the electronic recoil backgrounds to be 10\% of the $\gamma$ and $\beta$ counts. The total number of expected background events is the sum of these two, $N_\mathrm{BG} \approx 131.2$. The number of expected signal events is calculated from Eq.~\ref{eq:Nsig} (rescaled by 90\%). 

The standard 90\% upper limit on the total number of signal events $N_\mathrm{sig}^{90\%}$ is given by:~\cite{Feldman:1997qc}
\begin{equation}\label{eq:upperlimit}
\sum_{k = N_\mathrm{obs} + 1}^\infty P(k | N_\mathrm{BG} + N_\mathrm{sig}^{90\%}) = 90\%\,,
\end{equation}
where $P(k | N)$ is the Poisson probability of observing $k$ events when an average of $N$ events is expected. Equation~\ref{eq:upperlimit} is solved numerically for $N_\mathrm{sig}^{90\%}$, which can then be converted into an upper limit on the DM-nucleon coupling (or cross section) for a given DM mass $m_\chi$. We have checked the resulting limits against the reported SI cross section limits; our limits are within a factor of 2 of the reported ones at high mass.

\subsection{CRESST-II}
\label{app:CRESST-II}

The CRESST-II detector \cite{Angloher:2015ewa} uses scintillating $\mathrm{CaWO}_4$ crystals as the target material. In calculating the signal rate, we include only the contribution from Oxygen ($X_\mathrm{O} = 0.22$) and Calcium ($X_\mathrm{Ca} = 0.14$) recoils, for which nuclear response functions are available. The complicated nuclear structure of Tungsten would require dedicated calculations of the nuclear form factors which are beyond the scope of this work. In this case, Oxygen and Calcium recoils are expected to dominate the rate for low mass DM (see Fig.~7 of Ref.~\cite{Angloher:2015ewa}).  

We use a threshold energy of $E_\mathrm{th} = 300 \, \, \mathrm{eV}$. The detector response is assumed to be a Gaussian with width $\sigma_E = 62 \, \, \mathrm{eV}$, multiplied by the detector efficiency $\eta(E')$ (extracted from Fig.~2 of Ref.~\cite{Angloher:2015ewa}):
\begin{equation}
\epsilon(E_R, E') = 0.5 \frac{\eta(E')}{\sqrt{2 \pi \sigma_E^2}} \exp \left[ -\frac{(E_R - E')^2}{2 \sigma_E^2} \right]\,.
\end{equation}
The factor of $0.5$ accounts for the fact that the acceptance region is only the lower half of the Oxygen nuclear recoil band. We have verified that we obtain the correct number of signal events for SI interactions by comparing with Fig.~7 of Ref.~\cite{Angloher:2015ewa}.

In the CRESST-II analysis, all events inside the acceptance region are conservatively considered as possible signal events. These events are shown in bins of 100 eV in Fig.~6 of Ref.~\cite{Angloher:2015ewa}. Because there is no background estimation ($N_\mathrm{BG} = 0$), we consider only those `signal' events which could feasibly have been due DM-nucleus scattering. We therefore include only events with $E_R < E_\mathrm{max}$, where $E_\mathrm{max} = 2 \mu_{\chi T}^2 (v_e + v_\mathrm{esc})^2/m_T + 3 \sigma_E$. This is the maximum energy recoil which could be produced by a DM particle of mass $m_\chi$ (plus a small factor of $3\sigma_E$ accounting for the finite energy resolution of the detector). The number of observed events $N_\mathrm{obs}$ is therefore taken as the sum of all events with energies between $E_\mathrm{th} = 300 \,\, \mathrm{eV}$ and $E_\mathrm{max}$.

The $90\%$ Poisson upper limit on the DM-nucleon couplings are then calculated as for LUX (see Eq.~\ref{eq:upperlimit}), but for $N_\mathrm{BG} = 0$ and with $N_\mathrm{obs}$ depending on the DM mass. We have checked the resulting limits against the reported limits on SI interactions. Our limits are roughly a factor of 4 higher in the DM-nucleon cross section.

\bibliographystyle{JHEP}
\bibliography{SingleScatter.bib,ref_rc.bib,Chris.bib}

\providecommand{\href}[2]{#2}\begingroup\raggedright\begin{thebibliography}{10}

\bibitem{Mitsou:2013rwa}
V.~A. Mitsou, \emph{{Shedding Light on Dark Matter at Colliders}},
  \href{http://dx.doi.org/10.1142/S0217751X13300524}{\emph{Int. J. Mod. Phys.}
  {\bf A28} (2013) 1330052}, [\href{https://arxiv.org/abs/1310.1072}{{\tt
  1310.1072}}].

\bibitem{Gaskins:2016cha}
J.~M. Gaskins, \emph{{A review of indirect searches for particle dark matter}},
   \href{https://arxiv.org/abs/1604.00014}{{\tt 1604.00014}}.

\bibitem{Goodman:1984dc}
M.~W. Goodman and E.~Witten, \emph{{Detectability of Certain Dark Matter
  Candidates}}, \href{http://dx.doi.org/10.1103/PhysRevD.31.3059}{\emph{Phys.
  Rev.} {\bf D31} (1985) 3059}.

\bibitem{Drukier:1986tm}
A.~Drukier, K.~Freese and D.~Spergel, \emph{{Detecting Cold Dark Matter
  Candidates}}, \href{http://dx.doi.org/10.1103/PhysRevD.33.3495}{\emph{Phys.
  Rev.} {\bf D33} (1986) 3495--3508}.

\bibitem{Agnese:2015nto}
{\scshape SuperCDMS} collaboration, R.~Agnese et~al., \emph{{New Results from
  the Search for Low-Mass Weakly Interacting Massive Particles with the CDMS
  Low Ionization Threshold Experiment}},
  \href{http://dx.doi.org/10.1103/PhysRevLett.116.071301}{\emph{Phys. Rev.
  Lett.} {\bf 116} (2016) 071301},
  [\href{https://arxiv.org/abs/1509.02448}{{\tt 1509.02448}}].

\bibitem{Angloher:2015ewa}
{\scshape CRESST} collaboration, G.~Angloher et~al., \emph{{Results on light
  dark matter particles with a low-threshold CRESST-II detector}},
  \href{http://dx.doi.org/10.1140/epjc/s10052-016-3877-3}{\emph{Eur. Phys. J.}
  {\bf C76} (2016) 25}, [\href{https://arxiv.org/abs/1509.01515}{{\tt
  1509.01515}}].

\bibitem{Akerib:2016vxi}
D.~S. Akerib et~al., \emph{{Results from a search for dark matter in the
  complete LUX exposure}},  \href{https://arxiv.org/abs/1608.07648}{{\tt
  1608.07648}}.

\bibitem{Tan:2016zwf}
{\scshape PandaX-II} collaboration, A.~Tan et~al., \emph{{Dark Matter Results
  from First 98.7 Days of Data from the PandaX-II Experiment}},
  \href{http://dx.doi.org/10.1103/PhysRevLett.117.121303}{\emph{Phys. Rev.
  Lett.} {\bf 117} (2016) 121303},
  [\href{https://arxiv.org/abs/1607.07400}{{\tt 1607.07400}}].

\bibitem{Essig:2011nj}
R.~Essig, J.~Mardon and T.~Volansky, \emph{{Direct Detection of Sub-GeV Dark
  Matter}}, \href{http://dx.doi.org/10.1103/PhysRevD.85.076007}{\emph{Phys.
  Rev.} {\bf D85} (2012) 076007}, [\href{https://arxiv.org/abs/1108.5383}{{\tt
  1108.5383}}].

\bibitem{Pollack:2014rja}
J.~Pollack, D.~N. Spergel and P.~J. Steinhardt, \emph{{Supermassive Black Holes
  from Ultra-Strongly Self-Interacting Dark Matter}},
  \href{http://dx.doi.org/10.1088/0004-637X/804/2/131}{\emph{Astrophys. J.}
  {\bf 804} (2015) 131}, [\href{https://arxiv.org/abs/1501.00017}{{\tt
  1501.00017}}].

\bibitem{Collar:1992qc}
J.~I. Collar and F.~T. Avignone, \emph{{Diurnal modulation effects in cold dark
  matter experiments}},
  \href{http://dx.doi.org/10.1016/0370-2693(92)90873-3}{\emph{Phys. Lett.} {\bf
  B275} (1992) 181--185}.

\bibitem{Collar:1993ss}
J.~I. Collar and F.~T. Avignone, III, \emph{{The Effect of elastic scattering
  in the Earth on cold dark matter experiments}},
  \href{http://dx.doi.org/10.1103/PhysRevD.47.5238}{\emph{Phys. Rev.} {\bf D47}
  (1993) 5238--5246}.

\bibitem{Hasenbalg:1997hs}
F.~Hasenbalg, D.~Abriola, F.~T. Avignone, J.~I. Collar, D.~E. Di~Gregorio,
  A.~O. Gattone et~al., \emph{{Cold dark matter identification: Diurnal
  modulation revisited}},
  \href{http://dx.doi.org/10.1103/PhysRevD.55.7350}{\emph{Phys. Rev.} {\bf D55}
  (1997) 7350--7355}, [\href{https://arxiv.org/abs/astro-ph/9702165}{{\tt
  astro-ph/9702165}}].

\bibitem{Foot:2003iv}
R.~Foot, \emph{{Implications of the DAMA and CRESST experiments for mirror
  matter type dark matter}},
  \href{http://dx.doi.org/10.1103/PhysRevD.69.036001}{\emph{Phys. Rev.} {\bf
  D69} (2004) 036001}, [\href{https://arxiv.org/abs/hep-ph/0308254}{{\tt
  hep-ph/0308254}}].

\bibitem{Zaharijas:2004jv}
G.~Zaharijas and G.~R. Farrar, \emph{{A Window in the dark matter exclusion
  limits}}, \href{http://dx.doi.org/10.1103/PhysRevD.72.083502}{\emph{Phys.
  Rev.} {\bf D72} (2005) 083502},
  [\href{https://arxiv.org/abs/astro-ph/0406531}{{\tt astro-ph/0406531}}].

\bibitem{Sigurdson:2004zp}
K.~Sigurdson, M.~Doran, A.~Kurylov, R.~R. Caldwell and M.~Kamionkowski,
  \emph{{Dark-matter electric and magnetic dipole moments}},
  \href{http://dx.doi.org/10.1103/PhysRevD.70.083501,
  10.1103/PhysRevD.73.089903}{\emph{Phys. Rev.} {\bf D70} (2004) 083501},
  [\href{https://arxiv.org/abs/astro-ph/0406355}{{\tt astro-ph/0406355}}].

\bibitem{Mack:2007xj}
G.~D. Mack, J.~F. Beacom and G.~Bertone, \emph{{Towards Closing the Window on
  Strongly Interacting Dark Matter: Far-Reaching Constraints from Earth's Heat
  Flow}}, \href{http://dx.doi.org/10.1103/PhysRevD.76.043523}{\emph{Phys. Rev.}
  {\bf D76} (2007) 043523}, [\href{https://arxiv.org/abs/0705.4298}{{\tt
  0705.4298}}].

\bibitem{Cline:2012is}
J.~M. Cline, Z.~Liu and W.~Xue, \emph{{Millicharged Atomic Dark Matter}},
  \href{http://dx.doi.org/10.1103/PhysRevD.85.101302}{\emph{Phys. Rev.} {\bf
  D85} (2012) 101302}, [\href{https://arxiv.org/abs/1201.4858}{{\tt
  1201.4858}}].

\bibitem{Daci:2015hca}
N.~Daci, I.~De~Bruyn, S.~Lowette, M.~H.~G. Tytgat and B.~Zaldivar,
  \emph{{Simplified SIMPs and the LHC}},
  \href{http://dx.doi.org/10.1007/JHEP11(2015)108}{\emph{JHEP} {\bf 11} (2015)
  108}, [\href{https://arxiv.org/abs/1503.05505}{{\tt 1503.05505}}].

\bibitem{Lee:2015qva}
S.~K. Lee, M.~Lisanti, S.~Mishra-Sharma and B.~R. Safdi, \emph{{Modulation
  Effects in Dark Matter-Electron Scattering Experiments}},
  \href{http://dx.doi.org/10.1103/PhysRevD.92.083517}{\emph{Phys. Rev.} {\bf
  D92} (2015) 083517}, [\href{https://arxiv.org/abs/1508.07361}{{\tt
  1508.07361}}].

\bibitem{Kouvaris:2014lpa}
C.~Kouvaris and I.~M. Shoemaker, \emph{{Daily modulation as a smoking gun of
  dark matter with significant stopping rate}},
  \href{http://dx.doi.org/10.1103/PhysRevD.90.095011}{\emph{Phys. Rev.} {\bf
  D90} (2014) 095011}, [\href{https://arxiv.org/abs/1405.1729}{{\tt
  1405.1729}}].

\bibitem{Kouvaris:2015laa}
C.~Kouvaris, \emph{{Earth’s stopping effect in directional dark matter
  detectors}}, \href{http://dx.doi.org/10.1103/PhysRevD.93.035023}{\emph{Phys.
  Rev.} {\bf D93} (2016) 035023}, [\href{https://arxiv.org/abs/1509.08720}{{\tt
  1509.08720}}].

\bibitem{Foot:2011fh}
R.~Foot, \emph{{Diurnal modulation due to self-interacting mirror and hidden
  sector dark matter}},
  \href{http://dx.doi.org/10.1088/1475-7516/2012/04/014}{\emph{JCAP} {\bf 1204}
  (2012) 014}, [\href{https://arxiv.org/abs/1110.2908}{{\tt 1110.2908}}].

\bibitem{Foot:2014osa}
R.~Foot and S.~Vagnozzi, \emph{{Diurnal modulation signal from dissipative
  hidden sector dark matter}},
  \href{http://dx.doi.org/10.1016/j.physletb.2015.06.063}{\emph{Phys. Lett.}
  {\bf B748} (2015) 61--66}, [\href{https://arxiv.org/abs/1412.0762}{{\tt
  1412.0762}}].

\bibitem{Clarke:2015gqw}
J.~D. Clarke and R.~Foot, \emph{{Plasma dark matter direct detection}},
  \href{http://dx.doi.org/10.1088/1475-7516/2016/01/029}{\emph{JCAP} {\bf 1601}
  (2016) 029}, [\href{https://arxiv.org/abs/1512.06471}{{\tt 1512.06471}}].

\bibitem{Bernabei:2015nia}
R.~Bernabei et~al., \emph{{Investigating Earth shadowing effect with
  DAMA/LIBRA-phase1}},
  \href{http://dx.doi.org/10.1140/epjc/s10052-015-3473-y}{\emph{Eur. Phys. J.}
  {\bf C75} (2015) 239}, [\href{https://arxiv.org/abs/1505.05336}{{\tt
  1505.05336}}].

\bibitem{Fitzpatrick:2012ix}
A.~L. Fitzpatrick, W.~Haxton, E.~Katz, N.~Lubbers and Y.~Xu, \emph{{The
  Effective Field Theory of Dark Matter Direct Detection}},
  \href{http://dx.doi.org/10.1088/1475-7516/2013/02/004}{\emph{JCAP} {\bf 1302}
  (2013) 004}, [\href{https://arxiv.org/abs/1203.3542}{{\tt 1203.3542}}].

\bibitem{EarthShadow}
B.~J. Kavanagh, R.~Catena and C.~Kouvaris, \emph{\textsc{EarthShadow} v1.0},
  {\emph{\href{http://ascl.net/1611.012}{Astrophysics Source Code Library,
  record ascl:1611.012}} (2016)}. Available at \href{https://github.com/bradkav/EarthShadow}{https://github.com/bradkav/EarthShadow}.

\bibitem{Chang:2009yt}
S.~Chang, A.~Pierce and N.~Weiner, \emph{{Momentum Dependent Dark Matter
  Scattering}},
  \href{http://dx.doi.org/10.1088/1475-7516/2010/01/006}{\emph{JCAP} {\bf 1001}
  (2010) 006}, [\href{https://arxiv.org/abs/0908.3192}{{\tt 0908.3192}}].

\bibitem{Fan:2010gt}
J.~Fan, M.~Reece and L.-T. Wang, \emph{{Non-relativistic effective theory of
  dark matter direct detection}},
  \href{http://dx.doi.org/10.1088/1475-7516/2010/11/042}{\emph{JCAP} {\bf 1011}
  (2010) 042}, [\href{https://arxiv.org/abs/1008.1591}{{\tt 1008.1591}}].

\bibitem{Fitzpatrick:2012ib}
A.~L. Fitzpatrick, W.~Haxton, E.~Katz, N.~Lubbers and Y.~Xu, \emph{{Model
  Independent Direct Detection Analyses}},
  \href{https://arxiv.org/abs/1211.2818}{{\tt 1211.2818}}.

\bibitem{Anand:2013yka}
N.~Anand, A.~L. Fitzpatrick and W.~Haxton, \emph{{Model-independent WIMP
  Scattering Responses and Event Rates: A Mathematica Package for Experimental
  Analysis}},
  \href{http://dx.doi.org/10.1103/PhysRevC.89.065501}{\emph{Phys.Rev.} {\bf
  C89} (2014) 065501}, [\href{https://arxiv.org/abs/1308.6288}{{\tt
  1308.6288}}].

\bibitem{Menendez:2012tm}
J.~Menendez, D.~Gazit and A.~Schwenk, \emph{{Spin-dependent WIMP scattering off
  nuclei}},
  \href{http://dx.doi.org/10.1103/PhysRevD.86.103511}{\emph{Phys.Rev.} {\bf
  D86} (2012) 103511}, [\href{https://arxiv.org/abs/1208.1094}{{\tt
  1208.1094}}].

\bibitem{Cirigliano:2012pq}
V.~Cirigliano, M.~L. Graesser and G.~Ovanesyan, \emph{{WIMP-nucleus scattering
  in chiral effective theory}},
  \href{http://dx.doi.org/10.1007/JHEP10(2012)025}{\emph{JHEP} {\bf 1210}
  (2012) 025}, [\href{https://arxiv.org/abs/1205.2695}{{\tt 1205.2695}}].

\bibitem{DelNobile:2013sia}
M.~Cirelli, E.~Del~Nobile and P.~Panci, \emph{{Tools for model-independent
  bounds in direct dark matter searches}},
  \href{http://dx.doi.org/10.1088/1475-7516/2013/10/019}{\emph{JCAP} {\bf 1310}
  (2013) 019}, [\href{https://arxiv.org/abs/1307.5955}{{\tt 1307.5955}}].

\bibitem{Klos:2013rwa}
P.~Klos, J.~Menéndez, D.~Gazit and A.~Schwenk, \emph{{Large-scale nuclear
  structure calculations for spin-dependent WIMP scattering with chiral
  effective field theory currents}},
  \href{http://dx.doi.org/10.1103/PhysRevD.89.029901,
  10.1103/PhysRevD.88.083516}{\emph{Phys.Rev.} {\bf D88} (2013) 083516},
  [\href{https://arxiv.org/abs/1304.7684}{{\tt 1304.7684}}].

\bibitem{Peter:2013aha}
A.~H. Peter, V.~Gluscevic, A.~M. Green, B.~J. Kavanagh and S.~K. Lee,
  \emph{{WIMP physics with ensembles of direct-detection experiments}},
  \href{http://dx.doi.org/10.1016/j.dark.2014.10.006}{\emph{Phys.Dark Univ.}
  {\bf 5-6} (2014) 45--74}, [\href{https://arxiv.org/abs/1310.7039}{{\tt
  1310.7039}}].

\bibitem{Hill:2013hoa}
R.~J. Hill and M.~P. Solon, \emph{{WIMP-nucleon scattering with heavy WIMP
  effective theory}},
  \href{http://dx.doi.org/10.1103/PhysRevLett.112.211602}{\emph{Phys.Rev.Lett.}
  {\bf 112} (2014) 211602}, [\href{https://arxiv.org/abs/1309.4092}{{\tt
  1309.4092}}].

\bibitem{Catena:2014uqa}
R.~Catena and P.~Gondolo, \emph{{Global fits of the dark matter-nucleon
  effective interactions}},
  \href{http://dx.doi.org/10.1088/1475-7516/2014/09/045}{\emph{JCAP} {\bf 1409}
  (2014) 045}, [\href{https://arxiv.org/abs/1405.2637}{{\tt 1405.2637}}].

\bibitem{Catena:2014hla}
R.~Catena, \emph{{Analysis of the theoretical bias in dark matter direct
  detection}},
  \href{http://dx.doi.org/10.1088/1475-7516/2014/09/049}{\emph{JCAP} {\bf 1409}
  (2014) 049}, [\href{https://arxiv.org/abs/1407.0127}{{\tt 1407.0127}}].

\bibitem{Catena:2014epa}
R.~Catena, \emph{{Prospects for direct detection of dark matter in an effective
  theory approach}},
  \href{http://dx.doi.org/10.1088/1475-7516/2014/07/055}{\emph{JCAP} {\bf 1407}
  (2014) 055}, [\href{https://arxiv.org/abs/1406.0524}{{\tt 1406.0524}}].

\bibitem{Gluscevic:2014vga}
V.~Gluscevic and A.~H.~G. Peter, \emph{{Understanding WIMP-baryon interactions
  with direct detection: A Roadmap}},
  \href{http://dx.doi.org/10.1088/1475-7516/2014/09/040}{\emph{JCAP} {\bf 1409}
  (2014) 040}, [\href{https://arxiv.org/abs/1406.7008}{{\tt 1406.7008}}].

\bibitem{Panci:2014gga}
P.~Panci, \emph{{New Directions in Direct Dark Matter Searches}},
  \href{http://dx.doi.org/10.1155/2014/681312}{\emph{Adv.High Energy Phys.}
  {\bf 2014} (2014) 681312}, [\href{https://arxiv.org/abs/1402.1507}{{\tt
  1402.1507}}].

\bibitem{Vietze:2014vsa}
L.~Vietze, P.~Klos, J.~Menéndez, W.~Haxton and A.~Schwenk, \emph{{Nuclear
  structure aspects of spin-independent WIMP scattering off xenon}},
  \href{https://arxiv.org/abs/1412.6091}{{\tt 1412.6091}}.

\bibitem{Barello:2014uda}
G.~Barello, S.~Chang and C.~A. Newby, \emph{{A Model Independent Approach to
  Inelastic Dark Matter Scattering}},
  \href{http://dx.doi.org/10.1103/PhysRevD.90.094027}{\emph{Phys.Rev.} {\bf
  D90} (2014) 094027}, [\href{https://arxiv.org/abs/1409.0536}{{\tt
  1409.0536}}].

\bibitem{Catena:2015uua}
R.~Catena and P.~Gondolo, \emph{{Global limits and interference patterns in
  dark matter direct detection}},
  \href{http://dx.doi.org/10.1088/1475-7516/2015/08/022}{\emph{JCAP} {\bf 1508}
  (2015) 022}, [\href{https://arxiv.org/abs/1504.06554}{{\tt 1504.06554}}].

\bibitem{Schneck:2015eqa}
{\scshape SuperCDMS} collaboration, K.~Schneck et~al., \emph{{Dark matter
  effective field theory scattering in direct detection experiments}},
  \href{http://dx.doi.org/10.1103/PhysRevD.91.092004}{\emph{Phys. Rev.} {\bf
  D91} (2015) 092004}, [\href{https://arxiv.org/abs/1503.03379}{{\tt
  1503.03379}}].

\bibitem{Dent:2015zpa}
J.~B. Dent, L.~M. Krauss, J.~L. Newstead and S.~Sabharwal, \emph{{General
  analysis of direct dark matter detection: From microphysics to observational
  signatures}}, \href{http://dx.doi.org/10.1103/PhysRevD.92.063515}{\emph{Phys.
  Rev.} {\bf D92} (2015) 063515}, [\href{https://arxiv.org/abs/1505.03117}{{\tt
  1505.03117}}].

\bibitem{Catena:2015vpa}
R.~Catena, \emph{{Dark matter directional detection in non-relativistic
  effective theories}},
  \href{http://dx.doi.org/10.1088/1475-7516/2015/07/026}{\emph{JCAP} {\bf 1507}
  (2015) 026}, [\href{https://arxiv.org/abs/1505.06441}{{\tt 1505.06441}}].

\bibitem{Kavanagh:2015jma}
B.~J. Kavanagh, \emph{{New directional signatures from the nonrelativistic
  effective field theory of dark matter}},
  \href{http://dx.doi.org/10.1103/PhysRevD.92.023513}{\emph{Phys. Rev.} {\bf
  D92} (2015) 023513}, [\href{https://arxiv.org/abs/1505.07406}{{\tt
  1505.07406}}].

\bibitem{D'Eramo:2016atc}
F.~D'Eramo, B.~J. Kavanagh and P.~Panci, \emph{{You can hide but you have to
  run: direct detection with vector mediators}},
  \href{http://dx.doi.org/10.1007/JHEP08(2016)111}{\emph{JHEP} {\bf 08} (2016)
  111}, [\href{https://arxiv.org/abs/1605.04917}{{\tt 1605.04917}}].

\bibitem{Catena:2016hoj}
R.~Catena, A.~Ibarra and S.~Wild, \emph{{DAMA confronts null searches in the
  effective theory of dark matter-nucleon interactions}},
  \href{http://dx.doi.org/10.1088/1475-7516/2016/05/039}{\emph{JCAP} {\bf 1605}
  (2016) 039}, [\href{https://arxiv.org/abs/1602.04074}{{\tt 1602.04074}}].

\bibitem{Kahlhoefer:2016eds}
F.~Kahlhoefer and S.~Wild, \emph{{Studying generalised dark matter interactions
  with extended halo-independent methods}},
  \href{https://arxiv.org/abs/1607.04418}{{\tt 1607.04418}}.

\bibitem{Toivanen:2008zz}
P.~Toivanen, M.~Kortelainen, J.~Suhonen and J.~Toivanen, \emph{{Dark-matter
  detection by elastic and inelastic LSP scattering on Xe-129 and Xe-131}},
  \href{http://dx.doi.org/10.1016/j.physletb.2008.06.057}{\emph{Phys. Lett.}
  {\bf B666} (2008) 1--4}.

\bibitem{Hoferichter:2015ipa}
M.~Hoferichter, P.~Klos and A.~Schwenk, \emph{{Chiral power counting of one-
  and two-body currents in direct detection of dark matter}},
  \href{http://dx.doi.org/10.1016/j.physletb.2015.05.041}{\emph{Phys. Lett.}
  {\bf B746} (2015) 410--416}, [\href{https://arxiv.org/abs/1503.04811}{{\tt
  1503.04811}}].

\bibitem{Hoferichter:2016nvd}
M.~Hoferichter, P.~Klos, J.~Menéndez and A.~Schwenk, \emph{{Analysis
  strategies for general spin-independent WIMP-nucleus scattering}},
  \href{http://dx.doi.org/10.1103/PhysRevD.94.063505}{\emph{Phys. Rev.} {\bf
  D94} (2016) 063505}, [\href{https://arxiv.org/abs/1605.08043}{{\tt
  1605.08043}}].

\bibitem{Brown:2014}
B.~A. Brown and W.~D.~M. Rae, \emph{{The Shell-Model Code NuShellX@MSU}},
  {\emph{Nuclear Data Sheets} {\bf 120} (2014) 115--118}.

\bibitem{Brown:2001zz}
B.~Brown, \emph{{The nuclear shell model towards the drip lines}},
  \href{http://dx.doi.org/10.1016/S0146-6410(01)00159-4}{\emph{Prog.Part.Nucl.Phys.}
  {\bf 47} (2001) 517--599}.

\bibitem{Lee:2013wza}
S.~K. Lee, M.~Lisanti, A.~H.~G. Peter and B.~R. Safdi, \emph{{Effect of
  Gravitational Focusing on Annual Modulation in Dark-Matter Direct-Detection
  Experiments}},
  \href{http://dx.doi.org/10.1103/PhysRevLett.112.011301}{\emph{Phys. Rev.
  Lett.} {\bf 112} (2014) 011301}, [\href{https://arxiv.org/abs/1308.1953}{{\tt
  1308.1953}}].

\bibitem{Kouvaris:2015xga}
C.~Kouvaris and N.~G. Nielsen, \emph{{Daily modulation and gravitational
  focusing in direct dark matter search experiments}},
  \href{http://dx.doi.org/10.1103/PhysRevD.92.075016}{\emph{Phys. Rev.} {\bf
  D92} (2015) 075016}, [\href{https://arxiv.org/abs/1505.02615}{{\tt
  1505.02615}}].

\bibitem{Green:2011bv}
A.~M. Green, \emph{{Astrophysical uncertainties on direct detection
  experiments}}, \href{http://dx.doi.org/10.1142/S0217732312300042}{\emph{Mod.
  Phys. Lett.} {\bf A27} (2012) 1230004},
  [\href{https://arxiv.org/abs/1112.0524}{{\tt 1112.0524}}].

\bibitem{Piffl:2013mla}
T.~Piffl et~al., \emph{{The RAVE survey: the Galactic escape speed and the mass
  of the Milky Way}},
  \href{http://dx.doi.org/10.1051/0004-6361/201322531}{\emph{Astron.
  Astrophys.} {\bf 562} (2014) A91},
  [\href{https://arxiv.org/abs/1309.4293}{{\tt 1309.4293}}].

\bibitem{Lundberg:2004dn}
J.~Lundberg and J.~Edsjo, \emph{{WIMP diffusion in the solar system including
  solar depletion and its effect on earth capture rates}},
  \href{http://dx.doi.org/10.1103/PhysRevD.69.123505}{\emph{Phys. Rev.} {\bf
  D69} (2004) 123505}, [\href{https://arxiv.org/abs/astro-ph/0401113}{{\tt
  astro-ph/0401113}}].

\bibitem{Geochemistry}
W.~Mcdonough, \emph{Treatise on Geochemistry}, vol.~2.
\newblock Elsevier, 2003.

\bibitem{Britannica}
``{The Earth: its properties, composition, and structure}.'' Britannica CD,
  Version 99, {Encyclop{\ae}dia Britannica, Inc.} \copyright, 1994--1999.

\bibitem{Morokoff1995}
W.~J. Morokoff and R.~E. Caflisch, \emph{Quasi-monte carlo integration},
  \href{http://dx.doi.org/10.1006/jcph.1995.1209}{\emph{Journal of
  Computational Physics} {\bf 122} (dec, 1995) 218--230}.

\bibitem{Li:2014rca}
J.~Li, X.~Ji, W.~Haxton and J.~S.~Y. Wang, \emph{{The second-phase development
  of the China JinPing underground Laboratory}},
  \href{http://dx.doi.org/10.1016/j.phpro.2014.12.055}{\emph{Phys. Procedia}
  {\bf 61} (2015) 576--585}, [\href{https://arxiv.org/abs/1404.2651}{{\tt
  1404.2651}}].

\bibitem{Indumathi:2015hfa}
{\scshape INO} collaboration, D.~Indumathi, \emph{{India-based neutrino
  observatory (INO): Physics reach and status report}},
  \href{http://dx.doi.org/10.1063/1.4915571}{\emph{AIP Conf. Proc.} {\bf 1666}
  (2015) 100003}.

\bibitem{Urquijo:2016dxd}
P.~Urquijo, \emph{{Searching for Dark Matter at the Stawell Underground Physics
  Laboratory}}, \href{http://dx.doi.org/10.1051/epjconf/201612304002}{\emph{EPJ
  Web Conf.} {\bf 123} (2016) 04002},
  [\href{https://arxiv.org/abs/1605.03299}{{\tt 1605.03299}}].

\bibitem{Strauss:2016sxp}
R.~Strauss et~al., \emph{{The CRESST-III low-mass WIMP detector}},
  \href{http://dx.doi.org/10.1088/1742-6596/718/4/042048}{\emph{J. Phys. Conf.
  Ser.} {\bf 718} (2016) 042048}.

\bibitem{Agnese:2016cpb}
{\scshape SuperCDMS} collaboration, R.~Agnese et~al., \emph{{Projected
  Sensitivity of the SuperCDMS SNOLAB experiment}}, {\emph{Submitted to: Phys.
  Rev. D} (2016) }, [\href{https://arxiv.org/abs/1610.00006}{{\tt
  1610.00006}}].

\bibitem{Aguilar-Arevalo:2016ndq}
{\scshape DAMIC} collaboration, A.~Aguilar-Arevalo et~al., \emph{{Search for
  low-mass WIMPs in a 0.6 kg day exposure of the DAMIC experiment at SNOLAB}},
  \href{http://dx.doi.org/10.1103/PhysRevD.94.082006}{\emph{Phys. Rev.} {\bf
  D94} (2016) 082006}, [\href{https://arxiv.org/abs/1607.07410}{{\tt
  1607.07410}}].

\bibitem{Civitarese:2016uuc}
O.~Civitarese, K.~J. Fushimi and M.~E. Mosquera, \emph{{Calculated WIMP signals
  at the ANDES laboratory: comparison with northern and southern located dark
  matter detectors}},  \href{https://arxiv.org/abs/1611.00802}{{\tt
  1611.00802}}.

\bibitem{Emken}
T.~Emken and C.~Kouvaris. \textit{in preparation}.

\bibitem{Cerdeno:2010jj}
D.~G. Cerdeno and A.~M. Green, \emph{{Direct detection of WIMPs}},
  \href{https://arxiv.org/abs/1002.1912}{{\tt 1002.1912}}.

\bibitem{Feldman:1997qc}
G.~J. Feldman and R.~D. Cousins, \emph{{A Unified approach to the classical
  statistical analysis of small signals}},
  \href{http://dx.doi.org/10.1103/PhysRevD.57.3873}{\emph{Phys. Rev.} {\bf D57}
  (1998) 3873--3889}, [\href{https://arxiv.org/abs/physics/9711021}{{\tt
  physics/9711021}}].

\end{thebibliography}\endgroup

\end{document}